\documentclass{article}

\usepackage{arxiv}

\usepackage[utf8]{inputenc} % allow utf-8 input
\usepackage[T1]{fontenc}    % use 8-bit T1 fonts
\usepackage{hyperref}       % hyperlinks
\usepackage{url}            % simple URL typesetting
\usepackage{booktabs}       % professional-quality tables
\usepackage{amsfonts}       % blackboard math symbols
\usepackage{nicefrac}       % compact symbols for 1/2, etc.
\usepackage{microtype}      % microtypography
\usepackage{cleveref}       % smart cross-referencing
\usepackage{lipsum}         % Can be removed after putting your text content
\usepackage{graphicx}
\usepackage[numbers]{natbib}
\usepackage{doi}

\usepackage{forest}
\usetikzlibrary{shadows.blur}
\usetikzlibrary{arrows.meta,shapes,positioning,shadows,trees}
\usepackage{rotating}
\usepackage{lscape}
\usepackage{tabularx}
\usepackage{longtable,pdflscape,booktabs}
\usepackage{caption}
\usepackage{array}

\usepackage{colortbl}
\usepackage{xcolor}
\colorlet{updated}{black}

\tikzset{
    basic/.style  = {draw, text width=2cm, drop shadow, font=\sffamily, rectangle},
    root/.style   = {basic, thin, align=center, fill=black, text=white},
    onode/.style = {basic, thin, align=center, fill=gray!60,text width=5cm,},
    tnode/.style = {basic, thin, align=left, fill=white, text width=15.5em},
    edge from parent/.style={draw=black, edge from parent fork right}
}

\title{The Landscape of User-centered Misinformation Interventions - A Systematic Literature Review}

\author{Katrin Hartwig\\
	Science and Technology for Peace and Security (PEASEC)\\
	Technical University of Darmstadt\\
	Pankratiusstraße 2, 64285 Darmstadt, Germany \\
	\texttt{hartwig@peasec.tu-darmstadt.de} \\
	\And
	Frederic Doell \\
	Science and Technology for Peace and Security (PEASEC)\\
	Technical University of Darmstadt\\
	Pankratiusstraße 2, 64285 Darmstadt, Germany \\
	\texttt{frederic.doell@stud.tu-darmstadt.de} \\
    \AND
    Christian Reuter\\
    Science and Technology for Peace and Security (PEASEC)\\
	Technical University of Darmstadt\\
	Pankratiusstraße 2, 64285 Darmstadt, Germany \\
	\texttt{reuter@peasec.tu-darmstadt.de} \\
}

\renewcommand{\shorttitle}{\textit{arXiv} Template}

\hypersetup{
pdftitle={A template for the arxiv style},
pdfsubject={q-bio.NC, q-bio.QM},
pdfauthor={David S.~Hippocampus, Elias D.~Striatum},
pdfkeywords={First keyword, Second keyword, More},
}

\begin{document}
\maketitle

\begin{abstract}
Misinformation is one of the key challenges facing society today.
User-centered misinformation interventions as digital countermeasures that exert a direct influence on users represent a promising means to deal with the large amounts of information available.
While an extensive body of research on this topic exists, researchers are confronted with a diverse research landscape spanning multiple disciplines. 
This review systematizes the landscape of user-centered misinformation interventions to facilitate knowledge transfer, identify trends, and enable informed decision-making.
Over 5,700 scholarly publications were screened and a systematic literature review ($N=163$) was conducted. A taxonomy was derived regarding intervention design (e.g., (binary) label), user interaction (active or passive), and timing (e.g., post exposure to misinformation). 
We provide a structured overview of approaches across multiple disciplines, and derive six overarching challenges for future research.
\end{abstract}

% keywords can be removed
\keywords{misinformation, disinformation, fake news, user intervention, countermeasure, media literacy}

\section{Introduction}
\label{sec:intro}
The fast spread of misinformation is an enormous challenge both for society and individuals, with a great impact on democracy.
Severe and fatal consequences can be observed in relation to misinformation shared on social media related to COVID-19, with mistrust sowed in health measures required for combating a pandemic. With this in mind, \citet{pennycook_fighting_2020} even go so far as calling it a `matter of life and death'. 
In light of the grave consequences, the need for digital misinformation interventions to slow down its propagation is evident. Those technical approaches can be divided roughly into two main steps \citep{potthast_stylometric_2018}: The (automatic) detection of misinformation and, second, the implementation of countermeasures as a concrete decision on what to do after successful detection.
A great deal of research exists on detecting misinformation, which is often based on machine learning algorithms (e.g., \citep{helmstetter_weakly_2018, brasoveanu_semantic_2019, shu_fake_2017}). Such algorithms are typically so-called `black box' algorithms, which -- while producing promising results with regard to detection accuracy -- are not transparent in their reasoning. In order to make an algorithm's decisions transparent to users, interventions may profit from using `white box' algorithms/explainable AI, which give greater insights into how the algorithm behaves and what variables influence the model \citep{chen_spread_2022}. The reliance on these automatic detection measures is increasing. For example, because of the COVID-related increase in traffic, Twitter (now X) has increased its use of machine learning and automation against misinformation \citep{gadde_update_2020}.
Complementary research has been done on the implementation of concrete interventions after successful automatic detection (e.g., \citep{bhuiyan_nudgecred_2021, roozenbeek_fake_2019,saltz_encounters_2021}). Those interventions are available in a wide range: Some aim at efficiently and automatically deleting content before exposure, while others try to educate users by showing corrections or flagging problematic content. While there is a large and heterogeneous field of interventions, they have a direct impact on end users of social media, as they focus on whether and how to communicate their output and findings, for instance, via information visualization. Although promising approaches have been established, the ongoing challenge of users being confronted and influenced by misinformation on  diverse social media platforms such as TikTok, Twitter/X, Instagram, Facebook, and Co. suggests a need for further systematic design, implementation, and evaluation of effective digital interventions.

This review study aims to systematize knowledge on digital user-centered misinformation interventions.
The term `misinformation' is often used as an umbrella term for better readability, encompassing misleading information that has been created deliberately (frequently referred to as `disinformation' or `fake news') as well as unintentionally (frequently referred to as `misinformation') \citep{chen_spread_2022, wang_systematic_2019, almaliki_online_2019}. 
For example, \citet{li_health_2022} justify the use of misinformation as an umbrella term by stating that the majority of studies use this term to encompass different types of misleading information in general without denying the proper distinction between disinformation and misinformation.
Indeed, misinformation and related phenomena such as rumors and conspiracy theories can all lead to severe consequences, even if those were not intended. Thus, in accordance with other research and systematic reviews \citep{chen_spread_2022, li_health_2022}, in this paper, we will use `misinformation' for better readability and to allow for a broader perspective on different kinds of misleading information while not denying the significant differences of phenomena.
While the term `misinformation intervention' has already been established by other researchers \citep{bak-coleman_combining_2022,saltz_encounters_2021,saltz_misinformation_2021}, we define \textit{user-centered misinformation interventions} as digital countermeasures that go beyond a purely algorithmic back-end solution and exert a direct influence on the user in the form of information presentation or information withholding. Accordingly, we do not include approaches that deal exclusively with the automatic detection of misinformation without describing the subsequent communication to the user. 
 
We provide a taxonomy that classifies and aggregates interventions regarding multiple relevant dimensions, such as time of intervention, addressed platform, and the thorough differentiation between intervention categories (e.g., correction, (binary) labeling, transparent indicators) to help identify promising research directions and encourage cross-disciplinary transferability. Researchers are faced with a very diverse research landscape on user-centered countermeasures, which is spread across multiple disciplines, such as computer science, human-computer interaction, information systems, psychology, communication sciences, journalism, and even medical research. Hence, we address the challenge of gaining an overview and help build on existing research while considering and learning from current insights of different relevant disciplines as well as research on different social media platforms. Thereby, we seek to facilitate informed decision-making of researchers and practitioners when analyzing, designing, and evaluating (novel) digital countermeasures to combat misinformation.
We are especially interested in approaches communicating to users how an algorithm arrives at its results (e.g., white box algorithms) instead of giving a top-down answer (e.g., misleading, not misleading) without explanation. In our paper, we understand a `transparent' intervention as an intervention that allows for informed decisions and the ability to comprehend why the content potentially contains misinformation, for example, via explanations of varying degrees, and can, thus, be considered as more user-centered than top-down interventions. There is evidence that transparently assisting users in their own assessment of misinformation is more promising than a top-down approach that provides social media posts solely with a label stating `This is/isn't misinformation' without cues to help comprehend the decision \citep{kirchner_countering_2020} or simply removes misinformation \citep{atreja_remove_2023}. Research indicated that giving explanations or comprehensible cues can be significant to establish trust in the intervention \citep{kirchner_countering_2020}, and counteract feelings of reactance or related backfire effects \citep{nyhan_when_2010} that are controversially discussed in research \citep{wood_elusive_2019}.

While the topic of misinformation has been studied in systematic reviews, e.g. regarding specific contexts such as health \citep{li_health_2022} or political misinformation \citep{jerit_political_2020}, existing literature reviews on interventions against misinformation and similar phenomena focus on more general overviews. For example, a related literature map by \citet{almaliki_online_2019} focuses on the research field of misinformation. It provides a general overview rather than analyzing the characteristics of concrete interventions and comparing the different approaches. They state that ``less than 2\% of the selected papers proposed digital intervention techniques'', while our focus lies on those studies that fall into the 2\% as a promising subgroup of interventions with growing research interest. Furthermore, when literature reviews, or meta-analyses deal with concrete interventions, they often focus on the detection step and machine learning interventions (e.g., \citep{caled_digital_2021, rana_review_2018,xu_unified_2021,guo_future_2021,zhou_survey_2021,zubiaga_detection_2019,mirsky_creation_2022}) instead of user-centered interventions or focus on a specific subgroup like corrections \citep{chan_meta-analysis_2023, prike_effective_2023}, warnings \citep{mende_fighting_2024}, accuracy prompts \citep{pennycook_accuracy_2022}, or contexts like COVID-19 \citep{janmohamed_interventions_2021}. A first systematic overview of strategies against misinformation, including countermeasures with a direct influence on end users, was given by \citet{chen_spread_2022}, who differentiate between five broad categories of solutions according to communication elements: message-based, source-based, network-based, policy-based, and education-based approaches. This is in contrast to our approach which is not based on communication elements (e.g., message versus source) but rather considers interventions more in terms of their in-depth design. This design can be applied to the content itself, within a network-based approach, or to sources (e.g., by highlighting components in color as a passive intervention during exposure to misinformation). The authors give an overview of exemplary implementations within the four clusters. We build on that by providing an in-depth analysis of the design, interaction type, and timing of interventions as central aspects for user-centered implementations. In addition, we provide an overview of the methodological characteristics of intervention studies. 
Furthermore, \citet{aghajari_reviewing_2023} reviewed misinformation interventions with a focus on underlying driving factors of misinformation like social contexts and beliefs. Thus, in a constrained search process around the term `misinformation', they categorize interventions according to content-based, source-based, individual user-based, and community-based strategies. Our study complements the analysis of strategies in terms of their driving factors (e.g., content-based strategies like corrections) with an HCI perspective detached from individualistic or community-based emphasis. 
Differentiating misinformation interventions on an individual trial system level, \citet{roozenbeek_countering_2023} review boosting interventions, nudging, debunking, and content labeling in comparison to interventions that rely on algorithms, business models, legislation, and politics.
We complement the findings of related reviews by (a) shedding light on user-centered aspects of concrete misinformation interventions by performing an in-depth analysis of their design, implementation, and methodological evaluation for a broad perspective that offers a more comprehensive understanding of misinformation interventions. Thereby, we (b) specifically discuss and categorize characteristics impacting end users, such as the intervention design, user interaction, and timing of the intervention. We further complement the existing research landscape by (c) performing a review on publications of diverse disciplines and not limited to a specific period until 2024, searching three major databases. In doing so, we address calls for future work on a review including multiple disciplines and phenomena \citep{aghajari_reviewing_2023}, and, when combined with findings of existing reviews, we provide a different perspective and more nuanced understanding of the research landscape. To our knowledge, a systematization of knowledge on specific user-centered misinformation interventions has yet not been conducted to this extent and with this perspective.

Our overarching goal is to deeply examine and classify misinformation intervention studies in terms of methodological characteristics in study design and evaluation, content characteristics of user interventions, and derived trends and challenges for future research.
Addressing that goal, all our considerations lead us to the following research questions:

\begin{enumerate}
    \item[RQ1:] \textit{What are the typical methodological characteristics of existing studies on misinformation interventions?}
    \item[RQ2:] \textit{How do existing forms of user-centered misinformation interventions assist users in dealing with misinformation online?}
    \item[RQ3:] \textit{Which trends and chances for future research can be derived from the existing literature?}
\end{enumerate}

The paper is structured as follows: First, we present our methodology of a systematic literature review and the procedure of deriving a taxonomy of user-centered misinformation interventions (see Section \ref{sec:meth}). Then, we present our results, including methodological aspects of analyzed publications such as addressed formats and platforms, applied methods of user studies, sample size, and participant details (see Section \ref{sec:methodologicaloverviews}). Then we present our taxonomy (see Section \ref{sec:taxo}), distinguishing nine intervention designs, active versus passive user interaction, and five points in time at which an intervention can be applied. We further discuss transparency as a specific measure to facilitate users in dealing autonomously with misinformation (see Section \ref{sec:transparen}). Lastly, we present `nudging' as a concept applied in many extracted publications (see Section \ref{sec:digitnudge}). 
In Section \ref{sec:disc}, we answer our research questions regarding the design of user-centered misinformation interventions, methodological characteristics of existing studies, and the derived trends, open questions, and challenges for future research.

\section{Methodology}
\label{sec:meth}
In this section, we present our methodological approach of performing a systematic literature review, comprising of the identification and screening of relevant literature (Section \ref{sec:identificationlit}) and the thorough analysis and structuring of publications and interventions included therein (Section \ref{sec:develtaxo}).
\subsection{Identification of Literature}
\label{sec:identificationlit}
\begin{figure}
    \includegraphics[width=0.6\textwidth]{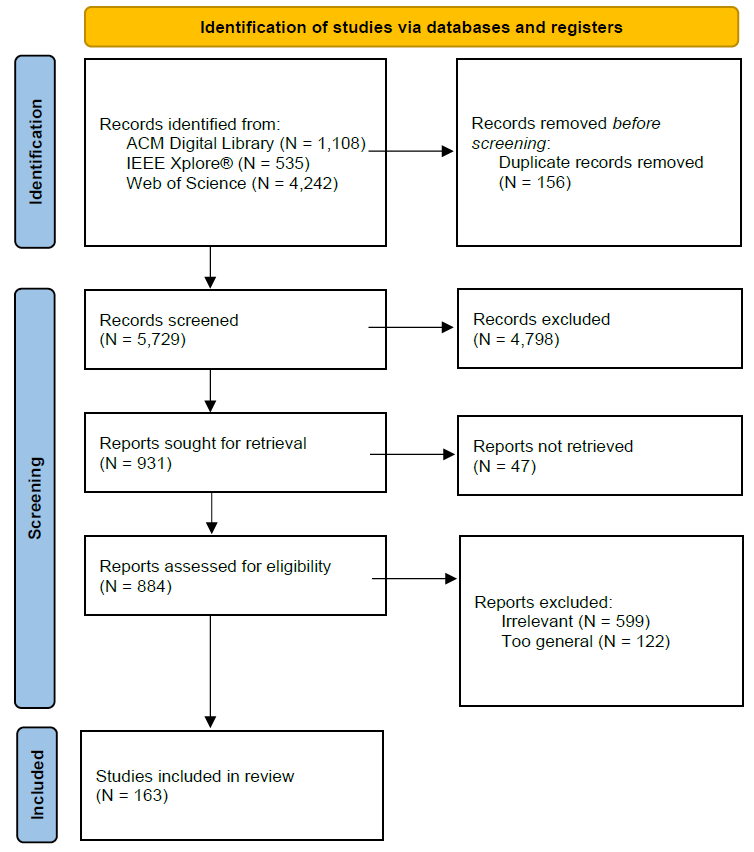}
    \caption{PRISMA Diagram demonstrating our data flow within the systematic literature review (created with the template from \citep{page_prisma_2021})}
    \label{fig:prisma_diagram}
\end{figure}

To identify and categorize relevant literature on misinformation interventions, we performed a systematic literature review, following the PRISMA guidelines \citep{page_prisma_2021} (see Figure \ref{fig:prisma_diagram}). \citet{schryen_knowledge_2020} state that literature reviews are important for ``developing domain knowledge'' and to identify knowledge-building activities, such as synthesizing, aggregating evidence, criticizing, theory building, identifying research gaps, and developing a research agenda. In accordance with these principles, we set up our literature search as follows: The initial search spans the ACM Digital Library, Web of Science, as well as the IEEE Xplore database. With this set of databases, we encompass a broad corpus of diverse literature as well as the ten conferences and journals listed by Google Scholar as the best regarding human-computer interaction: ACM Conference on Human Factors in Computing Systems (CHI), IEEE Transactions on Affective Computing, Proceedings of the ACM on Interactive Mobile, Wearable and Ubiquitous Technologies (IMWUT),  Proceedings of the ACM on Human-Computer Interaction (PACM), International Journal of Human-Computer Studies (IJHCS), ACM/IEEE International Conference on Human Robot Interaction (HRI), ACM Conference on Computer-Supported Cooperative Work \& Social Computing (CSCW), IEEE Transactions on Human-Machine Systems, Behaviour \& Information Technology (BIT), and the ACM Symposium on User Interface Software and Technology (UIST). The search took place in {\color{updated}February 2024}, and only papers that were published until that date could be considered. All publication years up to this date have been included, however, there were no relevant publications for our final set concerning content-wise inclusion and exclusion criteria before 2011.

The search term consists of two parts: The first part are terms included in or related to our umbrella term of `misleading information'. The second part includes synonyms for `intervention' and related concepts addressing user-centered measures. Only papers containing at least one term of each part in their title or abstract were included in our search. We did not filter for a publication year. The complete search term is the following:

\textit{((rumour* OR rumor* OR "misleading
information" OR "fake news" OR "false news"
OR misinformation OR disinformation OR "news credibility")
AND (combat* OR correct* OR interven* OR countermeasur* OR
counteract* OR treatment OR relief OR educat* OR warning OR nudg* OR
user-centered OR "media literacy"))
}

The term was modified to adhere to database requirements and to run comparable searches. Furthermore, because Web of Science returned many results, the term was adjusted to exclude obviously irrelevant disciplines (e.g., chemistry). The broad interdisciplinary nature of Web of Science explains its large amount of `false positives' during the initial search in comparison to the other two databases that already focus on disciplines relevant to digital misinformation interventions (e.g., computing and information technology). The search returned 1,108 results from the ACM Digital Library, 535 results from IEEE Xplore and 4,242 results from Web of Science, in total 5,885 results. After removing 156 duplicates, we screened 5,729 records, of which we excluded 4,798.

Records were removed for a multitude of reasons: First, some words of the search term had multiple meanings, which is why papers using a different interpretation were excluded. While we decided to take a broad perspective on diverse kinds of misleading information, including misinformation, disinformation, rumors, and related phenomena (e.g., conspiracy theories), as all types can have severe consequences, papers were excluded when the phenomenon investigated was not referring to our definition of the umbrella term `misinformation' (see Section \ref{sec:intro}). This was particularly the case for misinformation referring to eyewitnesses remembering something inaccurately (e.g., due to suggestion) during a testimony in court. For phenomena included in our broad definition, there was a variety of terms included in our sample (see row `Concept' in Table \ref{tab:sokfinaltable}). Furthermore, some technical terms have different meanings in different fields, for example, network science uses the term rumor in the context of nodes spreading information (e.g., \citep{buchegger_effect_2003}).
Second, when papers concentrated solely on the technical detection step with no involvement of the user at all, e.g., machine learning approaches focusing on increasing the detection rate, they were excluded.
Furthermore, interventions that took a network-based approach, for example, by simulating which nodes to delete in order to reduce the spread of misinformation, were also excluded.
In addition, we excluded psychological experiments without concrete reference to misinformation as well as surveys and questionnaires exploring background information (e.g., Which demographics are susceptible to misinformation?).
Additionally, we decided to exclude reviews.

931 papers were sought for retrieval, of which 47 were removed because they could not be accessed, leaving us with a total of 884 publications that were assessed for eligibility. In the last step, a total of 721 papers were finally sorted out, 599 thereof because of the aforementioned criteria. Another 122 papers were excluded because they were too general, including papers that did not meet this review's focus because the intervention occurred long before the actual usage of social media, like educational school lectures, trainings, or serious games, but also implementations that focus exclusively on psychological phenomena (e.g., Do corrections of content generate reactance?). 
Particularly in the context of corrections or debunking of misinformation (e.g., in comment sections), there is a lot of in-depth research on factors impacting user reactions and interactions. Often, these studies focus on psychological or social phenomena that are particularly valuable to consider when designing interventions tailored to a specific persona. To receive a reasonable number of publications and thus allow for a thorough focus on research regarding the design and evaluation of digital interventions, we decided to exclude studies that rather address (psychological or social) impact factors without a particular focus on intervention design and evaluation.
The final set of papers contained 163 items which were included in our analysis and were categorized according to our taxonomy.
\subsection{Development of a Taxonomy}
\label{sec:develtaxo}
For the development of the taxonomy, we first collected different relevant dimensions to compare and differentiate studies on user interventions within the context of misinformation. We developed and applied those categories in an iterative process of brainstorming sessions with two researchers with expertise in computer science, psychology, and human-computer interaction based on already familiar studies within the field of interest (e.g., \citep{bhuiyan_nudgecred_2021, kirchner_countering_2020}). The coding process was initiated by a training phase where a common understanding of each category was obtained. When disagreeing on a categorization during the coding phase, the study was discussed to achieve a consensus. This approach of consensus coding is commonly applied in other research \citep{wilson_assembling_2018}. First, we defined our target group: researchers and practitioners interested in analyzing, designing, and evaluating digital countermeasures to combat misinformation that may potentially benefit from our taxonomy. To assist the target groups, several characteristics are particularly relevant as they provide information on (1) the intervention design, (2) the form of user interaction, and (3) the timing of the intervention. Categories were complemented and adjusted iteratively while identifying and reading new papers. For instance, when reading multiple papers that differed regarding the time of intervention, this category was included, and all relevant papers were categorized accordingly. Additionally, minor modifications to the categories were made during the process of reading and categorizing the articles when deemed necessary. Further, we looked at how user-centered interventions were categorized in other contexts with sensible information (e.g., cybersecurity) in systematic reviews \citep{franz_sok_2021}. The resulting final table can be found in the {color{updated}electronic supplement} (see Table \ref{tab:sokfinaltable}). 
A study can be sorted into several categories, and the subcategories are generally not mutually exclusive (e.g., some interventions may combine the intervention categories `highlighting design' and `(binary) label' and others compare a `correction' with `showing indicators').

\section{Results: The Landscape of User-centered Misinformation Interventions}
\label{sec:res}
In this section, a detailed analysis of the literature review is presented. First, we give an overview regarding \textit{methodologies} used by studies on user interventions (see Section \ref{sec:methodologicaloverviews}). We then provide a \textit{taxonomy of interventions to assist users in dealing with misinformation} by categorizing and clustering the identified research sample in distinct dimensions (see Section \ref{sec:taxo}). 
Furthermore, we highlight how \textit{transparency} (see Section \ref{sec:transparen}) is used to assist users in dealing autonomously with misinformation, present the concept of \textit{digital nudging} (see Section \ref{sec:digitnudge}) as a trending digital countermeasure, and finally discuss the impact and perceptions of reviewed misinformation interventions (see Section \ref{sec:impactperceptions}).

\subsection{Methodological Characteristics}
\label{sec:methodologicaloverviews}
\begin{figure}
    \includegraphics[width=\textwidth]{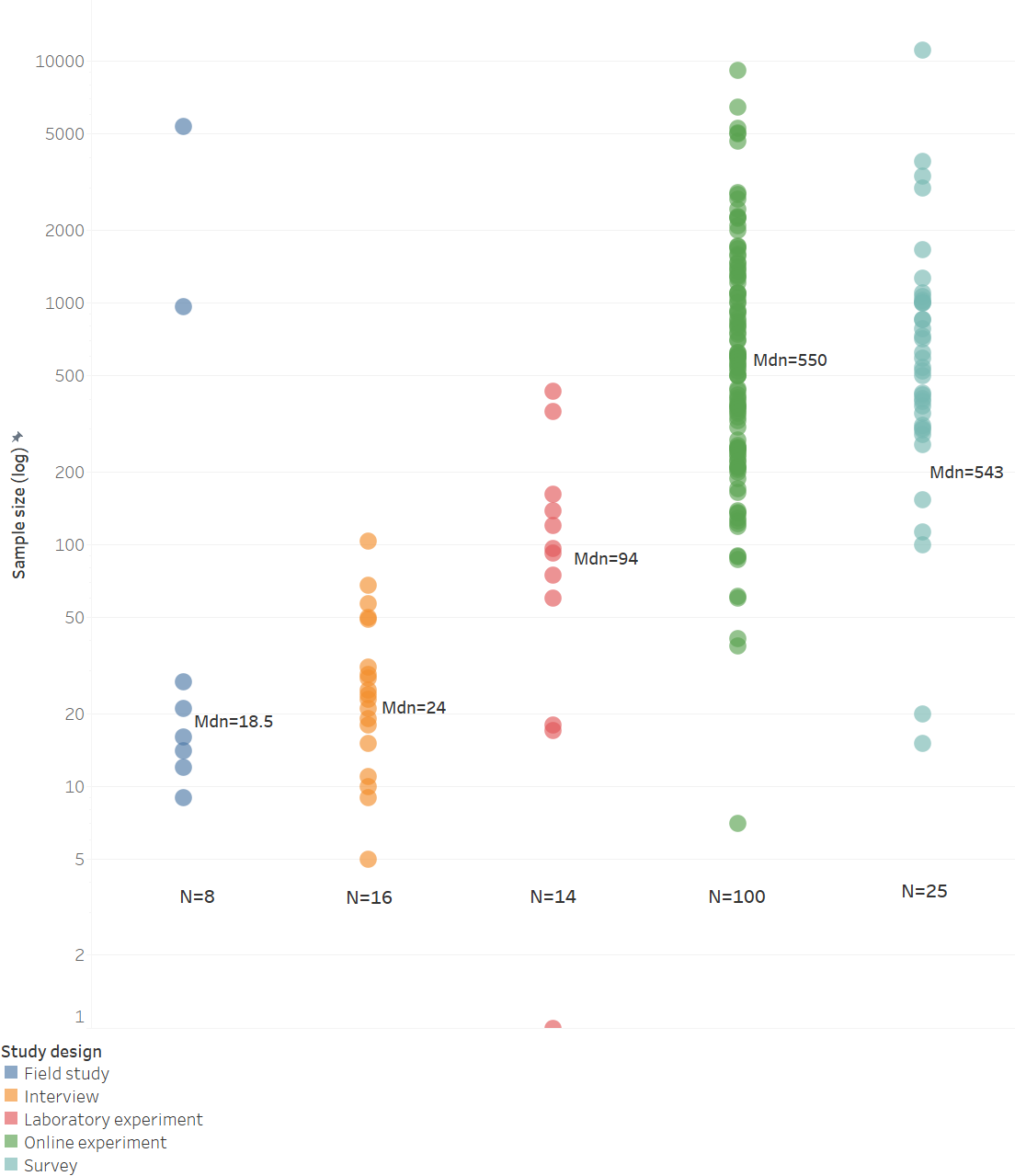}
    \caption{Sample sizes (log) of the individual studies broken down by study type including the median (Mdn). Note that a publication often contains multiple user studies.}
    \label{fig:samplesize}
\end{figure}

To provide an overview of research methods typically used in the field of user-centered interventions to assist in dealing with misinformation, details of the respective study designs were collected. 
All studies were published between 2011 and 2024.
First, we were interested in the different concepts included under the umbrella term `misleading information'. In total, 141 publications referred to either \textit{misinformation, disinformation or misleading information}. Furthermore, 5 publications were specifically interested in \textit{rumors}, and 10 publications that addressed the concept of \textit{news credibility}. Other publications referred to myths, propaganda, or controversial topics.
Out of a total of 163 included papers, 17 present exclusively \textit{conceptual} ideas of interventions. In contrast, the remaining studies collected empirical data in the form of \textit{laboratory experiments} (14 publications), \textit{{\color{updated}online} experiments} (100 publications), \textit{field studies} ({\color{updated}8} publications), \textit{surveys} ({\color{updated}25} publications), and \textit{interviews} ({\color{updated}16} publications). {\color{updated}In our study, we understand a field study as an evaluation type that specifically observes the natural behavior of participants in a real-world scenario, in contrast to experiments that encompass a controlled setting designed by the researchers. In the context of misinformation, research experiments rarely take place within an actual lab of the researcher (lab experiment) but typically remotely in an online setting (online experiment), for instance, as a link to the researcher's experimental website, as these experiments often do not require physical presence for additional hardware items.} In some cases, there is a combination, e.g., of survey and interview or of laboratory experiment and {\color{updated}online} experiment within one publication.
Regarding sample size, the empirical studies range from small groups of participants (<20 e.g., \citep{brashier_timing_2021,challenger_covid-19_2022,gao_label_2018,komendantova_value-driven_2021}) to large-scaled representative groups with far over 1,000 participants (e.g., \citep{autry_correcting_2021,jia_understanding_2022,seo_trust_2019}). You can find a visualization of sample sizes in Figure \ref{fig:samplesize}. {\color{updated}A closer look at the participants reveals a clear bias, with the majority of students reporting having U.S. adults and college students as participants.} However, there are also isolated studies that either address a very specific (vulnerable) target group (e.g., teenagers \citep{axelsson_learning_2021}, marginalized communities \citep{velez_latino-targeted_2023}, or {\color{updated}blind/low vision social media users \citep{sharevski_i_2023}}) or make a comparison between several countries (e.g., \citep{almaliki_misinformation-aware_2019}).

\begin{figure}
    \includegraphics[width=\textwidth]{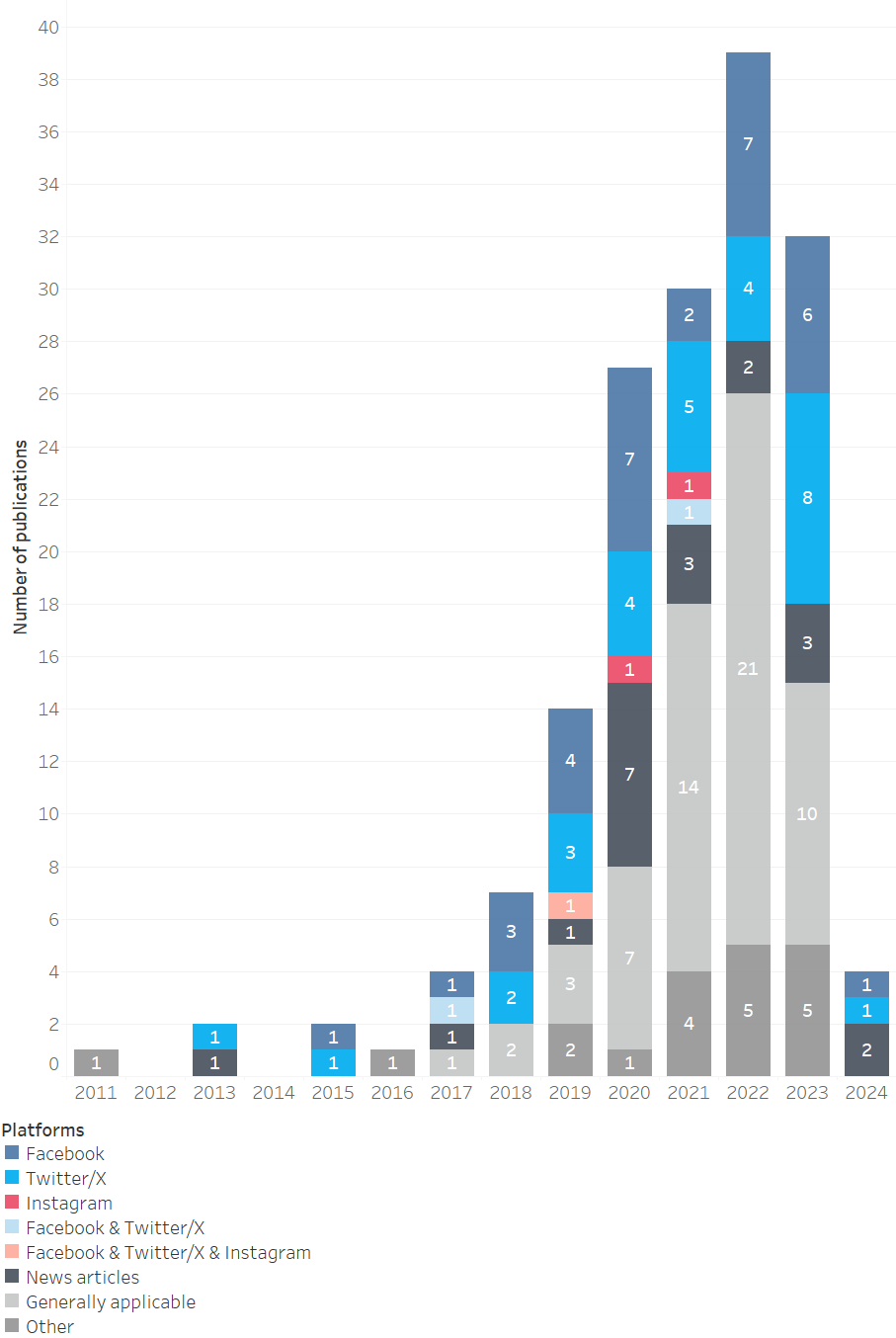}
    \caption{Number of papers published according to their addressed platform.}
    \label{fig:yearanzahl}
\end{figure}

While {\color{updated}65} publications generate their interventions or concepts generically for all online content and platforms (category \textit{General}), others are developed and evaluated for specific platforms (see Figure \ref{fig:yearanzahl} for temporal distribution regarding platforms). Nevertheless, transferability to other platforms is often not excluded. {\color{updated}36} publications address interventions for \textit{Facebook}, {\color{updated}32} publications for \textit{Twitter/X}, and 3 publications for \textit{Instagram}. Another {\color{updated}19} publications deal with platforms that do not fall into one of the categories already mentioned (e.g., {\color{updated}Reddit \citep{bozarth_wisdom_2023, waltenberger_reddit_2023}, TikTok \citep{guo_seeing_2023}, websites \citep{lee_misvis_2022}, arguments over an audio speaker \citep{danry_wearable_2020}, text documents \citep{furuta_fact-checking_2021}, messenger forwards \citep{pasquetto_social_2022}}) and therefore were categorized as \textit{Other}. This corresponds to known research biases that show a focus on much-researched platforms such as Twitter/X. This is often justified by the already developed data situation and easier linkage to existing literature. Especially the great relevance of misinformation on newer social media platforms like TikTok in crises like the Russian-Ukrainian war shows that there is still a great need for research. When looking more closely at the addressed content format {\color{updated}also see Figure \ref{fig:format} in the Appendix)}, we can see that most publications focus on \textit{social media posts} ({\color{updated}87} publications), {\color{updated}46} on \textit{articles or text in general} and only a few on \textit{images} {\color{updated}(8 publications)} and \textit{videos} ({\color{updated}8} publications) while we observe a growing relevance of misinformation of exactly these formats. Additionally, there are a few exceptions that address a very specific format, such as \textit{audio} (e.g., \citep{danry_wearable_2020}) {\color{updated}or misleading graphs \citep{wijnker_debunking_2022}}.

A minority of the publications describe interventions that go beyond a low-fidelity prototype (e.g., in the form of screenshots) to include an actual implementation. Thus, \textit{no implementations} can be read from {\color{updated}117} publications, while {\color{updated}13} publications deal with \textit{browser plugins or browser extensions} (e.g., \citep{bhuiyan_nudgecred_2021,khivasara_fake_2020}). {\color{updated}Thirteen} publications describe the implementation of a \textit{custom platform} (e.g., \citep{amin_visual_2021}), and 3 publications show a \textit{game-based implementation}.
Elaborate tools are an important part of mitigating the spread of misinformation and can be part of a holistic solution. 
An example is `Verifi!' \citep{karduni_vulnerable_2019}, which provides an interface for dealing with misinformation on Twitter {\color{updated}(now X)}. The system consists of five display options, allowing for easy comparison between how real and questionable news sources report on a subject, for example, by comparing the words or images used. Another example would be `Prta' \citep{martino_prta_2020}, which provides the user with a tool that takes a text or URL as input and highlights propaganda techniques.

\subsection{A Taxonomy of User-centered Misinformation Interventions}
\label{sec:taxo} 
The wide range of addressed concepts, platforms, and research areas shows that, on the one hand, a large number of conceptual ideas and empirical findings already exist for digitally supporting users in dealing with misinformation; on the other hand, these often differ fundamentally. In order to distinguish existing approaches from each other and to cluster commonalities, we have derived a taxonomy based on the identified literature. Therefore, we performed an in-depth analysis of interventions. For the interpretation of the following results, it is important to notice that publications can contain multiple interventions -- in total, there were {\color{updated}228} interventions within the {\color{updated}163} publications. Those interventions were analyzed individually regarding the taxonomy characteristics, while previously reported methodological findings are valid for the entire publication and, therefore, did not distinguish between individual interventions.
In the following, the literature-based categories of the taxonomy are explained in detail (see also Table \ref{tab:taxonomy} and Figure \ref{fig:taxonomy}):

\begin{table*}[]
\caption{A taxonomy of user-centered misinformation interventions.}
\label{tab:taxonomy}
\resizebox{\textwidth}{!}{%
\begin{tabular}{|l|lll|}
\hline
\rowcolor[HTML]{000000} 
{\color[HTML]{FFFFFF} \textbf{Category}}          & \multicolumn{1}{l|}{\cellcolor[HTML]{000000}{\color[HTML]{FFFFFF} \textbf{Definition}}}                                                                                          & \multicolumn{1}{l|}{\cellcolor[HTML]{000000}{\color[HTML]{FFFFFF} \textbf{Intervention examples}}}            & {\color[HTML]{FFFFFF} \textbf{Publications}} \\ \hline
\rowcolor[HTML]{656565} 
{\color[HTML]{FFFFFF} \textbf{Intervention design}} & \multicolumn{3}{l|}{\cellcolor[HTML]{656565}{\color[HTML]{FFFFFF} \parbox{22.5cm}{\vspace{.2\baselineskip}The intervention design distinguishes different countermeasures after the successful detection of misinformation. This includes general actions such as deleting content as well as concrete design decisions to encourage a learning effect.\vspace{.2\baselineskip}}}}                                                                                                        \\ \hline
Warning                                           & \multicolumn{1}{l|}{\parbox{8cm}{\vspace{.2\baselineskip}Interventions that give an explicit warning that the content is (potentially) misleading\vspace{.2\baselineskip}}} & \multicolumn{1}{l|}{\parbox{8cm}{\vspace{.2\baselineskip}Warning label; stop sign; ``This post was disputed''\vspace{.2\baselineskip}}} &     {\parbox{6cm}{\citep{andi_nudging_2020,ardevol-abreu_intentional_2020,bhuiyan_nudgecred_2021,bhuiyan_feedreflect_2018,clayton_real_2020,folkvord_effect_2022,freeze_fake_2021,gao_label_2018,grady_nevertheless_2021,hameleers_separating_2022,jennings_asymmetric_2021,kirchner_countering_2020,kreps_covid-19_2022,mena_cleaning_2020,moravec_appealing_2020,nekmat_nudge_2020,ozturk_combating_2015,park_experimental_2021,pennycook_implied_2020,von_der_weth_nudging_2020,sherman_designing_2021,seo_trust_2019,saltz_encounters_2021,porter_political_2022,sharevski_meaningful_2022,papakyriakopoulos_impact_2022,sharevski_i_2023,guo_seeing_2023,lim_effects_2023,barman_does_2023,wijnker_debunking_2022,scharrer_watch_2022-1,wang_pm_2022,zhang_investigation_2022,koch_effects_2023,lee_user_2023,figl_symbol_2023,van_der_meer_can_2023,hameleers_striking_2023,buczel_how_2024,dobber_effect_2023,pretus_imisleadingi_2024}}}                                         \\ \hline
Correction/debunking                              & \multicolumn{1}{l|}{\parbox{8cm}{\vspace{.2\baselineskip}Interventions that correct/debunk misinformation\vspace{.2\baselineskip}}}                                                                                                                                                            & \multicolumn{1}{l|}{\parbox{8cm}{\vspace{.2\baselineskip}Naturally occurring or artificially generated user comments or comments from officials that correct misinformation; Links to fact-checking websites; expert sources; {\color{updated}corrected headlines by users}\vspace{.2\baselineskip}}}                                                                                & {\parbox{6cm}{\vspace{.2\baselineskip}\citep{aird_does_2018,autry_correcting_2021,bode_related_2015,bode_see_2018,challenger_covid-19_2022,dai_effects_2021,ecker_correcting_2011,ecker_reminders_2017,ecker_can_2020,ecker_you_2020,gesser-edelsburg_correcting_2018,hameleers_picture_2020,hameleers_separating_2022,huang_when_2022,jahng_they_2021,kessler_debunking_2022,kim_eye_2021,kreps_covid-19_2022,kirchner_countering_2020,lee_effect_2022,liaw_maater_2013,lo_all_2021,martel_youre_2021, moon_how_2023-1,ozturk_combating_2015,pasquetto_social_2022,pluviano_misinformation_2017,porter_political_2022,rich_correcting_2020,sakhnini_review_2022,sangalang_potential_2019,schaewitz_combating_2020,smith_correcting_2019,sotirakou_toward_2022,sullivan_leveraging_2019,tanaka_exposure_2019,tanaka_toward_2013,thornhill_digital_2019,tseng_investigating_2022,tully_mobilizing_2020,van_der_meer_seeking_2020,vraga_i_2018,vraga_testing_2019,vraga_testing_2020,vraga_assessing_2021,vraga_creating_2022,vraga_effects_2021,vraga_using_2017,wahlheim_reminders_2020,wang_countering_2021,zhao_misinformation_2019, jahanbakhsh_our_2022,guo_seeing_2023,sheikh_ali_tahaqqaq_2023,jahanbakhsh_our_2022,dai_ask_2022,drolsbach_diffusion_2023,bozarth_wisdom_2023,barman_does_2023,caramancion_quantification_2022,wijnker_debunking_2022,irving_correcting_2022,wang_pm_2022,zhang_investigation_2022,song_fighting_2022,denner_effects_2023,porter_factual_2022,desai_does_2023,schmid_benefits_2022,lee_user_2023,westbrook_mechanisms_2023,bachmann_studying_2023,craig_one_2023,velez_latino-targeted_2023,wood_promoting_2023,van_der_meer_can_2023,liu_checking_2023,lee_correcting_2024,tao_hope_2023,tulin_beyond_2024}\vspace{.2\baselineskip}}}                                            \\ \hline
Showing indicators                                & \multicolumn{1}{l|}{\parbox{8cm}{\vspace{.2\baselineskip}Interventions that display indicators for misinformation to achieve transparency\vspace{.2\baselineskip}}}                                                                                                                                                            & \multicolumn{1}{l|}{\parbox{8cm}{\vspace{.2\baselineskip}showing how old a video actually is; color relevant words for misinformation classification; generic tips to detect misinformation; infographic\vspace{.2\baselineskip}}}                                                                                &                {\parbox{6cm}{\vspace{.2\baselineskip}\citep{ayoub_combat_2021,bhuiyan_designing_2021,chiang_impact_2022,domgaard_combating_2021,furuta_fact-checking_2021,grandhi_by_2021,guess_digital_2020,hameleers_separating_2022,karduni_vulnerable_2019,martino_prta_2020,sakhnini_review_2022,schmid_digital_2022,sherman_designing_2021,von_der_weth_nudging_2020,yong_explainability_2023,heuer_comparative_2022,lee_misvis_2022,sheikh_ali_tahaqqaq_2023,zade_tweet_2023,kim_exoskeleton_2023,ruffin_explaining_2022,waltenberger_reddit_2023,feng_examining_2023,bozarth_wisdom_2023,joshi_explainable_2023,caramancion_quantification_2022,wijnker_debunking_2022,koch_effects_2023,bachmann_studying_2023,hameleers_striking_2023,pretus_imisleadingi_2024}\vspace{.2\baselineskip}}}                             \\ \hline
{\color{updated}(Binary) labels}                                     & \multicolumn{1}{l|}{\parbox{8cm}{\vspace{.2\baselineskip}Interventions that label content as misinformation or true information; {\color{updated}often binary}\vspace{.2\baselineskip}}}                                                                                                                                                            & \multicolumn{1}{l|}{\parbox{8cm}{\vspace{.2\baselineskip}Tagging post as true or false; thumbs up/thumbs down; ``Prediction: It is Fake News!''; {\color{updated}traffic light symbols}\vspace{.2\baselineskip}}} &                  {\parbox{6cm}{\vspace{.2\baselineskip}\citep{aslett_news_2022,barua_f-nad_2019,brashier_timing_2021,ecker_can_2020,grady_nevertheless_2021,hawa_combating_2021,jahanbakhsh_leveraging_2022,jia_understanding_2022,khivasara_fake_2020,kreps_covid-19_2022,lee_effect_2022,lu_effects_2022,moravec_appealing_2020,pasquetto_social_2022,pennycook_implied_2020,porter_political_2022,sakhnini_review_2022,saltz_encounters_2021,schmid_digital_2022,sherman_designing_2021,sotirakou_toward_2022,thornhill_digital_2019,papakyriakopoulos_impact_2022,jahanbakhsh_exploring_2023,sharevski_i_2023,lee_misvis_2022,sheikh_ali_tahaqqaq_2023,lim_effects_2023,waltenberger_reddit_2023,bozarth_wisdom_2023,velasco_factit_2023,scharrer_watch_2022-1,zhang_investigation_2022,lee_user_2023,figl_symbol_2023,dobber_effect_2023,tulin_beyond_2024}\vspace{.2\baselineskip}}}                            \\ \hline
Highlighting design                               & \multicolumn{1}{l|}{\parbox{8cm}{\vspace{.2\baselineskip}Interventions that visually highlight relevant parts of a post for misinformation classification\vspace{.2\baselineskip}}} & \multicolumn{1}{l|}{\parbox{8cm}{\vspace{.2\baselineskip}highlight relevant words by color or size; color code tweets according to accuracy; highlight propaganda techniques using colors \vspace{.2\baselineskip}}}                                                                                &                    {\parbox{6cm}{\vspace{.2\baselineskip}\citep{amin_visual_2021,aslett_news_2022,bhuiyan_designing_2021,bhuiyan_feedreflect_2018,bhuiyan_nudgecred_2021,furuta_fact-checking_2021,jahanbakhsh_leveraging_2022,liaw_maater_2013,martino_prta_2020,sherman_designing_2021,thornhill_digital_2019,jahanbakhsh_exploring_2023,yong_explainability_2023,heuer_comparative_2022,lee_misvis_2022,sheikh_ali_tahaqqaq_2023,kim_exoskeleton_2023,ruffin_explaining_2022,joshi_explainable_2023,zheng_evaluating_2022,wijnker_debunking_2022}\vspace{.2\baselineskip}}}                          \\ \hline
{\parbox{3cm}{\vspace{.2\baselineskip}Visibility reduction\vspace{.2\baselineskip}}}                              & \multicolumn{1}{l|}{\parbox{8cm}{\vspace{.2\baselineskip}Interventions that reduce the visibility of misinformation {\color{updated}visually}\vspace{.2\baselineskip}}}                                                                                                                                                            & \multicolumn{1}{l|}{\parbox{8cm}{\vspace{.2\baselineskip}Reducing opacity or size\vspace{.2\baselineskip}}}                                                                                &                            {\parbox{6cm}{\vspace{.2\baselineskip}\citep{ardevol-abreu_intentional_2020,bhuiyan_nudgecred_2021,bhuiyan_feedreflect_2018,kirchner_countering_2020,kreps_covid-19_2022,porter_political_2022,saltz_encounters_2021,sherman_designing_2021,guo_seeing_2023,kim_exoskeleton_2023}\vspace{.2\baselineskip}}}                  \\ \hline
Removal                                           & \multicolumn{1}{l|}{\parbox{8cm}{\vspace{.2\baselineskip}Interventions that hide or remove misinformation\vspace{.2\baselineskip}}}                                                                                                                                                            & \multicolumn{1}{l|}{\parbox{8cm}{\vspace{.2\baselineskip}deleting misinformation\vspace{.2\baselineskip}}}                                                                                &                  {\parbox{6cm}{\vspace{.2\baselineskip}\citep{saltz_encounters_2021}\vspace{.2\baselineskip}}}                            \\ \hline
Complicate sharing                                & \multicolumn{1}{l|}{\parbox{8cm}{\vspace{.2\baselineskip}Interventions that include additional user effort before allowing to share misinformation\vspace{.2\baselineskip}}}                                                                                                                                                            & \multicolumn{1}{l|}{\parbox{8cm}{\vspace{.2\baselineskip}additional confirmation before sharing; require users to assess accuracy before sharing\vspace{.2\baselineskip}}} &                                {\parbox{6cm}{\citep{amin_visual_2021,jahanbakhsh_exploring_2021,kirchner_countering_2020,von_der_weth_nudging_2020,kim_exoskeleton_2023}\\}}              \\ \hline
Specific visualization                            & \multicolumn{1}{l|}{\parbox{8cm}{\vspace{.2\baselineskip}Interventions that use creative visualizations of relevant information\vspace{.2\baselineskip}}}                                                                                                                                                            & \multicolumn{1}{l|}{\parbox{8cm}{\vspace{.2\baselineskip}visualizing sentiment and controversy score of news articles; visualizing fact-checker decisions; platform based on social network analysis visualization; aggregate authentication measures; visualization of number of unvaccinated children with measles as fear correction; visualizing fact-checker decisions; infographic\vspace{.2\baselineskip}}}                                                                                &                                    {\parbox{6cm}{\vspace{.2\baselineskip}\citep{domgaard_combating_2021,jahanbakhsh_leveraging_2022,karduni_vulnerable_2019,kessler_debunking_2022,kim_controversy_2019,park_experimental_2021,pluviano_misinformation_2017,pourghomi_how_2017,safieddine_corporate_2016,schmid_digital_2022,sherman_designing_2021, jahanbakhsh_our_2022,wang_rumorlens_2022,lee_misvis_2022,zade_tweet_2023,jahanbakhsh_our_2022,chen_visualbubble_2022,zheng_evaluating_2022,zhang_investigation_2022,qian_fighting_2023}\vspace{.2\baselineskip}}}          \\ \hline
\rowcolor[HTML]{656565} 
{\color[HTML]{FFFFFF} \textbf{User interaction}}  & \multicolumn{3}{l|}{\cellcolor[HTML]{656565}{\color[HTML]{FFFFFF} \parbox{15cm}{\vspace{.2\baselineskip}Interventions require varying degrees of interaction with the countermeasure\vspace{.2\baselineskip}}}}                                                                                                        \\ \hline
Active                                            & \multicolumn{1}{l|}{\parbox{8cm}{\vspace{.2\baselineskip}Active interventions require users to actively interact with a countermeasure\vspace{.2\baselineskip}}} & \multicolumn{1}{l|}{\parbox{8cm}{\vspace{.2\baselineskip}click to confirm before sharing; overlay on Facebook/Twitter {\color{updated}(now X)} post; pop-up\vspace{.2\baselineskip}}} &                                  {\parbox{6cm}{\vspace{.2\baselineskip}\citep{agley_intervening_2021,amin_visual_2021,andi_nudging_2020,ardevol-abreu_intentional_2020,chen_examining_2022,folkvord_effect_2022,grady_nevertheless_2021,jahanbakhsh_exploring_2021,jahanbakhsh_leveraging_2022,khivasara_fake_2020,kirchner_countering_2020,kreps_covid-19_2022,pennycook_fighting_2020,porter_political_2022,saltz_encounters_2021,sharevski_two_2021,tsipursky_fighting_2018,von_der_weth_nudging_2020,vraga_testing_2020,guo_seeing_2023,kim_exoskeleton_2023,caramancion_quantification_2022,qian_fighting_2023,schmid_benefits_2022,lee_user_2023,hameleers_striking_2023,buczel_how_2024,dobber_effect_2023}\vspace{.2\baselineskip}}}            \\ \hline
Passive                                           & \multicolumn{1}{l|}{\parbox{8cm}{\vspace{.2\baselineskip}Passive interventions can be potentially ignored while using social media\vspace{.2\baselineskip}}}                                                                                                                                                            & \multicolumn{1}{l|}{\parbox{8cm}{\vspace{.2\baselineskip}A label below a social media post; tooltip; correction in comment section; warning next to a post\vspace{.2\baselineskip}}}                                                                                &       {\parbox{6cm}{\vspace{.2\baselineskip}\citep{aird_does_2018,aslett_news_2022,autry_correcting_2021,ayoub_combat_2021,bhuiyan_designing_2021,bhuiyan_feedreflect_2018,bhuiyan_nudgecred_2021,bode_related_2015,bode_see_2018,brashier_timing_2021,clayton_real_2020,dai_effects_2021,domgaard_combating_2021,duncan_whats_2020,ecker_reminders_2017,ecker_correcting_2011,ecker_can_2020,ecker_you_2020,freeze_fake_2021,furuta_fact-checking_2021,gao_label_2018,gesser-edelsburg_correcting_2018,grady_nevertheless_2021,guess_digital_2020,hameleers_picture_2020,hameleers_separating_2022,huang_when_2022,jahng_they_2021,jennings_asymmetric_2021,jia_understanding_2022,kessler_debunking_2022,kim_combating_2019,kim_eye_2021,kirchner_countering_2020,lee_effect_2022,liaw_maater_2013,lu_effects_2022,martel_youre_2021,mena_cleaning_2020,moon_how_2023-1,moravec_appealing_2020,nekmat_nudge_2020,ozturk_combating_2015,park_experimental_2021,pasquetto_social_2022,pennycook_implied_2020,pennycook_shifting_2021,pluviano_misinformation_2017,pourghomi_how_2017,rich_correcting_2020,safieddine_corporate_2016,saltz_encounters_2021,sangalang_potential_2019,schaewitz_combating_2020,seo_trust_2019,sharevski_two_2021,sherman_designing_2021,smith_correcting_2019,sullivan_leveraging_2019,tanaka_exposure_2019,tanaka_toward_2013,thornhill_digital_2019,tseng_investigating_2022,tully_designing_2020,tully_mobilizing_2020,van_der_meer_seeking_2020,vraga_using_2017,vraga_testing_2019,vraga_i_2018,vraga_testing_2019,vraga_testing_2020,vraga_creating_2022,vraga_effects_2021,wahlheim_reminders_2020,wang_countering_2021, jahanbakhsh_our_2022,sharevski_meaningful_2022,papakyriakopoulos_impact_2022,yong_explainability_2023,heuer_comparative_2022,sharevski_i_2023,guo_seeing_2023,sheikh_ali_tahaqqaq_2023,zade_tweet_2023,jahanbakhsh_our_2022,kim_exoskeleton_2023,ruffin_explaining_2022,lim_effects_2023,waltenberger_reddit_2023,feng_examining_2023,drolsbach_diffusion_2023,barman_does_2023,joshi_explainable_2023,wijnker_debunking_2022,irving_correcting_2022,scharrer_watch_2022-1,wang_pm_2022,song_fighting_2022,denner_effects_2023,porter_factual_2022,desai_does_2023,qian_fighting_2023,schmid_benefits_2022,koch_effects_2023,lee_user_2023,westbrook_mechanisms_2023,bachmann_studying_2023,craig_one_2023,velez_latino-targeted_2023,figl_symbol_2023,wood_promoting_2023,van_der_meer_can_2023,buczel_how_2024,pillai_explaining_2023,liu_checking_2023,lee_correcting_2024,dobber_effect_2023,tao_hope_2023,tulin_beyond_2024,pretus_imisleadingi_2024}\vspace{.2\baselineskip}}}                                       \\ \hline
\rowcolor[HTML]{656565} 
{\color[HTML]{FFFFFF} \textbf{Timing}}            & \multicolumn{3}{l|}{\cellcolor[HTML]{656565}{\color[HTML]{FFFFFF} \parbox{15cm}{\vspace{.2\baselineskip}Digital misinformation interventions can address varying points in time within the social media usage\vspace{.2\baselineskip}}}}                                                                                                        \\ \hline
Pre exposure                                      & \multicolumn{1}{l|}{\parbox{8cm}{\vspace{.2\baselineskip}Interventions that take place immediately before the exposure to (mis)information\vspace{.2\baselineskip}}}                                                                                                                                                            & \multicolumn{1}{l|}{\parbox{8cm}{\vspace{.2\baselineskip}accuracy nudge before a news-sharing task; Pro-Truth pledge to engage in more pro-social behavior; narrative fear appeal message to encourage health experts to correct health misinformation online; general warning message about misleading articles; generic infographic; protective message: ``Warning! Note: fake news can occur on Facebook. [...]''\vspace{.2\baselineskip}}} &                            {\parbox{6cm}{\vspace{.2\baselineskip}\citep{agley_intervening_2021,brashier_timing_2021,chen_examining_2022,clayton_real_2020,dai_effects_2021,domgaard_combating_2021,folkvord_effect_2022,grady_nevertheless_2021,guess_digital_2020,hameleers_separating_2022,karduni_vulnerable_2019,kirchner_countering_2020,moravec_appealing_2020,pennycook_fighting_2020,pennycook_shifting_2021,tanaka_toward_2013,tsipursky_fighting_2018,tully_designing_2020,vraga_assessing_2021,vraga_testing_2020,guo_seeing_2023,kim_exoskeleton_2023,caramancion_quantification_2022,scharrer_watch_2022-1,qian_fighting_2023,schmid_benefits_2022,lee_user_2023,craig_one_2023,wood_promoting_2023,hameleers_striking_2023,buczel_how_2024,dobber_effect_2023}\vspace{.2\baselineskip}}}                  \\ \hline
During exposure                                   & \multicolumn{1}{l|}{\parbox{8cm}{\vspace{.2\baselineskip}Interventions that take place during the usage of social media (and during the exposure to misinformation)\vspace{.2\baselineskip}}} & \multicolumn{1}{l|}{\parbox{8cm}{\vspace{.2\baselineskip}algorithmic corrections next to a post; user comments underneath a post; warnings; adding ``Rated False'' tag to article headline; credibility labels; highlighting indicators for propaganda; wearable reasoner giving AI-based feedback on claims\vspace{.2\baselineskip}}} &               {\parbox{6cm}{\vspace{.2\baselineskip}\citep{aird_does_2018,amin_visual_2021,andi_nudging_2020,ardevol-abreu_intentional_2020,aslett_news_2022,ayoub_combat_2021,bhuiyan_designing_2021,bhuiyan_feedreflect_2018,bhuiyan_nudgecred_2021,bode_related_2015,bode_see_2018,brashier_timing_2021,clayton_real_2020,danry_wearable_2020,duncan_whats_2020,ecker_can_2020,gao_label_2018,grady_nevertheless_2021,hameleers_picture_2020,hameleers_separating_2022,horne_trustworthy_2019,jahanbakhsh_leveraging_2022,jahng_they_2021,jennings_asymmetric_2021,jia_understanding_2022,kim_combating_2019,kim_eye_2021,kirchner_countering_2020,kreps_covid-19_2022,liaw_maater_2013,lo_all_2021,lu_effects_2022,mena_cleaning_2020,moon_how_2023-1,moravec_appealing_2020,nekmat_nudge_2020,ozturk_combating_2015,park_experimental_2021,pennycook_implied_2020,pluviano_misinformation_2017,porter_political_2022,seo_trust_2019,sharevski_two_2021,sherman_designing_2021,smith_correcting_2019,sullivan_leveraging_2019,thornhill_digital_2019,tully_mobilizing_2020,vraga_assessing_2021,vraga_i_2018,vraga_using_2017,vraga_creating_2022,vraga_effects_2021, jahanbakhsh_our_2022,sharevski_meaningful_2022,papakyriakopoulos_impact_2022,jahanbakhsh_exploring_2023,yong_explainability_2023,heuer_comparative_2022,sharevski_i_2023,guo_seeing_2023,zade_tweet_2023,jahanbakhsh_our_2022,ruffin_explaining_2022,dai_ask_2022,lim_effects_2023,waltenberger_reddit_2023,feng_examining_2023,drolsbach_diffusion_2023,bozarth_wisdom_2023,barman_does_2023,joshi_explainable_2023,zheng_evaluating_2022,wijnker_debunking_2022,irving_correcting_2022,scharrer_watch_2022-1,wang_pm_2022,zhang_investigation_2022,song_fighting_2022,porter_factual_2022,koch_effects_2023,lee_user_2023,bachmann_studying_2023,figl_symbol_2023,lee_correcting_2024,dobber_effect_2023,tulin_beyond_2024,pretus_imisleadingi_2024}\vspace{.2\baselineskip}}}                               \\ \hline
Post exposure                                     & \multicolumn{1}{l|}{\parbox{8cm}{\vspace{.2\baselineskip}Interventions occurring after seeing misinformation\vspace{.2\baselineskip}}}                                                                                                                                                            & \multicolumn{1}{l|}{\parbox{8cm}{\vspace{.2\baselineskip}warnings after exposure to misinformation; responses by official health authority; debunking text based on debunking handbook including plausible scientific explanations to close gaps in mental models\vspace{.2\baselineskip}}}                                                                                &            {\parbox{6cm}{\vspace{.2\baselineskip}\citep{autry_correcting_2021,brashier_timing_2021,dai_effects_2021,ecker_you_2020,ecker_reminders_2017,ecker_correcting_2011,freeze_fake_2021,gesser-edelsburg_correcting_2018,grady_nevertheless_2021,huang_when_2022,kessler_debunking_2022,lee_effect_2022,lu_effects_2022,martel_youre_2021,pasquetto_social_2022,pennycook_shifting_2021,rich_correcting_2020,sangalang_potential_2019,schaewitz_combating_2020,tanaka_exposure_2019,tanaka_toward_2013,tseng_investigating_2022,van_der_meer_seeking_2020,vraga_testing_2019,vraga_testing_2020,wahlheim_reminders_2020,wang_countering_2021,denner_effects_2023,desai_does_2023,schmid_benefits_2022,westbrook_mechanisms_2023,craig_one_2023,velez_latino-targeted_2023,wood_promoting_2023,van_der_meer_can_2023,buczel_how_2024,pillai_explaining_2023,liu_checking_2023,lee_correcting_2024,tao_hope_2023}\vspace{.2\baselineskip}}}                                  \\ \hline
At the moment of sharing                          & \multicolumn{1}{l|}{\parbox{8cm}{\vspace{.2\baselineskip}Interventions taking place directly at the moment of sharing misinformation\vspace{.2\baselineskip}}}                                                                                                                                                            & \multicolumn{1}{l|}{\parbox{8cm}{\vspace{.2\baselineskip}Encouraging to reflect on content before sharing; endorsing accuracy prompt: ``I think this news is accruate'' placed into sharing button; behavioral nudges using checkboxes to indicate whether a heading is accurate and to tag reasons via checklist at posting time; report of linguistic analysis as immediate feedback when sharing and possibility to cancel a tweet within 30 seconds\vspace{.2\baselineskip}}}                                                                                &         {\parbox{6cm}{\vspace{.2\baselineskip}\citep{amin_visual_2021,capraro_i_2022,jahanbakhsh_exploring_2021,jahanbakhsh_leveraging_2022,kirchner_countering_2020,von_der_weth_nudging_2020}\vspace{.2\baselineskip}}}                                     \\ \hline
On request of the user                            & \multicolumn{1}{l|}{\parbox{8cm}{\vspace{.2\baselineskip}Interventions that take place detached from social media platforms and have to be reached out to actively\vspace{.2\baselineskip}}}                                                                                                                                                            & \multicolumn{1}{l|}{\parbox{8cm}{\vspace{.2\baselineskip}web-app based on social network analysis for user exploration; Android application where user can enter URL or text for credibility assessment; system to analyze articles or URLs via interface or API\vspace{.2\baselineskip}}}                                                                                &          {\parbox{6cm}{\vspace{.2\baselineskip}\citep{barua_f-nad_2019,chiang_impact_2022,hawa_combating_2021,karduni_vulnerable_2019,khivasara_fake_2020,kim_controversy_2019,martino_prta_2020,pourghomi_how_2017,safieddine_corporate_2016,sakhnini_review_2022,schmid_digital_2022,sotirakou_toward_2022,wang_rumorlens_2022,lee_misvis_2022,sheikh_ali_tahaqqaq_2023,kim_exoskeleton_2023,chen_visualbubble_2022,velasco_factit_2023}\vspace{.2\baselineskip}}}                                    \\ \hline
\end{tabular}%
}
\end{table*}

\begin{figure}[h!]
    \centering
    \label{fig:taxo}
    
\begin{forest} %forked edges,
for tree={
    grow=east,
    growth parent anchor=east,
    parent anchor=east,
    child anchor=west,
    edge path={\noexpand\path[\forestoption{edge},->, >={latex}] 
         (!u.parent anchor) -- +(5pt,0pt) |- (.child anchor)
         \forestoption{edge label};}
         }
%for tree={draw,edge={-latex},fill=white,blur shadow, grow=0}%grow=0 macht es l->r, align=center
[Interventions, root
[\textbf{Time of intervention}, onode
    [On request of the user, tnode]
    [At the moment of sharing, tnode]
    [Post exposure, tnode]
    [During exposure, tnode]
    [Pre exposure, tnode]]
[\textbf{User interaction}, onode
     [Passive, tnode]
     [Active, tnode]]
[\textbf{Intervention design}, onode
     [Specific visualization, tnode]
     [Complicate sharing, tnode]
     [Removal, tnode]
     [Visibility reduction, tnode]
     [Highlighting design, tnode]
     [(Binary) labels, tnode]
     [Showing indicators, tnode]
     [Correction/debunking, tnode]
     [Warning, tnode]]
]
\end{forest}
\caption{A taxonomy for user-centered misinformation interventions.}
\end{figure}
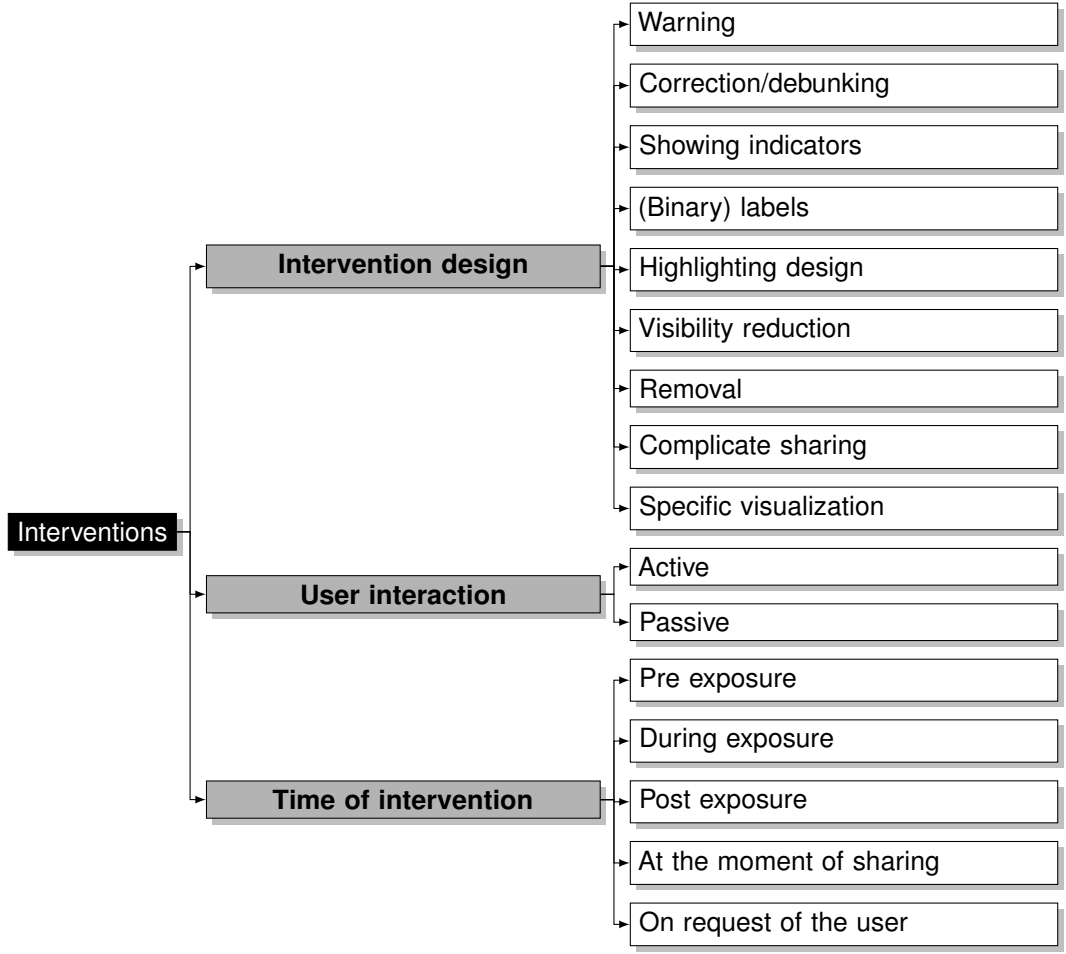
\label{fig:taxonomy}

\subsubsection{Intervention Design} 
\label{sec:interventiontype}
The identified interventions on user-centered misinformation interventions vary greatly in their starting point. The digital support approaches and concepts are as diverse as the possibilities for protecting users from the effects of misinformation (e.g., deleting problematic content, warning, or strengthening media literacy). In an iterative process, nine intervention designs were identified based on the literature. Interventions could be assigned to multiple intervention designs, as they often used combinations.
The majority of interventions propose or evaluate \textit{correction/debunking} of misleading contents ({\color{updated}65} stand-alone interventions and {\color{updated}30} interventions in combination with other intervention designs within {\color{updated}79} publications) and often represents a quite natural behavior of social media usage rather than an artificially generated technical countermeasure. For instance, many publications in that context evaluate whether corrections by users in the comment section of a post are effective in reducing belief in misleading content (e.g., \citep{martel_youre_2021}). Some of those interventions include a link to fact-checking websites, where the misleading content is debunked. This can be implemented both naturally by users in the comment section posting debunking links or digital interventions automatically exposing users to debunking (e.g., link to correcting source or user rebuttal within a comment/reply to a social media post or {\color{updated}exposure to automatically generated counterfactual explanations \citep{dai_ask_2022}}). 
{\color{updated}Thus, interventions vary in terms of who is the arbiter of credibility assessment. While some are expert-based or rely on algorithm decisions, others rely on crowdsourcing of the community \citep{drolsbach_diffusion_2023}. For instance, a browser extension allows users to suggest alternative headlines as a crowdsourced odd case for corrections, which are then presented to other users, empowering them to more actively participate in news consumption \citep{jahanbakhsh_our_2022}. The dimension of who decides what is wrong or right within corrections and other intervention types was not systematically covered by our taxonomy but constitutes a relevant research area that has already been addressed by several studies \citep{jahanbakhsh_our_2022, hannak_get_2014, vraga_using_2017}.}
Many interventions do not only expose users to content debunking or rebuttals but give an explicit \textit{warning} that the content is (potentially) misleading ({\color{updated}19} stand-alone interventions and {\color{updated}38} interventions in combination within {\color{updated}41} publications). Those warnings reach from warning labels like stop signs, to textual warnings, e.g., ``This post was disputed!''.

Misinformation interventions can have different objectives. One of these objectives is to strengthen media literacy. In these types of interventions, concrete assistance in the form of indicators, for example, is typical. By \textit{showing indicators} {\color{updated}that support users in evaluating the credibility of content, the aim is to achieve a learning effect.} {\color{updated}Thirty-three} interventions correspond to this intervention design (including {\color{updated}7} as stand-alone intervention; {\color{updated}31} publications), for example by showing how old a video actually is \citep{sherman_designing_2021} or by deriving words in the text that were particularly relevant for automatic detection as misleading \citep{ayoub_combat_2021}. Other intervention designs of this type compile more generic tips that users can apply to detect misleading content \citep{domgaard_combating_2021,guess_digital_2020,hameleers_separating_2022}. {\color{updated}Research on indicators for misinformation, for instance, from the perspective of journalists as annotators \citep{zhang_structured_2018}, can be considered a significant foundation to inform indicator-based interventions.} This intervention design is especially relevant, as studies have shown that users prefer transparent approaches where there is a potential learning effect \citep{kirchner_countering_2020}. 
However, in contrast to {\color{updated} showing indicators for misleading content in a nuanced way, {\color{updated}44} interventions take the approach of assigning \textit{(binary) labels} to contents (including {\color{updated}11} stand-alone interventions; {\color{updated}37} publications). This can be implemented, for example, by tagging content with a true or false tag, or thumbs up, thumbs down \citep{barua_f-nad_2019}. Other interventions give a probability in percent that content is misleading \citep{khivasara_fake_2020}, and thus extend the framework of binary labels with richer information or provide more nuanced labels e.g., using traffic light colors. Often, these labels have in common that they do not provide a transparent explanation. However, binary and more nuanced labels do not, per definition, rule out a user-centered approach as they may be sensibly combined with explanations and can be applied for simplification of a complex underlying rating system.}

Similarly, other ideas are concerned with increasing transparency. Some approaches use \textit{highlighting design} as an intervention design which aims at facilitating potential learning effects. {\color{updated}Twenty-one} interventions ({\color{updated}21} publications), for example, visually highlight relevant words for automatic classification within a social media post by color or size. For instance, \citet{bhuiyan_nudgecred_2021} color code tweets on Twitter {\color{updated}(now X)} according to their computed accuracy. In contrast, \citet{martino_prta_2020} highlight propaganda techniques (e.g., exaggeration, loaded language, or oversimplification) detected within a text using different colors.
{\color{updated}Often, interventions that show indicators are using some kind of highlighting design to do that, resulting in a common combination of both intervention designs.}
Indeed, despite highlighting components of content with colors, more \textit{specific visualizations} ({\color{updated}20} interventions; {\color{updated}19} publications) can be considered a distinct intervention design. Visualization is a very effective way to provide information as it provides \textit{``the highest bandwidth channel from computer to the human''} \citep[][p.~2]{ware_information_2012} and is used for interventions in different contexts \citep{hartwig_nudging_2022}. Within the literature sample, there is a diverse set of creative visualizations of information. For example, \citet{kim_controversy_2019} visualize the sentiment and controversy score of news articles within a conceptual study. In contrast, \citet{park_experimental_2021} visualize fact-checker decisions regarding textual rumors. Moreover, \citet{schmid_digital_2022} developed and evaluated a platform based on social network analysis of contents on Twitter/X for users to proactively assess misinformation through visualization, {\color{updated}and \citet{chen_visualbubble_2022} designed visualizations of filter bubbles, exposing users visually with topics, sources, and opinions outside of their own bubble.}

While previous intervention designs tended to provide a verified feedback together with the problematic content, one study evaluates the effect of the \textit{removal of misinformation} by hiding or removing a questionable post altogether \citep{saltz_encounters_2021}.
Similarly but less rigorous is the attempt of \textit{visibility reduction} ({\color{updated}14} interventions in {\color{updated}10} publications), e.g., by reducing opacity or size. {\color{updated}While we focus on visibility reduction that takes place visually, there are other (often network-based) approaches not covered by our more narrow understanding of user-centered misinformation interventions, as long as a user study demonstrating a direct impact on users is not included. For instance, studies reduce the visibility of misinformation in an algorithmic approach by reducing its flow \citep{jackson_learning_2022}. Similarly, \citet{epstein_will_2020} examine how layperson crowdsourcing of source credibility may be applied as input to social media ranking algorithms with promising results, leaving the potential for future research to investigate how this approach may be implemented regarding user feedback.}
Many intervention designs aim at preventing negative effects on people when confronted with misinformation or educating them to detect those contents themselves, as presented in the previous intervention designs. However, this can be extended to specifically preventing the spread of problematic content altogether, for instance, via \textit{complicating sharing} ({\color{updated}5} interventions in {\color{updated}5} publications). This may be implemented, for example, by including an additional confirmation before sharing or by nudging users to assess the accuracy of the content as they share it \citep{jahanbakhsh_exploring_2021}. These user-centered approaches stand in contrast to network-based approaches to prevent the spread of misinformation through computational techniques as described by \citet{chen_spread_2022}.

{\color{updated}Many publications used multiple intervention designs, for instance as combinations or comparing interventions of different types against each other \citep{hameleers_separating_2022}. 
For example, showing indicators for misinformation often comes with some sort of highlighting design or a specific visualization (of the indicators). }

\begin{figure}
    \centering
    \includegraphics[width=0.75\textwidth]{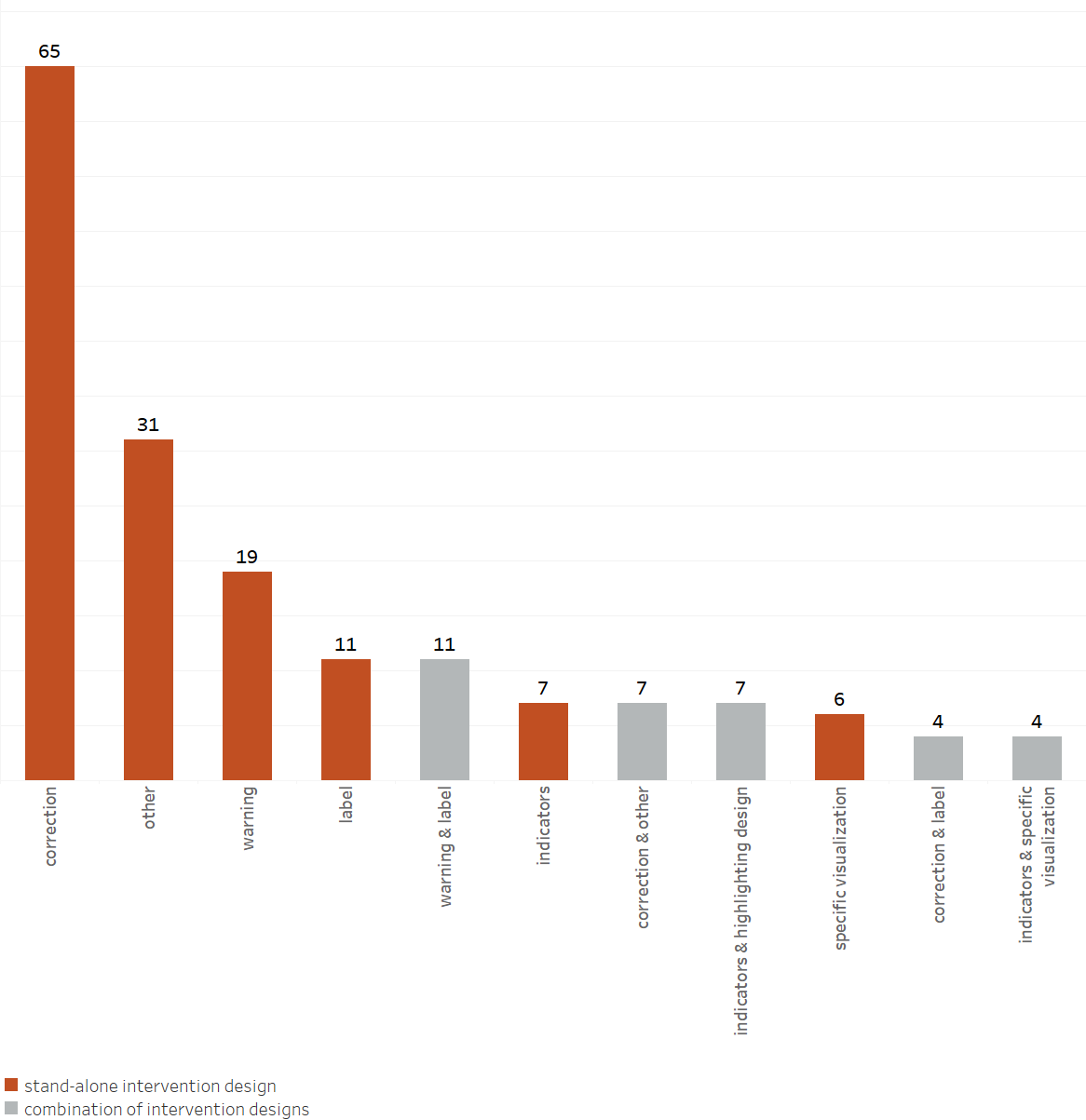}
    \caption{Number of interventions addressing common intervention designs while differentiating between approaches that use a single intervention design versus multiple types in combination. Instances <4 were excluded. A more detailed breakdown of the category `other' can be found in Section \ref{sec:interventiontype}.}
    \label{fig:interventiontypeInsgesamt}
\end{figure}

While most interventions could be assigned to one or more of the intervention designs listed above, {\color{updated}50} interventions additionally used an intervention design that did not clearly fit the scheme (including {\color{updated}30} stand-alone interventions; {\color{updated}39} publications) while not appearing often enough as a distinct intervention design to represent an own type within the taxonomy. There are particularly unusual approaches, such as the development of wearable glasses that provide audio feedback on the truth of content \citep{danry_wearable_2020} or a study that evaluates the effect of priming participants by letting them rate the accuracy of a headline before exposure to more potentially misleading content as a nudge to think more sufficiently \citep{pennycook_fighting_2020} {\color{updated}or on a similar basis, an explanation prompt that lets users explain why headlines were true or false \citep{pillai_explaining_2023}}.
Among the \textit{other} category, there are {\color{updated}nine} interventions giving diverse kinds of information about misinformation and its detection immediately before exposure, e.g., in form of an infographic \citep{agley_intervening_2021}, a video tutorial \citep{axelsson_learning_2021}, a text about negative consequences of misinformation \citep{clayton_real_2020}, a debiasing message \citep{dai_effects_2021}, an awareness training \citep{moravec_appealing_2020}, or a Pro-Truth Pledge \citep{tsipursky_fighting_2018}. Three interventions display a star rating or score, for instance, regarding credibility or sentiment \citep{duncan_whats_2020,kim_combating_2019,kim_controversy_2019}. Two interventions explicitly state to have integrated gamification elements \citep{almaliki_misinformation-aware_2019, sotirakou_toward_2022}.
You can find a visualization of intervention designs in Figure \ref{fig:interventiontypeInsgesamt}.

\subsubsection{User interaction} 
\label{sec:userinteraction}
\begin{figure}
    \centering
    \includegraphics[width=\textwidth]{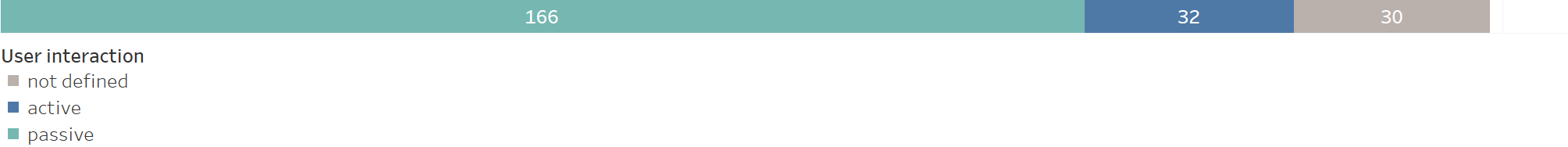}
    \caption{Number of interventions regarding each type of user interaction. Publications can contain multiple interventions with different user interaction types.}
    \label{fig:activepassive}
\end{figure}

We compared whether an intervention required users to \textit{actively} interact with the countermeasure (e.g., having to click to confirm sharing) or whether they could \textit{passively} ignore the countermeasure (e.g., a label below a post).
Furthermore, some interventions cannot clearly be labeled as active or passive as the actual form of implementation is not (yet) defined ({\color{updated}30} interventions in {\color{updated}27} publications). For instance, some approaches make a more generic proposal of a warning without stating if it can be ignored or not. Other approaches present an intervention which takes place only at the request of the user, for example, as a separate smartphone app (e.g., \citep{barua_f-nad_2019}) and, thus, is neither active nor passive during the actual usage of social media. In cases where the intervention is designed as a mandatory one-time experience (e.g., \citep{agley_intervening_2021, pennycook_fighting_2020, tsipursky_fighting_2018}), users do not have to interact with the intervention during the following usage of social media but definitely once before the usage. Thus, those interventions are classified as active.
We found that the majority of interventions are passive ({\color{updated}166} interventions in {\color{updated}120} publications) while only {\color{updated}32} interventions (in {\color{updated}28} publications) deal with active interventions. A representation of the number of interventions regarding different interaction types can be found in Figure \ref{fig:activepassive}.

\subsubsection{Time of intervention}
\label{sec:timeofintervention}
Misinformation interventions can address different points in time in the context of social media use. While countermeasures that take place long before the actual usage of social media (e.g., trainings, educational games for school lessons, inoculation) were excluded from analysis, we included countermeasures that take place immediately before the exposure (e.g., short messages when logging in to a social media platform). We found that {\color{updated}37} interventions take place \textit{pre-exposure} to misinformation (including {\color{updated}5} interventions coming with a combination of timings; {\color{updated}32} publications). For example, \citet{pennycook_fighting_2020} nudged participants of a large-scale online study to think about accuracy before getting a news-sharing task. They asked participants to rate a headline's accuracy before exposing them to multiple other headlines and measuring their sharing intentions. Indeed, they found that giving a simple accuracy induction resulted in increased sharing discernment \citep{pennycook_fighting_2020}.
The majority of interventions ({\color{updated}114} interventions in {\color{updated}86} publications) however, is designed to take place immediately \textit{during} the exposure to misinformation {\color{updated}while users are engaging with a social media platform or other content and encounter misinformation.}. This includes most of the corrections, warnings, labels, and highlighting approaches. For example, \citet{bode_see_2018} compare algorithmic corrections via related articles from Snopes with social corrections via user comments underneath a Facebook post referring to the same debunking link on Snopes.

\begin{figure}
    \centering
    \includegraphics[width=\textwidth]{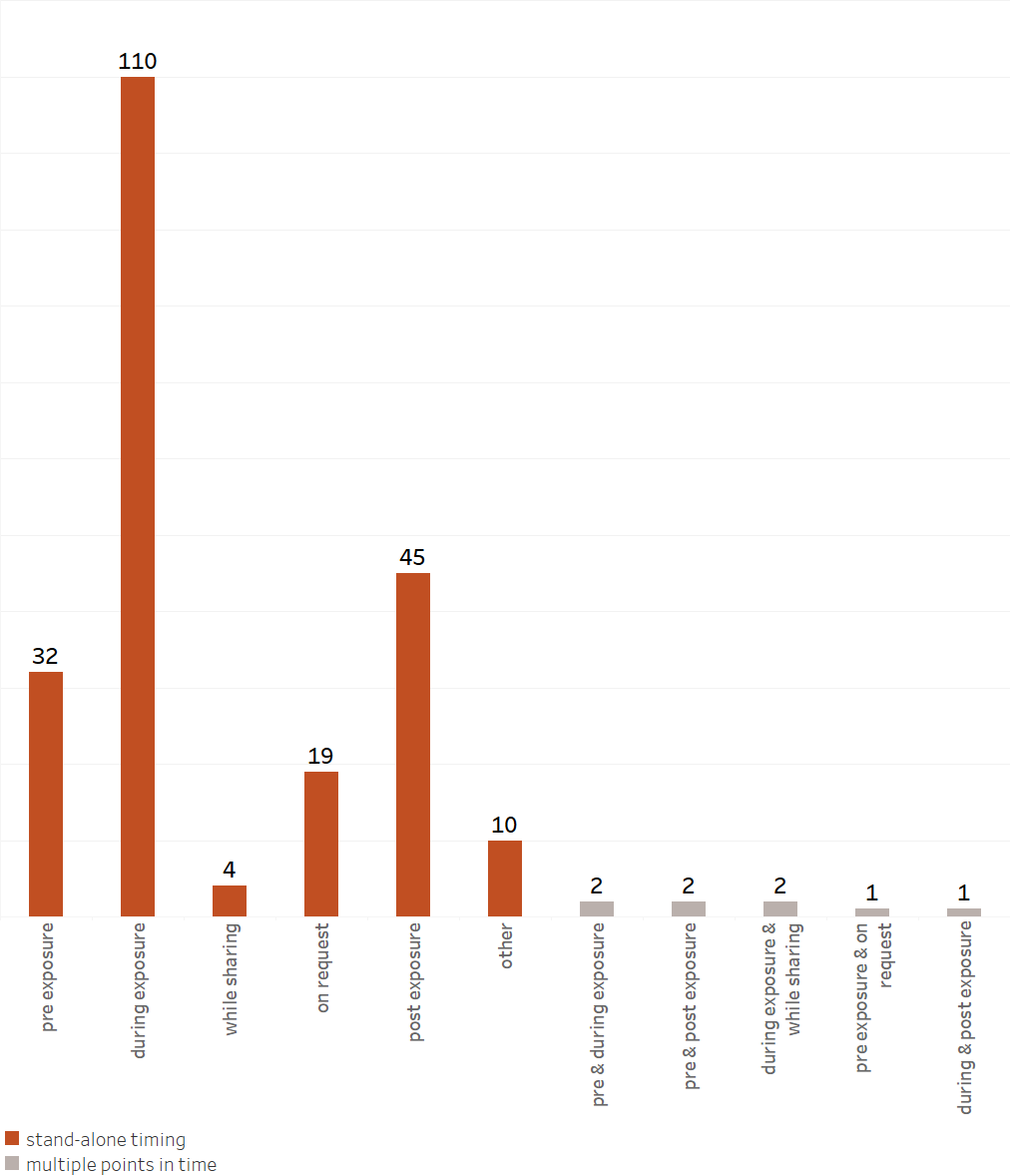}
    \caption{Number of interventions addressing different points in time.}
    \label{fig:timing}
\end{figure}

{\color{updated}Forty-eight} interventions in {\color{updated}41} publications deal with countermeasures \textit{post-exposure}, e.g., \citet{grady_nevertheless_2021} compared warnings after exposure to an article with warnings during other points in time.
To specifically combat the spread of misinformation, a few interventions intervene directly \textit{at the moment of sharing} (6 interventions in {\color{updated}6} publications; e.g., \citep{von_der_weth_nudging_2020}). {\color{updated}These presuppose that the user is about to share misinformation and he or she has already slightly passed the timing dimension `during exposure'.}
Detached from the actual social media platforms, some approaches offer their own platforms. Accordingly, the intervention here takes place \textit{at the request of the user} at any time ({\color{updated}20} interventions in {\color{updated}18} publications; e.g., \citep{barua_f-nad_2019}), when users have to actively reach out to the intervention on a separate platform.
A few interventions could not be assigned to one of those points in time (10 interventions in 8 publications). For example, \citet{furuta_fact-checking_2021} present a countermeasure to take place during article creation. 
While the majority of interventions specifically take place at one exact moment in time, {\color{updated}eight} interventions are designed to take place at multiple points in time, combining, for example, a pre-bunking message with a warning during exposure to misinformation. 
See Figure \ref{fig:timing} for a visualization of the number of publications for each timing of the interventions.

\subsection{How is transparency used to facilitate users autonomously dealing with misinformation?}
\label{sec:transparen}
\begin{figure}[ht]
\centering
    \includegraphics[width=\textwidth]{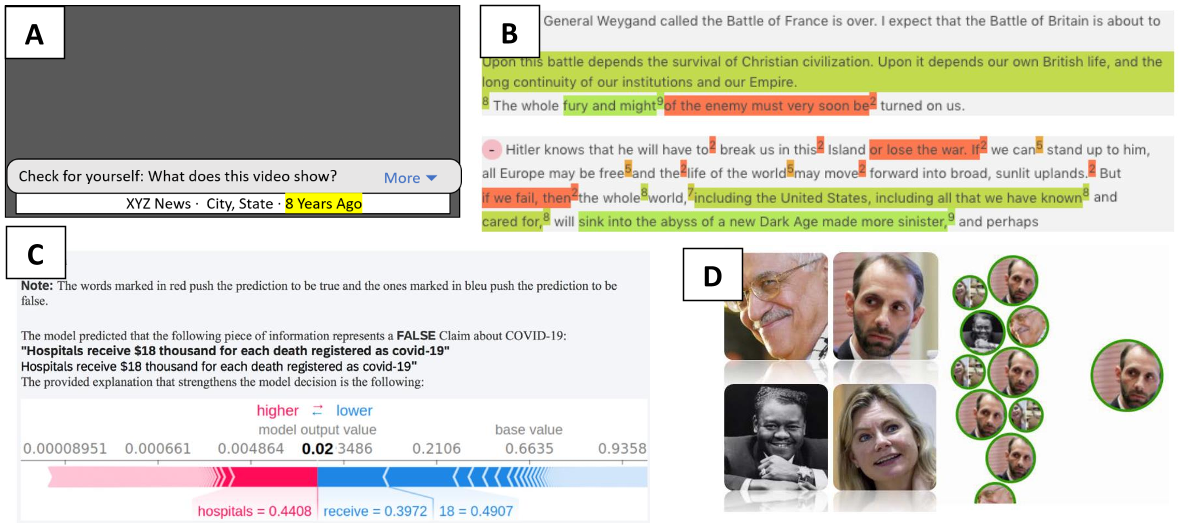}
    \caption{Four exemplary interventions using transparent design to various degrees. A: An intervention showing the publish date of a video as indicator \citet{sherman_designing_2021}. B: An intervention highlighting parts of a text containing propaganda techniques \citet{martino_prta_2020}. C: An intervention used to affirm and refute claims using explainable machine learning \citet{ayoub_combat_2021}. D: The image comparison view as part of a larger system designed by \citet{karduni_vulnerable_2019}. The screenshots were taken from the respective papers.}
    \label{fig:white-box-interventions}
\end{figure}

As highlighted in Section \ref{sec:interventiontype}, digital countermeasures in the context of misinformation can have different objectives. While some interventions aim to reduce the spread of such content itself, others aim to communicate the findings of the digital countermeasures to the end users, sometimes resulting in an environment that facilitates the strengthening of media literacy skills.
Within our systematic literature review, our special focus of interest is on transparent approaches that offer some form of explanation, as opposed to (binary) labels {\color{updated}without explanations} or deletion of problematic content.
Transparency can be achieved by different distinct intervention designs. One very common way of explanation is exposure to corrections or debunking. Indeed, the majority of publications within our scope deals with some sort of correction or debunking. While some approaches investigate the effect of user corrections within comment sections of social media posts, others focus on corrections of authorities. Interestingly, there is a very specific scientific discourse on the effectiveness of corrections, which is conducted in different disciplines. 
Corrections and debunking can be considered a central part of combating misinformation online. This type of intervention provides an opportunity to thoroughly confront officially refuted content with facts. Often, this takes the form of a more detailed article, which backs up its corrections with official sources. Official fact-checking websites, which are linked by the intervention, are usually used for this purpose. 
On the other hand, there are approaches that aim for transparency through media literacy training, for example, in the form of showing indicators and using a highlighting design of those (see examples in Figure \ref{fig:white-box-interventions}). Here it is examined which components of a social media post comprehensibly indicate that it is misleading content. We discussed this type of intervention in more detail in Section \ref{sec:interventiontype}. {\color{updated}In addition, while labeling content as false or true without explanations typically comes as a top-down approach not addressing users' needs for transparency, labeling interventions can indeed provide comprehensibility and transparency when applied as a simplification of an otherwise too complex rating system as a combination with additional explanations.}

\subsection{Nudging as an ambivalent trending countermeasure}
\label{sec:digitnudge}
During the content analysis of the identified relevant literature, one specific form of user intervention particularly caught the eye: {\color{updated}22} publications refer to their form of intervention as a digital nudge. A nudge is defined as an intervention that \textit{``alters people’s behavior in a predictable way without significantly changing their economic incentives''} \citep[][p. 6]{thaler_richard_h_nudge_2009}. It is a concept that has already been applied to many contexts, such as cybersecurity and health. The concept of nudging is controversially discussed in research. Thus, under certain circumstances, it represents a potential for subconscious manipulation also in harmful directions. Like many digital countermeasures, digital nudges, while promising, often offer little transparency and may run the risk of steering users in the wrong direction in ambiguous situations if they are not critically engaged with but rather trusted blindly. For instance, \citet{lu_effects_2022} show that AI-based credibility indicators can be used to steer participants in a certain direction, even if the AI is wrong. 
Since we did not specifically review which of the {\color{updated}228} interventions within {\color{updated}163} publications actually fit the definition of a nudge, we would nevertheless like to provide an overview of the publications that refer to their interventions as nudges themselves.
{\color{updated}Nudging has been applied by other reviews as a category of misinformation interventions on an individual level itself, complementing countermeasures like boosting, debunking, and content labeling \citep{roozenbeek_countering_2023}. In our review, we understand nudging as a concept that can be applied in diverse intervention designs.}
Some publications present intervention as ``accuracy nudges'' and introduce concepts in which users are specifically nudged to reflect on the accuracy of content and to act more thoughtfully accordingly (e.g., \citep{aslett_news_2022,jia_understanding_2022,pennycook_fighting_2020,capraro_i_2022, nekmat_nudge_2020}). For example, \citet{capraro_i_2022} report promising results that indicate positive effects on sharing behavior when using an accuracy prompt. Similarly, \citet{von_der_weth_nudging_2020} developed `ShareAware' as a nudge for more conscious posting and sharing behavior. In a different approach, \citet{andi_nudging_2020} developed a social norm-based nudge to effect sharing behavior by exposing participants to the message: ``\textit{[...] Most responsible people think twice before sharing content with their friends and followers}''.
In contrast, there are attempts to nudge users not only away from misinformation \citep{duncan_whats_2020} but towards the consumption of credible news (e.g., \citep{horne_trustworthy_2019,gao_label_2018} or a habit of assessing accuracy of information \citep{jahanbakhsh_exploring_2021, bhuiyan_feedreflect_2018,bhuiyan_nudgecred_2021}. In that context, \citet{thornhill_digital_2019} developed a nudge to steer users into fact-checking news online.

{\color{updated}\subsection{Impacts and perceptions of digital misinformation interventions}} 
\label{sec:impactperceptions}
Our systematic review revealed a wide range of intervention designs addressing various types of user interaction and timings. For future research, it is important to determine which interventions are most promising and should be given more consideration.
This paper aims to provide a comprehensive overview of misinformation interventions from diverse disciplines. These interventions strongly differ in their target of behavior change, such as improving credibility assessment, reducing the sharing of misinformation, helping users distinguish between misinformation and credible content, or decreasing the overall flow of misinformation on social media.
The focus is on the design characteristics of the intervention and the user-centered evaluation method, including qualitative and quantitative evaluations. Many studies evaluate the efficacy of an intervention within a specific context (social media platforms, user groups, format of content etc.) and in comparison to a specific condition (control group without intervention, state-of-art intervention of the specific platform etc.). Others derive rich qualitative insights, e.g., into how users perceive an intervention in terms of concepts like trust, reactance, or comprehensibility. 

While our review does not include a meta-analysis to derive statistical evidence of efficacy, we provide some initial insights into what appeared to be promising based on a qualitative overview, complemented by statistical effect size information that was explicitly stated in the corresponding publications. 
However, it is important to note that due to the nature of our review it does not allow for direct comparisons or objective evaluations of which interventions worked best or worst. Instead, it provides initial insights from a broad interdisciplinary perspective.
We present our overview of central findings regarding (positive and negative) impacts and perceptions of interventions in each publication in Table \ref{tab:effectperceptions} (electronic supplement). There we summarize beneficial effects of interventions that were mostly collected quantitatively, beneficial perceptions of interventions that were derived from qualitative studies, and insights on measures that were not effective or even resulted in counterproductive or unintended effects.
Taking a closer look at the studies, they identify measures and characteristics that do or do not impact efficacy - sometimes with contradictory findings that demonstrate a necessity for further investigations. For instance, the timing of the correction does sometimes but not always seem to matter \citep{dai_effects_2021, rich_correcting_2020}, and there are indications that efficacy is sometimes but not always impacted by whether the correction is narrative or non-narrative \citep{ecker_you_2020, lee_effect_2022, sangalang_potential_2019}.
There are further controversial findings on whether transparent information and explanations have a significant impact on efficacy (e.g., rather yes: \citep{gesser-edelsburg_correcting_2018, kirchner_countering_2020}; rather not: \citep{martel_youre_2021}). However, when considering findings on the role of transparency and explanations over all publications, the general tendency (including qualitative insights) indicates its impact on efficacy, user perception, and acceptance as promising. 

Further controversial findings discuss which intervention mechanism/source type (social versus algorithmic correction or warning by citizens versus news agency or (e.g., health) experts) matters in terms of efficacy \citep[e.g., ][]{huang_when_2022, bode_see_2018, hameleers_picture_2020, liu_checking_2023, lu_effects_2022}, for example, by emphasizing potential unintended over-reliance on AI predictions even if they are not correct \citep{lu_effects_2022}.
Other studies evaluate the impact of modality (e.g., images, videos, voice messages) of interventions on efficacy \citep{song_fighting_2022, pasquetto_social_2022,tseng_investigating_2022,wijnker_debunking_2022,zhang_investigation_2022}. For instance, \citet{pasquetto_social_2022} found that audio files were more effective in correcting beliefs than text or videos and \citet{karduni_vulnerable_2019} revealed that corrections using images are more effective in correcting myths than corrections without images, independent of the image type (machine-technical image, expert image, diagram).
When looking into the effect sizes (e.g., Cohen's d, (partial) $\eta^2$, Spearman's r and $\rho$) explicitly stated in the publications, they are small (e.g., (partial) $\eta^2$<0.06 or d<0.5) for the majority of publications \citep[e.g., ][]{axelsson_learning_2021,vraga_effects_2021}, and medium to large ((partial) $\eta^2$>0.06) in fewer cases \citep[e.g., ][]{vraga_assessing_2021}. As the interpretation is highly dependent on the research design, a future meta-analysis may complement our findings with statistical comparisons of efficacy that aim at controlling influencing factors revolving around the context of data in more narrowly defined domains.
Comparability is often not possible due to the very diverse settings in which studies take place, addressing different social media platforms, formats of content, and participants (e.g., students with potentially higher levels of media literacy, elderly, adolescents, representative studies in different countries), and applying a variety of research designs to measure efficacy, e.g., asking participants to state whether they would share specific content versus asking them to rate the credibility \citep{guay_how_2023}.
Indeed, finding a consensus in research to measure the efficacy of misinformation interventions has been emphasized as an important step towards more successful interventions, and possible frameworks have been proposed \citep{guay_how_2023}.
Some meta-analyses have already reported on the efficacy of specific misinformation interventions like corrections, where deriving a subgroup of studies with a similar research design and conditions is sometimes achievable, allowing for comparisons or comparable interventions in different contexts. 
For instance, \citet{chan_meta-analysis_2023} conducted a meta-analysis on the efficacy of corrections/debunking in the context of scientific misinformation, examining over 200 effect sizes and revealing that corrections are more successful when detailed. Still, in general, the debunking effect was not significant. Given the overall estimated lower impact of corrections/debunking and the strong research focus of the majority of studies on this type of intervention revealed in a related meta-analysis \citep{blair_interventions_2024}, which was confirmed in our review, this suggests scholars should not disregard other intervention types that might be less studied but more promising.

Due to the publication bias, most studies report statistically significant or qualitatively promising results. Only a few exceptions exclusively report what did not work in general \citep{kim_exoskeleton_2023,aslett_news_2022,caramancion_quantification_2022, sharevski_two_2021} or for specific user groups \citep{lee_user_2023, sharevski_i_2023}.
For instance, \citet{aslett_news_2022} report that dynamic source reliability labels placed in-feed did not reduce misperceptions, and \citet{caramancion_quantification_2022} demonstrates how preventive infographics had trivial to no effect.
Despite non-efficacy of interventions, in some cases, studies reveal other unintended or counterproductive effects such as over-correction and other spill-over effects on accurate content \citep{feng_examining_2023, hameleers_striking_2023}, over-reliance on interventions \citep{lu_effects_2022}, increased belief in misinformation under certain circumstances like subjective messages or repeated exposure to content \citep{autry_correcting_2021, pluviano_misinformation_2017,tanaka_exposure_2019}, priming of general mistrust in authentic content due to warnings \citep{van_der_meer_can_2023}, or lowered perception of extremeness due to stance labels on political ideology \citep{gao_label_2018}.

\section{Discussion}
\label{sec:disc}
In Section \ref{sec:res} we have systematically categorized a variety of existing misinformation interventions to assist in dealing with misinformation online, providing concrete examples of identified dimensions. 
The analysis of our systematic literature review underscores the impression that a variety of diverse approaches have emerged in recent years and continue to emerge. It can be observed that these often differ significantly in their characteristics. In this section, we discuss and summarize our findings regarding our research questions.

\subsection{RQ1: What are the typical methodological characteristics of existing studies on misinformation interventions?}
Methodologically, studies of misinformation interventions differ in various dimensions, although several emphases and typical patterns are also apparent. Due to the inclusion and exclusion criteria of our review, the sample contains mainly studies that collect empirical data and only a few publications with an exclusively conceptual approach. Typically, publications on misinformation interventions evaluate novel or already established interventions in user studies, often in comparison to other existing approaches. A particular focus is on {\color{updated}online} experiments with a collection of quantitative and qualitative data, as this method is suitable for large-scale samples and a controlled environment.  
In order to examine the interventions in a realistic environment and to minimize biases, more evaluation in the form of field studies would be desirable for future studies.
It is striking that mainly U.S. adults and college students are surveyed as study participants, while specific (vulnerable) target groups such as teenagers, persons of older age, or non-native speakers are largely neglected. This can be explained by the better accessibility of different user groups and represents a common problem known from other user studies in contexts of human-computer interaction and similar disciplines.

Not surprisingly, most publications deal with Facebook and Twitter/X as social media platforms or address news articles in general (see Section \ref{sec:methodologicaloverviews}). Considering current and emergent social media platforms like TikTok and Instagram as image- and video-based platforms is still largely missing within the research landscape. However, the impact of those platforms and content types has shown to be highly relevant. Looking closer at our publication sample, we can note that there are already isolated publications for addressing exceptional formats such as image, video and audio. {\color{updated}We can see that there is a positive correlation between the addressed formats `video' and `image'. Indeed, three out of five interventions for video formats are specifically addressing images as well (e.g., \citep{axelsson_learning_2021}).} 

\subsection{RQ2: How do existing forms of user-centered misinformation interventions assist users in dealing with misinformation online?}
Looking closely at misinformation interventions, one notices a publication emphasis on corrections and debunking. This form of intervention can be artificially controlled or occur naturally in the form of user comments. It is striking that corrections/debunking are examined in great detail in the literature from a wide variety of perspectives and with a focus on the smallest details{\color{updated}, e.g., regarding timing or repetition \citep{denner_effects_2023, craig_one_2023}}. This finding is also supported by the review study by \citet{chen_spread_2022}, who identified fact-based corrections as ``the most common type of corrective communication strategy'', classifying it as a part of message-based approaches.
{\color{updated}Often, corrections from the official side are based on thorough and elaborate journalistic work. For example, social media articles are linked to an official correction once the content has been thoroughly checked manually by experts. In the fast pace of social media and especially in emerging crisis situations, there is an overflow of accurate and misinformation that needs to be reacted to quickly. This is where expert-based corrections as digital countermeasures sometimes reach their limits as stand-alone interventions. Other approaches pursue corrections based on the assessment of users themselves. While there is no expert review here, active user participation in news consumption is facilitated and studies reveal promising findings. For instance, participants preferred suggested headlines by laypersons that corrected the original ones \citep{jahanbakhsh_our_2022}. Indeed, the effects of corrections by laypersons versus experts have been studied in prior work \citep{hannak_get_2014, vraga_using_2017} and constitute a relevant dimension of interventions that were beyond the scope of this work.} 
While many correction interventions provide users with additional (fact-based) knowledge and can thus create transparency, other types of interventions aim to increase transparency and thus media competence, for example, through linguistic or content-related indicators. 
An advantage of those interventions is the scalability of using automatic detection algorithms in real-time during emerging crisis situations and on large data sets, often based on machine learning approaches. However, when automatically showing indicators such as a missing verification seal or semantic propaganda techniques, the final decision on whether content is misinformation or not either lies with the user or is taken over by the algorithm based on (potentially biased) training data, missing the expert knowledge of professional fact-checkers.
Transparent misinformation interventions, independent of their implementation as correction or display of automatically detected indicators or other types, may offer the opportunity to counteract reactance of end users in contrast to approaches that lack an explanation and in some cases facilitate a feeling of censorship, paternalism, and loss of control.

In contrast to transparent approaches, there are also fewer educational interventions with the goal of reducing the consumption of misinformation through removal or visibility reduction. Both intervention designs can be considered helpful when taking into account the bias of people remembering content itself without a potentially shown correction or warning when exposed to misleading content \citep{grady_nevertheless_2021}. On the other hand, this intervention design may lead to (a feeling of) censorship and a resulting migration to other platforms or tools that take less rigorous action against problematic content. {\color{updated}While deleting/censoring dangerous or explicit content is a legitimate and important responsibility of social media platform operators, applying this solution of deletion to all problematic content, such as disinformation and misinformation, would not only lead to a migration of users to less restrictive platforms. In particular, it would represent a missed opportunity for media literacy education, some of which can be achieved through transparent digital countermeasures as a complement to school lessons.}

In order to develop misinformation interventions in a user-centered way and to be able to achieve an actual effect, the early inclusion of the needs and requirements of different target groups is indispensable. A particular challenge is the accessibility of people who have no trust in official bodies. In this context, limits certainly emerge as to who can be reached at all by the corresponding technical tools. In order to avoid reactance, the timing of the intervention certainly plays a role in addition to the transparency of approaches.  In our systematic literature review, we identified interventions that can be used at the request of the user and others that are permanently present during the normal use of social media. It is an interesting research question: which point in time or which regularity of an intervention is suitable for which target groups?
Considering the broad variety of misinformation interventions, we hope to provide a helpful overview of existing forms. We propose our taxonomy as a starting point to systematically capture intervention categories and identify relevant dimensions. It is intended to provide researchers with a framework to develop new interventions, to pool knowledge from different disciplines for the promotion of cross-disciplinary research, and to reveal promising research directions.

\subsection{RQ3: Which trends and chances for future research can be derived from the existing literature?}
In this paper, we have systematically analyzed {\color{updated}163} publications with {\color{updated}228} user interventions to assist in dealing with misinformation online. Our findings reveal current trends and movements in human-computer interaction, psychology, information systems, and communication sciences.
As potential avenues for future research, we propose the following questions and interests:

\textbf{(1) Are approaches for specific platforms transferable to other new platforms?} 
With regard to particularly relevant contexts of use, it is also necessary to consider current and emergent social media platforms. Social media platforms are constantly changing. For some time now, there has been a noticeable trend toward TikTok, and Facebook, in particular, is losing a great deal of its importance, especially among younger people. In order not to have to reinvent the wheel again and again, studies on the transferability of findings to new types of platforms are important. While the majority of studies surveyed much-researched platforms such as Twitter {\color{updated}(now X)} and Facebook (e.g., \citep{bhuiyan_feedreflect_2018,kim_eye_2021,ozturk_combating_2015,mena_cleaning_2020}; see Figure \ref{fig:yearanzahl}
and Section \ref{sec:methodologicaloverviews}), there has been little research on misinformation on image- and video-based platforms such as Instagram and TikTok. Given the usage rates of these media, particularly among youth, and the increasing relevance of the platforms for misinformation (e.g., concerning the Russian-Ukrainian war), addressing this research gap is considered particularly relevant. At the same time, there are major obstacles to overcome here, especially with regard to the availability of labeled datasets, {\color{updated}as they typically already exist for Twitter/X but are very time-consuming and complex to establish for video data.}

\textbf{(2) How can collected findings and technical approaches for text-based interventions be applied to emerging video- and image-based misinformation?} With regard to transferability to new platforms, transferability to other information channels is also particularly central. Can text-based user interventions (e.g., \citep{furuta_fact-checking_2021}) be adapted for video- and image-based channels? How can new indicators and measures for detecting misinformation (e.g., image reverse search) be integrated into misinformation interventions? {\color{updated}Challenges researchers are confronted with include the fast-evolving trends in social media. For instance, emotion-evoking features of content on TikTok might solely constitute a characteristic of the platform's content while it might be considered a more valuable indicator for misinformation in other modalities.}

\textbf{(3) How can chances of digital misinformation interventions be effectively combined with the advantages of human experts?} Fully automated mechanisms, e.g., for machine learning-based detection, can handle large amounts of data better than humans. In contrast, trained humans as experts (e.g., journalists \citep{sotirakou_toward_2022}) can handle specific case decisions better than algorithms when the boundary between true and false is blurred and information is missing. How can human expert knowledge be used within digital countermeasures without losing the performance of the automatic tool? In which steps of the countermeasure can human intervention be integrated? 

\textbf{(4) How can automatic detection be combined with user-centered feedback?} Automatic detection is often a black-box procedure and, therefore, cannot explain its decision-making. {\color{updated}While efforts in explainable artificial intelligence already reveal promising results to make detection approaches more transparent without the human directly in the loop, they are still challenged with how to present these explainable outputs in a way that is valuable for a layperson, especially for people with low media literacy.} How can the advantages of accurate automatic detection be combined with transparent and comprehensible explanations as user-centered feedback (see Section \ref{sec:transparen}; initial attempts e.g., by \citet{schmid_digital_2022})?

\textbf{(5) How can media literacy be encouraged without exposing users to misinformation?}
Approaches to increase media literacy are often based on a display of misinformation with additional reference to comprehensible indicators or debunking. Nevertheless, in this case, the user is still exposed to the misinformation. Studies suggest that even with a simultaneous warning, the misinformation content may be remembered at a later point in time, which speaks for less exposure to misinformation \citep{grady_nevertheless_2021} {\color{updated}and constitutes an effect that is controversially discussed in literature \citep{sanderson_listening_2023}}. However, users tend to feel reactance and paternalism when content is hidden or deleted \citep{kirchner_countering_2020}. How can media literacy be trained within misinformation interventions without continuing to expose users to misinformation?

\textbf{(6) Can vulnerable people profit from the general findings of participants with high media literacy? How can we reach vulnerable people with official tools?} The bias toward U.S. adults and college students as study participants continues to be striking. Since not all individuals are equally affected by misinformation \citep{shrestha_online_2020}, but rather particularly vulnerable groups exist, the inclusion of individuals with lower levels of media literacy in the iterative design and evaluation process of the interventions is essential. Initial approaches are already moving in this direction \citep{axelsson_learning_2021}. {\color{updated}Still, researchers are often challenged with conducting user studies outside of the university bubble with convenience samples of participants, as particularly studies with children and teenagers in the context of misinformation come with additional ethical questions and recruiting challenges. For instance, confronting adolescents with misleading information during a user study is a sensible task that needs thorough consideration.} How can the findings be applied to vulnerable people (see Section \ref{sec:methodologicaloverviews}; e.g., older people \citep{sakhnini_review_2022}, children, teenagers \citep{axelsson_learning_2021, skipper_but_2023}, non-native speakers)? How can these target groups be meaningfully integrated into the iterative design and evaluation process?

With our systematic literature review we hope to provide a starting point for cross-disciplinary debates and knowledge exchange, as well as an inspiration for future research.

\section{Limitations \& Conclusion}
\label{sec:limitationsconclusion}
\textit{First}, the large amount of publications in the area of interest, including a variety of different disciplines, was a challenge to deal with. Therefore, we focused specifically on approaches that take place within the real-time usage of social media and excluded approaches (especially educational trainings, games, and presentations) that take place long before the actual usage of social media \citep[e.g., ][]{skipper_but_2023}. However, those approaches may provide additional insights into effective and user-centered interventions and are, therefore, suggested for future research. {\color{updated}In addition, we excluded studies on psychological or social phenomena (e.g., norms) to receive a reasonable number of publications that allows for a thorough focus on research regarding the design and evaluation of digital interventions. These studies are valuable to consider when designing interventions tailored to specific persona and are suggested for future reviews.}

{\color{updated}\textit{Second}, our approach takes a broad perspective on types of misleading information referred to as `misinformation' as an umbrella term and encompassing unintentionally and intentionally misleading information as well as related phenomena (e.g., rumors, conspiracy theories). Table \ref{tab:sokfinaltable} roughly demonstrates in clusters which concept was used in each paper. While we excluded papers that contained a related term but understood it as a phenomenon not fitting within our broad definition (e.g., eyewitnesses remembering something inaccurately as `misinformation'), we did not perform an in-depth analysis of how each term was defined and utilized in each paper. This is a limitation that may have impacted our screening phase. In addition, there might be odd cases of papers not identified within our systematic screening phase as they use different terms to address the topic of misinformation not included in our search term. For instance, a study by \citet{ennals_highlighting_2010} was brought to our attention during the review phase that refers to `disputed claims' and was, thus, not detected.}

\textit{Third}, within our work, we thoroughly categorized the publications regarding multiple characteristics, involving two researchers with expert knowledge in that field of research. As the publications provide information on our categories in varying detail, we cannot exclude the possibility that some interventions were classified differently than they were intended by the authors themselves.

Misinformation remains a threat to the democratic order and the cohesion of society, and the fight against it remains important. It is a central goal to empower users in dealing with the overabundance of information online, especially during emerging crises. Digital misinformation interventions are one of several starting points to address that challenge, complementing professional journalistic work and media literacy training at schools. In this work, we have given an overview of existing countermeasures and have developed a taxonomy in order to systematize misinformation intervention research.   
Finally, we hope that this work – being a first step towards the systematization of misinformation intervention research – serves as an inspiration for future research and facilitates cross-disciplinary exchange of knowledge.

\section{Ackknowledgements}
This work was supported by the German Federal Ministry for Education and Research (BMBF) in the project NEBULA (13N16361).

\bibliographystyle{unsrtnat}
\bibliography{references}  %%% Uncomment this line and comment out the ``thebibliography'' section below to use the external .bib file (using bibtex) .


\newpage
\appendix
\section{Appendix}
\begin{figure}[hb]
    \includegraphics[width=0.5\textwidth]{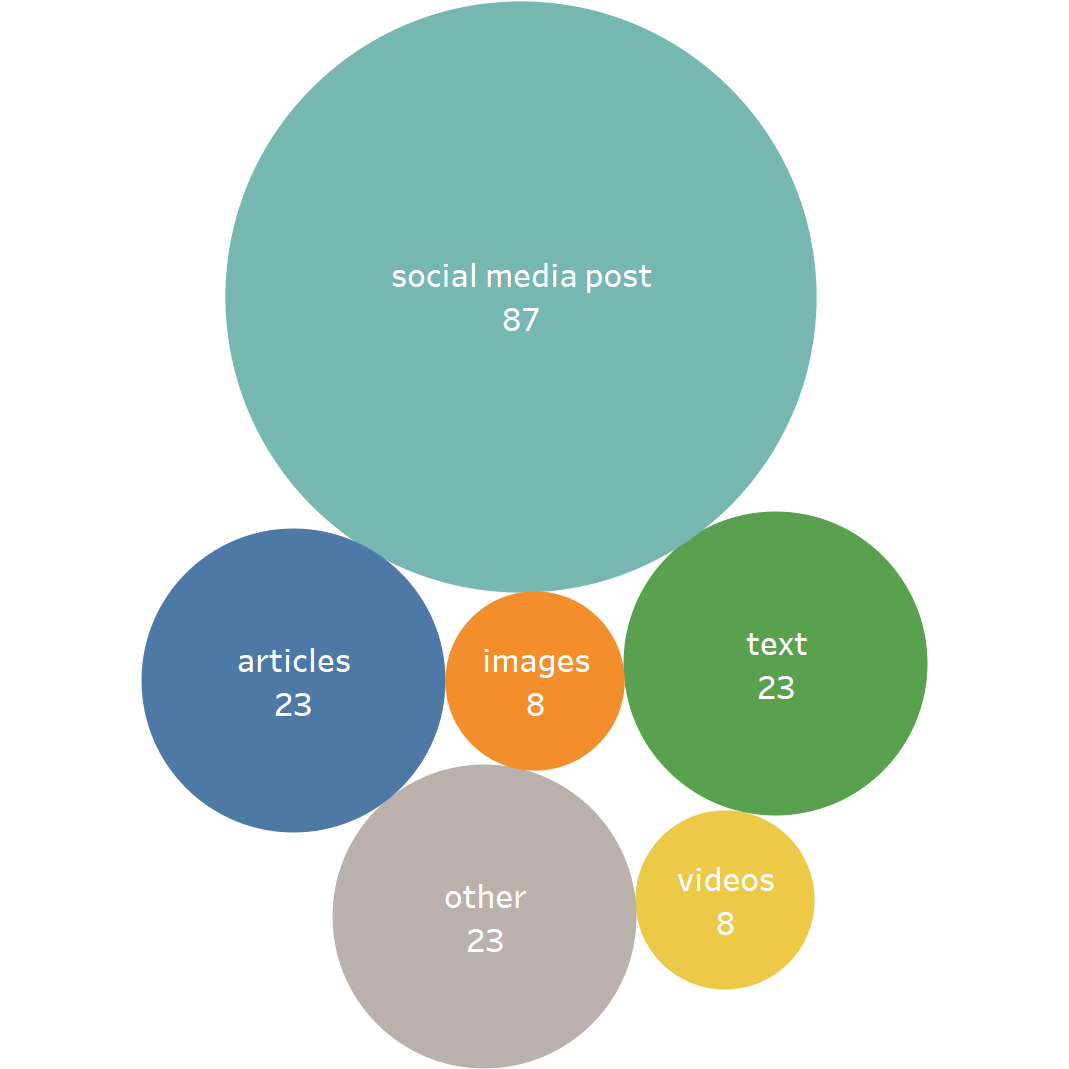}
    \caption{Number of papers addressing specific formats.}
    \label{fig:format}
\end{figure}
\newpage
\begin{landscape}
\section{Electronic Supplement}

\vspace{50px} % fügt vertikalen blank space ein
% Hier kann die Schriftgröße sowie Zeilenhöhe angepasst werden
\fontsize{6.95}{1.81}\selectfont

\setlength{\tabcolsep}{0.3em} % for the horizontal padding -> macht Zellen horizontal schmaler/breiter
{\renewcommand{\arraystretch}{3.65}% for the vertical padding -> macht Zellen vertikal flacher/höher
\centering 

\begin{longtable}{|m{1.57cm}|m{0.42cm}|m{1.0cm}|r|r|r|r|r|r|r|r|r|r|r|r|r|r|r|r|r|r|r|r|r|r|r|r|r|r|r|r|r|r|r|r|r|r|r|r|r|r|r|r|r|r|r|}
\caption{Categorization of {\color{updated}163} publications regarding user intervention characteristics and methodological aspects.}
\label{tab:sokfinaltable}\\

% define first head
\multicolumn{1}{|c|}{\rotatebox{90}{Author}} & \multicolumn{1}{|c|}{\rotatebox{90}{Year}} & \multicolumn{1}{|c|}{\rotatebox{90}{Sample Size}} & \multicolumn{1}{|c|}{\rotatebox{90}{Lab study}} & \multicolumn{1}{|c|}{\rotatebox{90}{\color{updated}Online experiment}} & \multicolumn{1}{|c|}{\rotatebox{90}{Field study}} & \multicolumn{1}{|c|}{\rotatebox{90}{Survey}} & \multicolumn{1}{|c|}{\rotatebox{90}{Interviews}} & \multicolumn{1}{|c|}{\rotatebox{90}{Conceptual}} & \multicolumn{1}{|c|}{\rotatebox{90}{Facebook}} & \multicolumn{1}{|c|}{\rotatebox{90}{Twitter/X}} & \multicolumn{1}{|c|}{\rotatebox{90}{General}} & \multicolumn{1}{|c|}{\rotatebox{90}{Other}} & \multicolumn{1}{|c|}{\rotatebox{90}{Social media posts}} & \multicolumn{1}{|c|}{\rotatebox{90}{Articles}} & \multicolumn{1}{|c|}{\rotatebox{90}{Text}} & \multicolumn{1}{|c|}{\rotatebox{90}{Images}} & \multicolumn{1}{|c|}{\rotatebox{90}{Video}} & \multicolumn{1}{|c|}{\rotatebox{90}{Other}} & \multicolumn{1}{|c|}{\rotatebox{90}{Warning}}& \multicolumn{1}{|c|}{\rotatebox{90}{Correction/debunking}} & \multicolumn{1}{|c|}{\rotatebox{90}{Showing indicators}} & \multicolumn{1}{|c|}{\rotatebox{90}{(Binary) label}} & \multicolumn{1}{|c|}{\rotatebox{90}{Highlighting design}} & \multicolumn{1}{|c|}{\rotatebox{90}{Visibility reduction}} & \multicolumn{1}{|c|}{\rotatebox{90}{Removal}} & \multicolumn{1}{|c|}{\rotatebox{90}{Complicate sharing}} & \multicolumn{1}{|c|}{\rotatebox{90}{Specific visualization}} & \multicolumn{1}{|c|}{\rotatebox{90}{Other}} & \multicolumn{1}{|c|}{\rotatebox{90}{Active}}& \multicolumn{1}{|c|}{\rotatebox{90}{Passive}} & \multicolumn{1}{|c|}{\rotatebox{90}{Neither/unclear}} & \multicolumn{1}{|c|}{\rotatebox{90}{Pre exposure}} & \multicolumn{1}{|c|}{\rotatebox{90}{During}} & \multicolumn{1}{|c|}{\rotatebox{90}{At the moment of sharing}} & \multicolumn{1}{|c|}{\rotatebox{90}{On request}} & \multicolumn{1}{|c|}{\rotatebox{90}{Post exposure}} & \multicolumn{1}{|c|}{\rotatebox{90}{Other}} & \multicolumn{1}{|c|}{\rotatebox{90}{Mis-/Disinformation}} & \multicolumn{1}{|c|}{\rotatebox{90}{Rumors}}& \multicolumn{1}{|c|}{\rotatebox{90}{News credibility}} & \multicolumn{1}{|c|}{\rotatebox{90}{Other}} & \multicolumn{1}{|c|}{\rotatebox{90}{Browser extension/Plugin}} & \multicolumn{1}{|c|}{\rotatebox{90}{Own platform}} & \multicolumn{1}{|c|}{\rotatebox{90}{Game}} & \multicolumn{1}{|c|}{\rotatebox{90}{Other}} \\ \hline
\textbf{Author} & \textbf{Year} & \textbf{Sample Size} & \multicolumn{6}{l |}{\textbf{Evaluation Type}} & \multicolumn{4}{l |}{\textbf{Platform}}  & \multicolumn{6}{l |}{\textbf{Format}}   & \multicolumn{10}{l |}{\textbf{Intervention Design}}  & \multicolumn{3}{l |}{\textbf{Intera.}} & \multicolumn{6}{l |}{\textbf{Timing}} & \multicolumn{4}{l |}{\textbf{Concept}} & \multicolumn{4}{l |}{\textbf{Implem.}}  \\ \hline
\endfirsthead
\multicolumn{4}{c}%
{Continued from previous page} \\
%\toprule
% define each head on following pages
\multicolumn{1}{|c|}{\rotatebox{90}{Author}} & \multicolumn{1}{|c|}{\rotatebox{90}{Year}} & \multicolumn{1}{|c|}{\rotatebox{90}{Sample Size}} & \multicolumn{1}{|c|}{\rotatebox{90}{Lab study}} & \multicolumn{1}{|c|}{\rotatebox{90}{\color{updated}Online experiment}} & \multicolumn{1}{|c|}{\rotatebox{90}{Field study}} & \multicolumn{1}{|c|}{\rotatebox{90}{Survey}} & \multicolumn{1}{|c|}{\rotatebox{90}{Interviews}} & \multicolumn{1}{|c|}{\rotatebox{90}{Conceptual}} & \multicolumn{1}{|c|}{\rotatebox{90}{Facebook}} & \multicolumn{1}{|c|}{\rotatebox{90}{Twitter/X}} & \multicolumn{1}{|c|}{\rotatebox{90}{General}} & \multicolumn{1}{|c|}{\rotatebox{90}{Other}} & \multicolumn{1}{|c|}{\rotatebox{90}{Social media posts}} & \multicolumn{1}{|c|}{\rotatebox{90}{Articles}} & \multicolumn{1}{|c|}{\rotatebox{90}{Text}} & \multicolumn{1}{|c|}{\rotatebox{90}{Images}} & \multicolumn{1}{|c|}{\rotatebox{90}{Video}} & \multicolumn{1}{|c|}{\rotatebox{90}{Other}} & \multicolumn{1}{|c|}{\rotatebox{90}{Warning}}& \multicolumn{1}{|c|}{\rotatebox{90}{Correction/debunking}} & \multicolumn{1}{|c|}{\rotatebox{90}{Showing indicators}} & \multicolumn{1}{|c|}{\rotatebox{90}{(Binary) label}} & \multicolumn{1}{|c|}{\rotatebox{90}{Highlighting design}} & \multicolumn{1}{|c|}{\rotatebox{90}{Visibility reduction}} & \multicolumn{1}{|c|}{\rotatebox{90}{Removal}} & \multicolumn{1}{|c|}{\rotatebox{90}{Complicate sharing}} & \multicolumn{1}{|c|}{\rotatebox{90}{Specific visualization}} & \multicolumn{1}{|c|}{\rotatebox{90}{Other}} & \multicolumn{1}{|c|}{\rotatebox{90}{Active}}& \multicolumn{1}{|c|}{\rotatebox{90}{Passive}} & \multicolumn{1}{|c|}{\rotatebox{90}{Neither/unclear}} & \multicolumn{1}{|c|}{\rotatebox{90}{Pre exposure}} & \multicolumn{1}{|c|}{\rotatebox{90}{During}} & \multicolumn{1}{|c|}{\rotatebox{90}{At the moment of sharing}} & \multicolumn{1}{|c|}{\rotatebox{90}{On request}} & \multicolumn{1}{|c|}{\rotatebox{90}{Post exposure}} & \multicolumn{1}{|c|}{\rotatebox{90}{Other}} & \multicolumn{1}{|c|}{\rotatebox{90}{Mis-/Disinformation}} & \multicolumn{1}{|c|}{\rotatebox{90}{Rumors}}& \multicolumn{1}{|c|}{\rotatebox{90}{News credibility}} & \multicolumn{1}{|c|}{\rotatebox{90}{Other}} & \multicolumn{1}{|c|}{\rotatebox{90}{Browser extension/Plugin}} & \multicolumn{1}{|c|}{\rotatebox{90}{Own platform}} & \multicolumn{1}{|c|}{\rotatebox{90}{Game}} & \multicolumn{1}{|c|}{\rotatebox{90}{Other}} \\ \hline
\textbf{Author} & \textbf{Year} & \textbf{Sample Size} & \multicolumn{6}{l |}{\textbf{Evaluation Type}} & \multicolumn{4}{l |}{\textbf{Platform}}  & \multicolumn{6}{l |}{\textbf{Format}}   & \multicolumn{10}{l |}{\textbf{Intervention Design}}  & \multicolumn{3}{l |}{\textbf{Intera.}} & \multicolumn{6}{l |}{\textbf{Timing}} & \multicolumn{4}{l |}{\textbf{Concept}} & \multicolumn{4}{l |}{\textbf{Implem.}}  \\ \hline
\endhead
% Content
% Authors 1-10
\citet{agley_intervening_2021} & 2021 & 1,000 & & $\bullet$ & & & &  & & & $\bullet$ & & & & & & & $\bullet$ & & & & & & & & & & $\bullet$ & $\bullet$ & & & $\bullet$ & & & & & & $\bullet$ & & & & & & $\bullet$ & \\ \hline
\citet{aird_does_2018} & 2018 & 370 & $\bullet$ & $\bullet$ & & & & & & & $\bullet$ & & & & & & & $\bullet$ & & $\bullet$ & & & & & & & & & & $\bullet$ & & & $\bullet$ & & & & & & & & $\bullet$ & & & $\bullet$ & \\ \hline
\citet{almaliki_misinformation-aware_2019} & 2019 & 100 & & & & $\bullet$ & & & & & $\bullet$ & & $\bullet$ & & & & & & & & & & & & & & & $\bullet$ & & $\bullet$ & & & & & & & $\bullet$ & $\bullet$ & & & & & & & $\bullet$ \\ \hline
\citet{amin_visual_2021} & 2021 & 38 & & $\bullet$ & & & & & &  & & $\bullet$ & & & & & & $\bullet$ & & & & & $\bullet$ & & & $\bullet$ & & & $\bullet$ & & & & $\bullet$ & $\bullet$ & & & & $\bullet$ & & & & & $\bullet$ & & \\ \hline
\citet{andi_nudging_2020} & 2020 & 1,003 & & & & $\bullet$ & & & & & $\bullet$ & & $\bullet$ & & & & & & $\bullet$ & & & & & & & & & & $\bullet$ & & & & $\bullet$ & & & & & $\bullet$ & & & & & & & $\bullet$ \\ \hline
\citet{ardevol-abreu_intentional_2020} & 2020 & \parbox{0.5cm}{\vspace{.2\baselineskip}N1=31\\N2=350\vspace{.2\baselineskip}} & & & & $\bullet$ & $\bullet$ & & & & $\bullet$ & & $\bullet$ & & & & & & $\bullet$ & & & & & $\bullet$ & & & & & $\bullet$ & & & & $\bullet$ & & & & & $\bullet$ & & & & & & & $\bullet$ \\ \hline
\citet{aslett_news_2022} & 2022 & \parbox{0.5cm}{\vspace{.2\baselineskip}N1=3,862\\N2=3,337\\N3=968\vspace{.2\baselineskip}} & & & $\bullet$ & $\bullet$ & & & & & & $\bullet$ & & & & & & $\bullet$ & & & & $\bullet$ & $\bullet$ & & & & & & & $\bullet$ & & & $\bullet$ & & & & & & & $\bullet$ & & $\bullet$ & & & \\ \hline
\citet{autry_correcting_2021} & 2021 & \parbox{0.5cm}{\vspace{.2\baselineskip}N1=357\\N2=75\vspace{.2\baselineskip}} & $\bullet$ & & & & & & & & $\bullet$ & & & & & & & $\bullet$ & & $\bullet$ & & & & & & & & & & $\bullet$ & & & & & & $\bullet$ & & $\bullet$ & & & & & & $\bullet$ & \\ \hline
\citet{axelsson_learning_2021} & 2021 & \parbox{0.5cm}{\vspace{.2\baselineskip}N1=90\\N2=119\vspace{.2\baselineskip}} & & $\bullet$ & & & &  & & & $\bullet$ & & $\bullet$ & & & $\bullet$ & $\bullet$ & & & & & & & & & & & $\bullet$ & & & $\bullet$ & & & & & & $\bullet$ & $\bullet$ & & & & & & & $\bullet$ \\ \hline
\citet{ayoub_combat_2021} & 2021 & 244 & & $\bullet$ & & & & & & & $\bullet$ & & & & $\bullet$ & & & & & & $\bullet$ & & & & & & & & & $\bullet$ & & & $\bullet$ & & & & & $\bullet$ & & & & & $\bullet$ & & \\ \hline
% Authors 11-20
\citet{bachmann_studying_2023} & 2023 & 1,472 &  & $\bullet$ &  &  &  &  & $\bullet$ &  &  &  & $\bullet$ &  &  &  &  &  &  & $\bullet$ & $\bullet$ &  &  &  &  &  &  &  &  & $\bullet$ &  &  & $\bullet$ &  &  &  &  & $\bullet$ &  &  &  &  &  & & $\bullet$ \\ \hline
\citet{barman_does_2023} & 2023 & 348 &  & $\bullet$ &  &  &  &  &  &  & $\bullet$ &  & $\bullet$ &  &  &  &  &  & $\bullet$ & $\bullet$ &  &  &  &  &  &  &  &  &  & $\bullet$ &  &  & $\bullet$ &  &  &  &  & $\bullet$ &  &  &  &  &  & & $\bullet$ \\ \hline
\citet{barua_f-nad_2019} & 2019 & - & & & & & & $\bullet$ & & & & $\bullet$ & & & $\bullet$ & & & & & & & $\bullet$ & & & & & & & & & $\bullet$ & & & & $\bullet$ & & & $\bullet$ & & & & & $\bullet$ & & \\ \hline
\citet{bhuiyan_nudgecred_2021} & 2021 & \parbox{0.5cm}{\vspace{.2\baselineskip}N1=430\\N2=12\vspace{.2\baselineskip}} & $\bullet$ & & $\bullet$ & & & & & $\bullet$ & & & $\bullet$ & & & & & & $\bullet$ & & & & $\bullet$ & $\bullet$ & & & & & & $\bullet$ & & & $\bullet$ & & & & & & & $\bullet$ & & $\bullet$ & & & \\ \hline
\citet{bhuiyan_designing_2021} & 2021 & 31 & $\bullet$ & & & & $\bullet$ & & & & & $\bullet$ & & $\bullet$ & & & & & & & $\bullet$ & & $\bullet$ & & & & & & & $\bullet$ & & & $\bullet$ & & & & & & & $\bullet$ & & & & & $\bullet$ \\ \hline
\citet{bhuiyan_feedreflect_2018} & 2018 & 16 & & & $\bullet$ & & & & & $\bullet$ & & & $\bullet$ & & & & & & $\bullet$ & & & & $\bullet$ & $\bullet$ & & & & & & $\bullet$ & & & $\bullet$ & & & & & & & $\bullet$ & & $\bullet$ & & & \\ \hline
\citet{bode_related_2015} & 2015 & \parbox{0.5cm}{\vspace{.2\baselineskip}N1=524\\N2=500\vspace{.2\baselineskip}} & & & & $\bullet$ & & & $\bullet$ & & & & $\bullet$ & & & & & & & $\bullet$ & & & & & & & & & & $\bullet$ & & & $\bullet$ & & & & & $\bullet$ & & & & & & & $\bullet$ \\ \hline
\citet{bode_see_2018} & 2018 & 136 & & $\bullet$ & & & & & $\bullet$ & & & & $\bullet$ & & & & & & & $\bullet$ & & & & & & & & & & $\bullet$ & & & $\bullet$ & & & & & $\bullet$ & & & & $\bullet$ & & & \\ \hline
\citet{bozarth_wisdom_2023} & 2023 & 18 &  &  &  &  & $\bullet$ &  &  &  &  & $\bullet$ & $\bullet$ &  &  &  &  &  &  & $\bullet$ & $\bullet$ & $\bullet$ &  &  &  &  &  &  &  &  & $\bullet$ &  & $\bullet$ &  &  &  &  & $\bullet$ &  &  &  &  &  & & $\bullet$ \\ \hline
\citet{brashier_timing_2021} & 2021 & 2,683 & & $\bullet$ & & & & & & & $\bullet$ & & $\bullet$ & & & & & & & & & $\bullet$ & & & & & & & & $\bullet$ & & $\bullet$ & $\bullet$ & & & $\bullet$ & & $\bullet$ & & & & & & & $\bullet$ \\ \hline
\citet{buczel_how_2024} & 2024 & 337 &  & $\bullet$ &  &  &  &  &  &  & $\bullet$ &  &  & $\bullet$ &  &  &  &  & $\bullet$ &  &  &  &  &  &  &  &  &  & $\bullet$ & $\bullet$ &  & $\bullet$ &  &  &  & $\bullet$ &  & $\bullet$ &  &  &  &  &  & & $\bullet$ \\ \hline
\citet{capraro_i_2022} & 2022 & \parbox{0.5cm}{\vspace{.2\baselineskip}N1=550\\N2=558\\N3=550\\N4=372\vspace{.2\baselineskip}} & & $\bullet$ & & & & & $\bullet$ & & & & $\bullet$ & & & & & & & & & & & & & & & $\bullet$ & & & $\bullet$ & & & $\bullet$ & & & & $\bullet$ & & & & & & & $\bullet$ \\ \hline
\citet{caramancion_quantification_2022} & 2022 & 327 &  & $\bullet$ &  &  &  &  & &  & $\bullet$ &  &  &  &  &  &  & $\bullet$ &  & $\bullet$ & $\bullet$ &  &  &  &  &  &  & $\bullet$ & $\bullet$ &  &  & $\bullet$ &  &  &  &  &  & $\bullet$ &  &  &  &  &  & & $\bullet$ \\ \hline
\citet{challenger_covid-19_2022} & 2022 & \parbox{0.5cm}{\vspace{.2\baselineskip}N1=1,291\\N2=2,084\vspace{.2\baselineskip}} & & $\bullet$ & & & & & & & $\bullet$ & & & & & & & $\bullet$ & & $\bullet$ & & & & & & & & & & & $\bullet$ & & & & & & $\bullet$ & & & & $\bullet$ & & & & $\bullet$ \\ \hline
\citet{chen_visualbubble_2022} & 2022 & 10 &  &  &  &  & $\bullet$ & & $\bullet$ &  &  &  & $\bullet$ &  &  &  &  &  &  &  &  &  &  &  &  &  & $\bullet$ &  &  &  & $\bullet$ &  &  &  & $\bullet$ &  &  & $\bullet$ &  &  &  &  &  & & $\bullet$ \\ \hline
\citet{chen_examining_2022} & 2022 & 415 & & $\bullet$ & & & & & & & $\bullet$ & & & & & & & $\bullet$ & & & & & & & & & & $\bullet$ & $\bullet$ & & & $\bullet$ & & & & & & $\bullet$ & & & & & & & $\bullet$ \\ \hline
\citet{chiang_impact_2022} & 2022 & 60 & & $\bullet$ & & & & & & & $\bullet$ & & & $\bullet$ & & & & & & & $\bullet$ & & & & & & & & & & $\bullet$ & & & & $\bullet$ & & & & & $\bullet$ & & & $\bullet$ & & \\ \hline
% Authors 21-30
\citet{clayton_real_2020} & 2020 & 2,994 & & & & $\bullet$ & & & $\bullet$ & & & & $\bullet$ & & & & & & $\bullet$ & & & & & & & & & $\bullet$ & & $\bullet$ & & $\bullet$ & $\bullet$ & & & & & $\bullet$ & & & & & & & $\bullet$ \\ \hline
\citet{craig_one_2023} & 2023 & 231 &  & $\bullet$ &  &  &  &  &  &  &  & $\bullet$ &  &  & $\bullet$ &  &  &  &  & $\bullet$ &  &  &  &  &  &  &  & $\bullet$ &  & $\bullet$ &  & $\bullet$ &  &  &  & $\bullet$ &  & $\bullet$ &  &  &  &  &  & & $\bullet$ \\ \hline
\citet{dai_effects_2021} & 2021 & 350 & & $\bullet$ & & & & & & & $\bullet$ & & & & $\bullet$ & & & & & $\bullet$ & & & & & & & & $\bullet$ & & $\bullet$ & & $\bullet$ & & & & $\bullet$ & & $\bullet$ & & & & & & & $\bullet$ \\ \hline
\citet{dai_ask_2022} & 2022 & \parbox{0.5cm}{\vspace{.2\baselineskip}N1=425\\N2=625\vspace{.2\baselineskip}} &  &  &  & $\bullet$ &  &  &  &  & $\bullet$ &  &  &  & $\bullet$ &  &  &  &  & $\bullet$ &  &  &  &  &  &  &  &  &  &  & $\bullet$ &  & $\bullet$ &  &  &  &  & $\bullet$ &  &  &  &  &  & & $\bullet$ \\ \hline
\citet{danry_wearable_2020} & 2020 & 18 & $\bullet$ & & & & & & & & & $\bullet$ & & & & & & $\bullet$ & & & & & & & & & & $\bullet$ & & & $\bullet$ & & $\bullet$ & & & & & $\bullet$ & & & & & & & $\bullet$ \\ \hline
\citet{denner_effects_2023} & 2023 & 211 &  & $\bullet$ & &  &  &  & $\bullet$ &  &  &  & $\bullet$ &  &  &  &  &  &  & $\bullet$ &  &  &  &  &  &  &  &  &  & $\bullet$ &  &  &  &  &  & $\bullet$ &  & $\bullet$ &  &  &  &  &  &  & $\bullet$ \\ \hline
\citet{desai_does_2023} & 2023 & 365 &  & $\bullet$ &   &  &  &  &  & $\bullet$ &  &  & $\bullet$ &  &  &  &  &  &  & $\bullet$ &  &  &  &  &  &  &  &  &  & $\bullet$ &  &  &  &  &  & $\bullet$ &  & $\bullet$ &  &  &  &  &  &  & $\bullet$ \\ \hline
\citet{dobber_effect_2023} & 2023 & 1,054 &  & $\bullet$ &  &  &  &   &  &  &  & $\bullet$ &  &  &  &  & $\bullet$ &  & $\bullet$ &  &  & $\bullet$ &  &  &  &  &  &  & $\bullet$ & $\bullet$ &  & $\bullet$ & $\bullet$ &  &  &  &  &  &  &  & $\bullet$ &  &  &  & $\bullet$ \\ \hline
\citet{domgaard_combating_2021} & 2021 & 250 & & $\bullet$ & & & & & &  & $\bullet$ & &  & $\bullet$ & & & & & & & $\bullet$ & & & & & & $\bullet$ & & & $\bullet$ & & $\bullet$ & & & & & & $\bullet$ & & & & & & & $\bullet$ \\ \hline
\citet{drolsbach_diffusion_2023} & 2023 & 7 &  & $\bullet$ &  &  &  &  &  & $\bullet$ &  &  & $\bullet$ &  &  &  &  &  &  & $\bullet$ &  &  &  &  &  &  &  &  &  & $\bullet$ &  &  & $\bullet$ &  &  &  &  & $\bullet$ &  &  &  &  &  &  & $\bullet$ \\ \hline
\citet{duncan_whats_2020} & 2020 & 390 & & & & $\bullet$ & & & & & & $\bullet$ & & & $\bullet$ & & & & & & & & & & & & & $\bullet$ & & $\bullet$ & & & $\bullet$ & & & & & & & $\bullet$ & & & & & $\bullet$ \\ \hline
\citet{ecker_you_2020} & 2020 & 2,279 & & $\bullet$ & & & & & &  & & $\bullet$ & & $\bullet$ & & & & & & $\bullet$ & & & & & & & & & & $\bullet$ & & & & & & $\bullet$ & & $\bullet$ & & & & & & & $\bullet$ \\ \hline
\citet{ecker_reminders_2017} & 2017 & 60 & $\bullet$ & & & &  & & & & & $\bullet$ & & $\bullet$ & & & & & & $\bullet$ & & & & & & & & & & $\bullet$ & & & & & & $\bullet$ & & & & & $\bullet$ & & & & $\bullet$ \\ \hline
\citet{ecker_correcting_2011} & 2011 & \parbox{0.5cm}{\vspace{.2\baselineskip}N1=161\\N2=138\vspace{.2\baselineskip}} & $\bullet$ & & & &  & & & & & $\bullet$ & & $\bullet$ & & & & & & $\bullet$ & & & & & & & & & & $\bullet$ & & & & & & $\bullet$ & & & & & $\bullet$ & & & & $\bullet$ \\ \hline
 \citet{ecker_can_2020} & 2020 & 1718 & & $\bullet$ & & & &  & & & $\bullet$ & & $\bullet$ & & & & & & & $\bullet$ & & $\bullet$ & & & & & & & & $\bullet$ & & & $\bullet$ & & & & & $\bullet$ & & & & & & & $\bullet$ \\ \hline
 \citet{feng_examining_2023} & 2023 & 595 &  & $\bullet$ &  &  &   &  &  & $\bullet$ &  &  &  &  &  & $\bullet$ & $\bullet$ &  &  &  & $\bullet$ &  &  &  &  &  &  &  &  & $\bullet$ &  &  & $\bullet$ &  &  &  &  & $\bullet$ &  &  &  &  &  &  & $\bullet$ \\ \hline
 \citet{figl_symbol_2023} & 2023 & 256 &  & $\bullet$ &  &  &   &  & $\bullet$ &  &  &  & $\bullet$ &  &  &  &  &  & $\bullet$ &  &  & $\bullet$ &  &  &  &  &  &  &  & $\bullet$ &  &  & $\bullet$ &  &  &  & & $\bullet$ &  &  &  &  &  &  & $\bullet$ \\ \hline
% Authors 31-40
\citet{folkvord_effect_2022} & 2022 & 307 & & $\bullet$ & & & & & $\bullet$ & & & & $\bullet$ & & & & & & $\bullet$ & & & & & & & & & $\bullet$ & $\bullet$ & & & $\bullet$ & & & & & & $\bullet$ & & & & & & & $\bullet$ \\ \hline
\citet{freeze_fake_2021} & 2021 & 434 & & $\bullet$ & & & & && & & $\bullet$ & & $\bullet$ & & & & & $\bullet$ & & & & & & & & & & & $\bullet$ & & & & & & $\bullet$ & & $\bullet$ & & & & & & & $\bullet$ \\ \hline
\citet{furuta_fact-checking_2021} & 2021 & - & & & & & & $\bullet$ & & & & $\bullet$ & & & $\bullet$ & & & & & & $\bullet$ & & $\bullet$ & & & & & & & $\bullet$ & & & & & & & $\bullet$ & $\bullet$ & & & & & & & $\bullet$ \\ \hline
\citet{gao_label_2018} & 2018 & 122 & & $\bullet$ & & & & & & & $\bullet$ & & $\bullet$ & & & & & & $\bullet$ & & & & & & & & & $\bullet$ & & $\bullet$ & & & $\bullet$ & & & & & $\bullet$ & & $\bullet$ & & & $\bullet$ & & \\ \hline
\citet{gesser-edelsburg_correcting_2018} & 2018 & 243 & $\bullet$ & $\bullet$ & & & & & $\bullet$ & & & & $\bullet$ & & & & & & & $\bullet$ & & & & & & & & & & $\bullet$ & & & & & & $\bullet$ & & & & & $\bullet$ & & & & $\bullet$ \\ \hline
\citet{grady_nevertheless_2021} & 2021 & 418 & & & & $\bullet$ & & & & & $\bullet$ & & $\bullet$ & & & & & & $\bullet$ & & & $\bullet$ & & & & & & & $\bullet$ & $\bullet$ & & $\bullet$ & $\bullet$ & & & $\bullet$ & & $\bullet$ & & & & & & & $\bullet$ \\ \hline
\citet{grandhi_by_2021} & 2021 & 376 & & & & $\bullet$ & & & & & $\bullet$ & & $\bullet$ & & & & & & & & $\bullet$ & & & & & & & & & & $\bullet$ & & & & & & $\bullet$ & & & $\bullet$ & & & & & $\bullet$ \\ \hline
\citet{guess_digital_2020} & 2020 & \parbox{0.5cm}{\vspace{.2\baselineskip}N1=9,190\\N2=4,669\\N3=6,439\vspace{.2\baselineskip}} & $\bullet$ & $\bullet$ & & & & & $\bullet$ & & & & $\bullet$ & & & & & & & & $\bullet$ & & & & & & & & & $\bullet$ & & $\bullet$ & & & & & & $\bullet$ & & & & & & & $\bullet$ \\ \hline
\citet{guo_seeing_2023} & 2023 & 28 &  &  &  &  & $\bullet$ &  &    &  &  & $\bullet$ &  &  &  &  & $\bullet$ &  & $\bullet$ & $\bullet$ &  &  &  & $\bullet$ &  &  &  &  & $\bullet$ & $\bullet$ &  & $\bullet$ & $\bullet$ &  &  &  &  & $\bullet$ &  &  &  &  &  &  & $\bullet$ \\ \hline
\citet{hameleers_separating_2022} & 2020 & 1,091 & & $\bullet$ & & & & &  & & $\bullet$ & & & $\bullet$ & & & & & $\bullet$ & $\bullet$ & $\bullet$ & & & & & & & & & $\bullet$ & & $\bullet$ & $\bullet$ & & & & & $\bullet$ & & & & & & & $\bullet$ \\ \hline
\citet{hameleers_picture_2020} & 2020 & 1,404 & & $\bullet$ & & &  & & & $\bullet$ & & & $\bullet$ & & & & & & & $\bullet$ & & & & & & & & & & $\bullet$ & & & $\bullet$ & & & & & $\bullet$ & & & & & & & $\bullet$ \\ \hline
\citet{hameleers_striking_2023} & 2023 & 1,105 &  & $\bullet$ &  & $\bullet$ &  &   &  &  & $\bullet$ &  &  &  & $\bullet$ &  &  &  & $\bullet$ &  & $\bullet$ &  &  &  &  &  &  &  & $\bullet$ &  &  & $\bullet$ &  &  &  &  &  & $\bullet$ &  &  &  &  &  &  & $\bullet$ \\ \hline
% Authors 41-50
\citet{hawa_combating_2021} & 2021 & - & & & & & & $\bullet$ & &  & & $\bullet$ & & & $\bullet$ & & & & & & & $\bullet$ & & & & & & & & & $\bullet$ & & & & $\bullet$ & & & $\bullet$ & & & & & $\bullet$ & & \\ \hline
\citet{heuer_comparative_2022} & 2022 & \parbox{0.5cm}{\vspace{.2\baselineskip}N1=188\\N2=208\vspace{.2\baselineskip}} &  & $\bullet$ &  &  &  &  &  &  & $\bullet$ &  &  & $\bullet$ &  &  &  &  &  &  & $\bullet$ &  & $\bullet$ &  &  &  &  & $\bullet$ &  & $\bullet$ &  &  & $\bullet$ &  &  &  &  & $\bullet$ &  &  &  &  &  &  & $\bullet$ \\ \hline
\citet{horne_trustworthy_2019} & 2019 & - & & & & & & $\bullet$ & & & $\bullet$ & & $\bullet$ & & & & & & & & & & & & & & & $\bullet$ & & & $\bullet$ & & $\bullet$ & & & & & $\bullet$ & & & & & & & $\bullet$ \\ \hline
\citet{huang_when_2022} & 2020 & \parbox{0.5cm}{\vspace{.2\baselineskip}N1=235\\N2=235\vspace{.2\baselineskip}} & & $\bullet$ & & & & & $\bullet$ & & & & $\bullet$ & & & & & & & $\bullet$ & & & & & & & & & & $\bullet$ & & & & & & $\bullet$ & & $\bullet$ & & & & & & & $\bullet$ \\ \hline
\citet{irving_correcting_2022} & 2022 & 129 &  & $\bullet$ &  &  &  &  &  &  & $\bullet$ &  &  &  & $\bullet$ &  &  &  &  & $\bullet$ &  &  &  &  &  &  &  &  &  & $\bullet$ &  &  & $\bullet$ &  &  &  &  & $\bullet$ &  &  &  &  &  &  & $\bullet$ \\ \hline
\citet{jahanbakhsh_leveraging_2022} & 2022 & \parbox{0.5cm}{\vspace{.2\baselineskip}N1=154\\N2=14\vspace{.2\baselineskip}} & & & $\bullet$ & $\bullet$ & & & & & $\bullet$ & & $\bullet$ & & & & & & & & & $\bullet$ & $\bullet$ & & & & $\bullet$ & & $\bullet$ & & & & $\bullet$ & $\bullet$ & & & & $\bullet$ & & & & & $\bullet$ & & \\ \hline
\citet{jahanbakhsh_exploring_2021} & 2021 & 1,668 & & & & $\bullet$ & & &  & & $\bullet$ & & $\bullet$ & & & & & & & & & & & & & $\bullet$ & & & $\bullet$ & & & & & $\bullet$ & & & & $\bullet$ & & & & & & & $\bullet$ \\ \hline
\citet{jahanbakhsh_exploring_2023} & 2023 & 61 &  & $\bullet$ &  &  &    &  &  & $\bullet$ &  &  & $\bullet$ &  &  &  &  &  &  &  &  & $\bullet$ & $\bullet$ &  &  &  &  & $\bullet$ &  &  & $\bullet$ &  & $\bullet$ &  &  &  &  & $\bullet$ &  &  &  &  &  &  & $\bullet$ \\ \hline
\citet{jahanbakhsh_our_2022} & 2022 & \parbox{0.5cm}{\vspace{.2\baselineskip}N1=27\\N2=312\vspace{.2\baselineskip}} &  &  & $\bullet$ & $\bullet$ &  &  &   &  & $\bullet$ &  &  & $\bullet$ &  &  &  &  &  & $\bullet$ &  &  &  &  &  &  & $\bullet$ &  &  & $\bullet$ &  &  & $\bullet$ &  &  &  &  & $\bullet$ &  &  &  & $\bullet$ &  &  &  \\ \hline
\citet{jahng_they_2021} & 2021 & 205 & & $\bullet$ & & & & && & $\bullet$ & & & $\bullet$ & & & & & & $\bullet$ & & & & & & & & & & $\bullet$ & & & $\bullet$ & & & & & $\bullet$ & & & & & & & $\bullet$ \\ \hline
\citet{jennings_asymmetric_2021} & 2021 & \parbox{0.5cm}{\vspace{.2\baselineskip}N1=1,262\\N2=1,586\vspace{.2\baselineskip}} & & $\bullet$ & && & & $\bullet$ & & & & $\bullet$ & & & & & & $\bullet$ & & & & & & & & & & & $\bullet$ & & & $\bullet$ & & & & & $\bullet$ & & & & & & & $\bullet$ \\ \hline
\citet{jia_understanding_2022} & 2022 & 1,677 & & $\bullet$ & & & & & & & $\bullet$ & & $\bullet$ & & & & & & & & & $\bullet$ & & & & & & & & $\bullet$ & & & $\bullet$ & & & & & $\bullet$ & & & & & & & $\bullet$ \\ \hline
\citet{joshi_explainable_2023} & 2023 & - &  &  &  &  &  & $\bullet$ &   &  & $\bullet$ &  & $\bullet$ &  &  &  &  &  & &  & $\bullet$ &  & $\bullet$ &  &  &  &  &  &  & $\bullet$ &  &  & $\bullet$ &  &  &  &  & $\bullet$ &  &  &  &  &  &  & $\bullet$ \\ \hline
\citet{karduni_vulnerable_2019} & 2019 & 5 & & & & & $\bullet$ & &  & $\bullet$ & & & $\bullet$ & & & & & & & & $\bullet$ & & & & & & $\bullet$ & & & & $\bullet$ & $\bullet$ & & & $\bullet$ & & & $\bullet$ & & $\bullet$ & & & $\bullet$ & & \\ \hline
\citet{kessler_debunking_2022} & 2022 & 700 & & $\bullet$ & & & & & & & $\bullet$ & & & & & & & $\bullet$ & & $\bullet$ & & & & & & & $\bullet$ & & & $\bullet$ & & & & & & $\bullet$ & & & & & $\bullet$ & & & & $\bullet$ \\ \hline
% Authors 51-60
\citet{khivasara_fake_2020} & 2020 & - & & & & & & $\bullet$ & & &  & $\bullet$ & & & $\bullet$ & & & & & & & $\bullet$ & & & & & & & $\bullet$ & & & & & & $\bullet$ & & & $\bullet$ & & & & $\bullet$ & & & \\ \hline
\citet{kim_combating_2019} & 2019 & \parbox{0.5cm}{\vspace{.2\baselineskip}N1=590\\N2=299\vspace{.2\baselineskip}} & & & & $\bullet$ & & &  $\bullet$ & & & & $\bullet$ & & & & & & & & & & & & & & & $\bullet$ & & $\bullet$ & & & $\bullet$ & & & & & $\bullet$ & & $\bullet$ & & & & & $\bullet$ \\ \hline
\citet{kim_controversy_2019} & 2019 & - & & & & & & $\bullet$ & & &  & $\bullet$ & $\bullet$ & & & & & & & & & & & & & & $\bullet$ & $\bullet$ & & & $\bullet$ & & & & $\bullet$ & & & & & & $\bullet$ & & $\bullet$ & & \\ \hline
\citet{kim_eye_2021} & 2021 & 92 & $\bullet$ & & & & &  & & $\bullet$ & & & $\bullet$ & & & & & & & $\bullet$ & & & & & & & & & & $\bullet$ & & & $\bullet$ & & & & & $\bullet$ & & & & & & & $\bullet$ \\ \hline
\citet{kim_exoskeleton_2023} & 2023 & \parbox{0.5cm}{\vspace{.2\baselineskip}N1=17\\N2=57\\N3=49\vspace{.2\baselineskip}} & $\bullet$ &  &  &  & $\bullet$  &  &  & $\bullet$ &  &  & $\bullet$ &  &  &  &  &  &  &  & $\bullet$ &  & $\bullet$ & $\bullet$ &  & $\bullet$ &  &  & $\bullet$ & $\bullet$ &  & $\bullet$ &  &  & $\bullet$ &  &  & $\bullet$ &  &  &  & $\bullet$ &  &  & \\ \hline
\citet{kirchner_countering_2020} & 2020 & \parbox{0.5cm}{\vspace{.2\baselineskip}N1=1,012\\N2=15\\N3=1,030\vspace{.2\baselineskip}} & & & & $\bullet$ & & & $\bullet$ & & & & $\bullet$ & & & & & & $\bullet$ & $\bullet$ & & & & $\bullet$ & & $\bullet$ & & & $\bullet$ & $\bullet$ & & $\bullet$ & $\bullet$ & $\bullet$ & & & & $\bullet$ & & & & & & & $\bullet$ \\ \hline
\citet{koch_effects_2023} & 2023 & 571 &  & $\bullet$ &  &  &  &    & $\bullet$ &  &  &  & $\bullet$ &  &  &  &  &  & $\bullet$ &  & $\bullet$ &  &  &  &  &  &  & $\bullet$ & $\bullet$ & $\bullet$ &  &  & $\bullet$ &  &  &  &  & $\bullet$ &  &  &  &  &  &  & $\bullet$ \\ \hline
\citet{komendantova_value-driven_2021} & 2021 & \parbox{0.5cm}{\vspace{.2\baselineskip}N1=103\\N2=68\\N3=50\vspace{.2\baselineskip}} & & & & & $\bullet$ & & & & $\bullet$ & & & & & & & $\bullet$ & & & & & & & & & & $\bullet$ & & & $\bullet$ & & & & & & $\bullet$ & $\bullet$ & & & & $\bullet$ & & & \\ \hline
\citet{kreps_covid-19_2022} & 2022 & 2,000 & & $\bullet$ & & & & & &  & $\bullet$ & & & & & & & $\bullet$ & $\bullet$ & $\bullet$ & & $\bullet$ & & $\bullet$ & & & & & $\bullet$ & & & & $\bullet$ & & & & & $\bullet$ & & & & & & & $\bullet$ \\ \hline
\citet{lee_effect_2022} & 2022 & 171 & & $\bullet$ & & & & & $\bullet$ & & & & $\bullet$ & & & & & & & $\bullet$ & & $\bullet$ & & & & & & & & $\bullet$ & & & & & & $\bullet$ & & $\bullet$ & & & & $\bullet$ & & & $\bullet$ \\ \hline
\citet{lee_misvis_2022} & 2022 & - &  &  & &  &  & $\bullet$ &  &  & & $\bullet$ &  &  &  &  &  & $\bullet$ & &  & $\bullet$ & $\bullet$ & $\bullet$ &  &  &  & $\bullet$ &  &  &  & $\bullet$ &  &  &  & $\bullet$ &  &  & $\bullet$ &  &  &  &  & $\bullet$ &  &  \\ \hline
\citet{lee_user_2023} & 2023 & 377 &  & $\bullet$ &  &  &  &   & $\bullet$ &  &  &  & $\bullet$ &  &  &  &  &  & $\bullet$ & $\bullet$ &  & $\bullet$ &  &  &  &  &  &  & $\bullet$ & $\bullet$ &  & $\bullet$ & $\bullet$ &  &  &  &  & $\bullet$ &  &  &  &  &  &  & $\bullet$ \\ \hline
\citet{lee_correcting_2024} & 2024 & 502 &  & $\bullet$ &  &  &  &  & $\bullet$ &  &  & & $\bullet$ &  &  &  &  &  &  & $\bullet$ &  &  &  &  &  &  &  &  &  & $\bullet$ &  &  & $\bullet$ &  &  & $\bullet$ &  & $\bullet$ &  &  &  &  &  &  & $\bullet$ \\ \hline
\citet{liaw_maater_2013} & 2013 & ? & & $\bullet$ & & & && & & & $\bullet$ & & $\bullet$ & & & & & & $\bullet$ & & & $\bullet$ & & & & & & & $\bullet$ & & & $\bullet$ & & & & & $\bullet$ & & & & $\bullet$ & & & \\ \hline
\citet{lim_effects_2023} & 2023 & 200 &  & $\bullet$ &  &  &  &   &  &  & $\bullet$ &  & $\bullet$ &  &  &  &  &  & $\bullet$ &  &  & $\bullet$ &  &  &  &  &  &  &  & $\bullet$ &  &  & $\bullet$ &  &  &  &  & $\bullet$ &  &  &  &  &  &  & $\bullet$ \\ \hline
\citet{liu_checking_2023} & 2023 & 859 &  & $\bullet$ &  &  &   &  &  & $\bullet$ &  &  & $\bullet$ &  &  &  &  &  &  & $\bullet$ &  &  &  &  &  &  &  &  &  & $\bullet$ &  &  &  &  &  & $\bullet$ &  & $\bullet$ &  &  &  &  &  &  & $\bullet$ \\ \hline
\citet{lo_all_2021} & 2021 & 89 & & $\bullet$ & & & & && & $\bullet$ & & & $\bullet$ & & & & & & $\bullet$ & & & & & & & & & & & $\bullet$ & & $\bullet$ & & & & & $\bullet$ & & & & & $\bullet$ & & \\ \hline
% Authors 61-70
\citet{lu_effects_2022} & 2022 & \parbox{0.5cm}{\vspace{.2\baselineskip}N1=538\\N2=1,098\vspace{.2\baselineskip}} & & $\bullet$ & & & & & & & $\bullet$ & & & $\bullet$ & & & & & & & & $\bullet$ & & & & & & $\bullet$ & & $\bullet$ & & & $\bullet$ & & & $\bullet$ & & $\bullet$ & & & & & & & $\bullet$ \\ \hline
\citet{martel_youre_2021} & 2021 & 2,228 & & $\bullet$ & & && & $\bullet$ & $\bullet$ & & & $\bullet$ & & & & & & & $\bullet$ & & & & & & & & & & $\bullet$ & & & & & & $\bullet$ & & $\bullet$ & & & & & & & $\bullet$ \\ \hline
\citet{martino_prta_2020} & 2020 & - & & & & & & $\bullet$ & & & & $\bullet$ & & & $\bullet$ & & & & & & $\bullet$ & & $\bullet$ & & & & & & & & $\bullet$ & & & & $\bullet$ & & & & & & $\bullet$ & & $\bullet$ & & \\ \hline
\citet{mena_cleaning_2020} & 2020 & 501 & & $\bullet$ & & & &  & $\bullet$ & & & & $\bullet$ & & & & & & $\bullet$ & & & & & & & & & & & $\bullet$ & & & $\bullet$ & & & & & $\bullet$ & & & & & & & $\bullet$ \\ \hline
\citet{moon_how_2023-1} & 2022 & 354 & & $\bullet$ & & & & & $\bullet$ & & & & $\bullet$ & & & & & & & $\bullet$ & & & & & & & & & & $\bullet$ & & & $\bullet$ & & & & & $\bullet$ & & & & & & & $\bullet$ \\ \hline
\citet{moravec_appealing_2020} & 2020 & 398 & & $\bullet$ & & & & & $\bullet$ & & & & $\bullet$ & & & & & & $\bullet$ & & & $\bullet$ & & & & & & $\bullet$ & & $\bullet$ & & $\bullet$ & $\bullet$ & & & & & $\bullet$ & & & & & & & $\bullet$ \\ \hline
\citet{nekmat_nudge_2020} & 2020 & 929 & & $\bullet$ & & & & & & & & $\bullet$ & & $\bullet$ & & & & & $\bullet$ & & & & & & & & & & & $\bullet$ & & & $\bullet$ & & & & & $\bullet$ & & & & & & & $\bullet$ \\ \hline
\citet{ozturk_combating_2015} & 2015 & 259 & & & & $\bullet$ & &  & & $\bullet$ & & & $\bullet$ & & & & & & $\bullet$ & $\bullet$ & & & & & & & & & & $\bullet$ & & & $\bullet$ & & & & & & $\bullet$ & & & & & & $\bullet$ \\ \hline
\citet{papakyriakopoulos_impact_2022} & 2022 & - &  &  &  &  &  & $\bullet$ &  & $\bullet$ &  &  & $\bullet$ &  &  &  &  &  & $\bullet$ &  &  & $\bullet$ &  &  &  &  &  &  &  & $\bullet$ &  &  & $\bullet$ &  &  &  &  & $\bullet$ &  &  &  &  &  & & $\bullet$ \\ \hline
\citet{park_experimental_2021} & 2021 & 11,145 & & & & $\bullet$ & & & & &  & $\bullet$ & & & $\bullet$ & & & & $\bullet$ & & & & & & & & $\bullet$ & & & $\bullet$ & & & $\bullet$ & & & & & & $\bullet$ & & & & & & $\bullet$ \\ \hline
\citet{pasquetto_social_2022} & 2022 & \parbox{0.5cm}{\vspace{.2\baselineskip}N1=2,805\\N2=25\vspace{.2\baselineskip}} & & $\bullet$ & & & $\bullet$ & & & & & $\bullet$ & & & & & & $\bullet$ & & $\bullet$ & & $\bullet$ & & & & & & $\bullet$ & & $\bullet$ & & & & & & $\bullet$ & & $\bullet$ & & & & & & & $\bullet$ \\ \hline
% Authors 71-80
\citet{pennycook_implied_2020} & 2020 & \parbox{0.5cm}{\vspace{.2\baselineskip}N1=5\\271\\N2=1,568\vspace{.2\baselineskip}} & & $\bullet$ & & & & & $\bullet$ & & & & $\bullet$ & & & & & & $\bullet$ & & & $\bullet$ & & & & & & & & $\bullet$ & & & $\bullet$ & & & & & $\bullet$ & & & & & & & $\bullet$ \\ \hline
\citet{pennycook_fighting_2020} & 2020 & \parbox{0.5cm}{\vspace{.2\baselineskip}N1=853\\N2=856\vspace{.2\baselineskip}} & & & & $\bullet$ & & &  & & $\bullet$ & & $\bullet$ & & & & & & & & & & & & & & & $\bullet$ & $\bullet$ & & & $\bullet$ & & & & & & $\bullet$ & & & & & & & $\bullet$ \\ \hline
\citet{pennycook_shifting_2021} & 2021 & N>5,000 & & & $\bullet$ & $\bullet$ & & & & $\bullet$ & & & $\bullet$ & & & & & & & & & & & & & & & $\bullet$ & & $\bullet$ & & $\bullet$ & & & & $\bullet$ & & $\bullet$ & & & & & & & $\bullet$ \\ \hline
\citet{pillai_explaining_2023} & 2023 & 499 &  & $\bullet$ &  &  &  &  &  &  &  &  &  &  &  &  &  & $\bullet$ &  &  &  &  &  &  &  &  &  & $\bullet$ &  & $\bullet$ &  &  &  &  &  & $\bullet$ &  & $\bullet$ &  &  &  &  &  &  & $\bullet$ \\ \hline
\citet{pluviano_misinformation_2017} & 2017 & 120 & $\bullet$ & & & & & & & & $\bullet$ & & & & $\bullet$ & & & & & $\bullet$ & & & & & & & $\bullet$ & & & $\bullet$ & & & $\bullet$ & & & & & $\bullet$ & & & $\bullet$ & & & & $\bullet$ \\ \hline
\citet{porter_political_2022} & 2022 & \parbox{0.5cm}{\vspace{.2\baselineskip}N1=5,000\\N2=2,000\vspace{.2\baselineskip}} & & $\bullet$ & & &  & & $\bullet$ & & & & $\bullet$ & & & & & & $\bullet$ & $\bullet$ & & $\bullet$ & & $\bullet$ & & & & & $\bullet$ & & & & $\bullet$ & & & & & $\bullet$ & & & & & & & $\bullet$ \\ \hline
\citet{porter_factual_2022} & 2022 & 5,075 &  & $\bullet$ &  & &  &   &  &  & $\bullet$ &  & $\bullet$ &  &  &  &  &  &  & $\bullet$ &  &  &  &  &  &  &  &  &  & $\bullet$ &  &  & $\bullet$ &  &  &  &  & $\bullet$ &  &  &  &  &  &  & $\bullet$ \\ \hline
\citet{pourghomi_how_2017} & 2017 & - & & & & & & $\bullet$ & $\bullet$ & & & & $\bullet$ & & & $\bullet$ & & & & & & & & & & & $\bullet$ & & & $\bullet$ & & & & & $\bullet$ & & & $\bullet$ & & & & & & & $\bullet$ \\ \hline
\citet{pretus_imisleadingi_2024} & 2024 & \parbox{0.5cm}{\vspace{.2\baselineskip}N1=1,709\\N2=804\vspace{.2\baselineskip}} &  & $\bullet$ &  &  & &  &  & $\bullet$ &  &  & $\bullet$ &  &  &  &  &  & $\bullet$ &  & $\bullet$ &  &  &  &  &  &  & $\bullet$ &  & $\bullet$ &  &  & $\bullet$ &  &  &  &  & $\bullet$ &  &  &  &  &  &  & $\bullet$ \\ \hline
\citet{qian_fighting_2023} & 2023 & 905 &  & $\bullet$ &  &  &  &  &  &  & $\bullet$ &  &  &  &  & $\bullet$ &  &  &  &  &  &  &  &  &  &  & $\bullet$ &  & $\bullet$ & $\bullet$ &  & $\bullet$ &  &  &  &  &  & $\bullet$ &  &  &  &  &  &  & $\bullet$ \\ \hline
\citet{rich_correcting_2020} & 2020 & \parbox{0.5cm}{\vspace{.2\baselineskip}N1=134\\N2=134\\N3=102\vspace{.2\baselineskip}} & & $\bullet$ & & & &  & & & & $\bullet$ & & $\bullet$ & & & & & & $\bullet$ & & & & & & & & & & $\bullet$ & & & & & & $\bullet$ & & $\bullet$ & & & & & & & $\bullet$ \\ \hline
\citet{ruffin_explaining_2022} & 2022 & \parbox{0.5cm}{\vspace{.2\baselineskip}N1=113\\N2=543\vspace{.2\baselineskip}} &  &  &  & $\bullet$ &  &  & &  & $\bullet$ &  &  &  &  & $\bullet$ &  &  &  &  & $\bullet$ &  & $\bullet$ &  &  &  &  &  &  & $\bullet$ &  &  & $\bullet$ &  &  &  &  &  &  &  & $\bullet$ &  &  &  & $\bullet$ \\ \hline
\citet{scharrer_watch_2022-1} & 2022 & 41 &  & $\bullet$ &  &  &   &  &  &  & $\bullet$ &  &  &  & $\bullet$ &  &  &  & $\bullet$ &  &  & $\bullet$ &  &  &  &  &  &  &  & $\bullet$ &  & $\bullet$ & $\bullet$ &  &  &  &  & $\bullet$ &  &  &  &  &  &  & $\bullet$ \\ \hline
\citet{safieddine_corporate_2016} & 2016 & - & & & & & & $\bullet$ & & & & $\bullet$ & & & $\bullet$ & $\bullet$ & & & & & & & & & & & $\bullet$ & & & $\bullet$ & & & & & $\bullet$ & & & $\bullet$ & & & & $\bullet$ & $\bullet$ & & \\ \hline
\citet{sakhnini_review_2022} & 2022 & 11 & & & & & $\bullet$ & & &  & $\bullet$ & & & & & & & $\bullet$ & & $\bullet$ & $\bullet$ & $\bullet$ & & & & & & & & & $\bullet$ & & & & $\bullet$ & & & $\bullet$ & & & & & $\bullet$ & & \\ \hline
\citet{saltz_encounters_2021} & 2021 & \parbox{0.5cm}{\vspace{.2\baselineskip}N1=15\\N2=23\vspace{.2\baselineskip}} & & & & & $\bullet$ & & & & $\bullet$ & & $\bullet$ & & & $\bullet$ & $\bullet$ & & $\bullet$ & & & $\bullet$ & & $\bullet$ & $\bullet$ & & & $\bullet$ & $\bullet$ & $\bullet$ & & & & & & & $\bullet$ & $\bullet$ & & & & & & & $\bullet$ \\ \hline
% Authors 81-90
\citet{sangalang_potential_2019} & 2019 & \parbox{0.5cm}{\vspace{.2\baselineskip}N1=385\\N2=586\vspace{.2\baselineskip}} & & $\bullet$ & & & & & & & & $\bullet$ & & & $\bullet$ & & & & & $\bullet$ & & & & & & & & & & $\bullet$ & & & & & & $\bullet$ & & $\bullet$ & & & & & & & $\bullet$ \\ \hline
\citet{schaewitz_combating_2020} & 2020 & 221 & & $\bullet$ & & & & &  & & & $\bullet$ & & $\bullet$ & & & & & & $\bullet$ & & & & & & & & & & $\bullet$ & & & & & & $\bullet$ & & $\bullet$ & & & & & & & $\bullet$ \\ \hline
\citet{schmid_digital_2022} & 2022 & \parbox{0.5cm}{\vspace{.2\baselineskip}N1=9\\N2=7\vspace{.2\baselineskip}} & & $\bullet$ & & & $\bullet$ & & & $\bullet$ & & & $\bullet$ & & & & & & & & $\bullet$ & $\bullet$ & & & & & $\bullet$ & & & & $\bullet$ & & & & $\bullet$ & & & $\bullet$ & & & & & $\bullet$ & & \\ \hline
\citet{schmid_benefits_2022} & 2022 & \parbox{0.5cm}{\vspace{.2\baselineskip}N1=2,444\\N2=817\vspace{.2\baselineskip}} &  & $\bullet$ &  &  &  &  &  &  & $\bullet$ &  &  &  & $\bullet$ &  &  &  &  & $\bullet$ &  &  &  &  &  &  &  &  & $\bullet$ &  &  & $\bullet$ &  &  &  & $\bullet$ &  & $\bullet$ &  &  &  &  &  &  & $\bullet$ \\ \hline
\citet{seo_trust_2019} & 2019 & \parbox{0.5cm}{\vspace{.2\baselineskip}N1=522\\N2=624\vspace{.2\baselineskip}} & & $\bullet$ & & &  & & & & $\bullet$ & & $\bullet$ & & & & & & $\bullet$ & & & & & & & & & & & $\bullet$ & & & $\bullet$ & & & & & $\bullet$ & & & & & & & $\bullet$ \\ \hline
\citet{sharevski_two_2021} & 2021 & 304 & & & & $\bullet$ & & & & $\bullet$ & & & & & & & & $\bullet$ & & & & & & & & & & $\bullet$ & $\bullet$ & $\bullet$ & & & $\bullet$ & & & & & $\bullet$ & & & & & & & $\bullet$ \\ \hline
\citet{sharevski_i_2023} & 2023 & 29 &  &  &  &  & $\bullet$ &  &  &   & $\bullet$ &  &  &  &  &  & $\bullet$ &  & $\bullet$ &  &  & $\bullet$ &  &  &  &  &  &  &  & $\bullet$ &  &  & $\bullet$ &  &  &  &  & $\bullet$ &  &  &  &  &  &  & $\bullet$ \\ \hline
\citet{sharevski_meaningful_2022} & 2022 & 337 &  & $\bullet$ &   &  &  &  &  & $\bullet$ &  &  & $\bullet$ &  &  &  &  &  & $\bullet$ &  &  &  &  &  &  &  &  &  &  & $\bullet$ &  &  & $\bullet$ &  &  &  &  & $\bullet$ &  &  &  &  &  &  & $\bullet$ \\ \hline
\citet{sheikh_ali_tahaqqaq_2023} & 2023 & - &  &  &  &  &  & $\bullet$ &   & $\bullet$ &  &  &  &  & $\bullet$ &  &  &  &  & $\bullet$ & $\bullet$ & $\bullet$ & $\bullet$ &  &  &  &  &  &  & $\bullet$ &  &  &  &  & $\bullet$ &  &  & $\bullet$ &  &  &  &  &  &  & $\bullet$ \\ \hline
\citet{sherman_designing_2021} & 2021 & \parbox{0.5cm}{\vspace{.2\baselineskip}N1=24\\N2=19\\N3=1,456\vspace{.2\baselineskip}} & & $\bullet$ & & & $\bullet$ & & & & $\bullet$ & & & & & & $\bullet$ & & $\bullet$ & & $\bullet$ & $\bullet$ & $\bullet$ & $\bullet$ & & & $\bullet$ & & & $\bullet$ & & & $\bullet$ & & & & & $\bullet$ & & & & & & & $\bullet$ \\ \hline
\citet{smith_correcting_2019} & 2019 & 744 & & $\bullet$ & & & & &  $\bullet$ & & & & & & & & & $\bullet$ & & $\bullet$ & & & & & & & & & & $\bullet$ & & & $\bullet$ & & & & & $\bullet$ & & & $\bullet$ & & & & $\bullet$ \\ \hline
\citet{song_fighting_2022} & 2022 & 610 &  & $\bullet$ &  &  &  &    & $\bullet$ &  &  &  & $\bullet$ &  &  &  &  & &  & $\bullet$ &  &  &  &  &  &  &  &  &  & $\bullet$ &  &  & $\bullet$ &  &  &  &  & $\bullet$ &  &  &  &  &  &  & $\bullet$ \\ \hline
\citet{sotirakou_toward_2022} & 2022 & - & & & & & & $\bullet$ & &  & $\bullet$ & & & & & & & $\bullet$ & & $\bullet$ & & $\bullet$ & & & & & & $\bullet$ & & & $\bullet$ & & & & $\bullet$ & & & $\bullet$ & & & & & $\bullet$ & & \\ \hline
\citet{sullivan_leveraging_2019} & 2019 & \parbox{0.5cm}{\vspace{.2\baselineskip}N1=625\\N2=600\vspace{.2\baselineskip}} & & $\bullet$ & & & & & $\bullet$ & & & & $\bullet$ & & & & & & & $\bullet$ & & & & & & & & & & $\bullet$ & & & $\bullet$ & & & & & $\bullet$ & & & & & & & $\bullet$ \\ \hline
\citet{tanaka_exposure_2019} & 2019 & 164 & & $\bullet$ & & & & & & $\bullet$ & & & $\bullet$ & & & & & & & $\bullet$ & & & & & & & & & & $\bullet$ & & & & & & $\bullet$ & & & $\bullet$ & & & & & & $\bullet$ \\ \hline
% Authors 91-100
\citet{tanaka_toward_2013} & 2013 & 87 & & $\bullet$ & & & &  & & $\bullet$ & & & $\bullet$ & & & & & & & $\bullet$ & & & & & & & & & & $\bullet$ & & $\bullet$ & & & & $\bullet$ & & & $\bullet$ & & & & & & $\bullet$ \\ \hline
\citet{tao_hope_2023} & 2023 & 836 &  & $\bullet$ &  &  &  &   &  &  & $\bullet$ &  & $\bullet$ &  &  &  &  &  &  & $\bullet$ &  &  &  &  &  &  &  &  &  & $\bullet$ &  &  &  &  &  & $\bullet$ &  & $\bullet$ &  &  &  &  &  &  & $\bullet$ \\ \hline
\citet{thornhill_digital_2019} & 2019 & 20 & & & & $\bullet$ & & &  & $\bullet$ & & & & & $\bullet$ & & & & & $\bullet$ & & $\bullet$ & $\bullet$ & & & & & & & $\bullet$ & & & $\bullet$ & & & & & $\bullet$ & & & & & $\bullet$ & & \\ \hline
\citet{tseng_investigating_2022} & 2022 & 210 & & $\bullet$ & & & & & & & $\bullet$ & & & & $\bullet$ & $\bullet$ & $\bullet$ & & & $\bullet$ & & & & & & & & & & $\bullet$ & & & & & & $\bullet$ & & $\bullet$ & & & & & & & $\bullet$ \\ \hline
\citet{tsipursky_fighting_2018} & 2018 & 21 & & & $\bullet$ & & &  & $\bullet$ & & & & $\bullet$ & & & & & & & & & & & & & & & $\bullet$ & $\bullet$ & & & $\bullet$ & & & & & & $\bullet$ & & & & & & & $\bullet$ \\ \hline
\citet{tulin_beyond_2024} & 2024 & 752 &  & $\bullet$ &  &  &   &  &  &  & $\bullet$ &  &  & $\bullet$ &  &  &  &  &  & $\bullet$ &  & $\bullet$ &  &  &  &  &  & &  & $\bullet$ &  &  & $\bullet$ &  &  &  &  & $\bullet$ &  &  &  &  &  &  & $\bullet$ \\ \hline
\citet{tully_mobilizing_2020} & 2020 & 610 & & $\bullet$ & & & & & & $\bullet$ & & & $\bullet$ & & & & & & & $\bullet$ & & & & & & & & & & $\bullet$ & & & $\bullet$ & & & & & $\bullet$ & & & & & & & $\bullet$ \\ \hline
\citet{tully_designing_2020} & 2020 & \parbox{0.5cm}{\vspace{.2\baselineskip}N1=702\\N2=787\vspace{.2\baselineskip}} & & $\bullet$ & & & & & & $\bullet$ & & & $\bullet$ & & & & & & & & & & & & & & & $\bullet$ & & $\bullet$ & & $\bullet$ & & & & & & $\bullet$ & & & & & & & $\bullet$ \\ \hline
\citet{van_der_meer_seeking_2020} & 2020 & 700 & & $\bullet$ & & & &  & & & $\bullet$ & & & $\bullet$ & & & & & & $\bullet$ & & & & & & & & & & $\bullet$ & & & & & & $\bullet$ & & $\bullet$ & & & & & & & $\bullet$ \\ \hline
\citet{van_der_meer_can_2023} & 2023 & 1,305 &  & $\bullet$ &  &    &  &  &  &  & $\bullet$ &  & & $\bullet$ &  &  &  &  & $\bullet$ & $\bullet$ &  &  &  &  &  &  &  & $\bullet$ &  & $\bullet$ &  &  &  &  &  & $\bullet$ &  & $\bullet$ &  &  &  &  &  &  & $\bullet$ \\ \hline
\citet{velasco_factit_2023} & 2023 & 285 &  &  &  & $\bullet$ &  &  &  &   & $\bullet$ &  &  &  & $\bullet$ &  &  &  &  &  &  & $\bullet$ &  &  &  &  &  &  &  &  & $\bullet$ &  &  &  & $\bullet$ &  &  & $\bullet$ &  &  &  & $\bullet$ &  &  &  \\ \hline
\citet{velez_latino-targeted_2023} & 2023 & 2,869 &  & $\bullet$ &  &  &  &    &  &  & $\bullet$ &  & $\bullet$ &  &  &  &  &  &  & $\bullet$ &  &  &  &  &  &  &  &  &  & $\bullet$ &  &  &  &  &  & $\bullet$ &  & $\bullet$ &  &  &  &  &  &  & $\bullet$ \\ \hline
\citet{von_der_weth_nudging_2020} & 2020 & - & & & & & & $\bullet$ & & $\bullet$ & & & $\bullet$ & & & & & & $\bullet$ & & $\bullet$ & & & & & $\bullet$ & & $\bullet$ & $\bullet$ & & & & & $\bullet$ & & & & $\bullet$ & & & & $\bullet$ & & & \\ \hline
\citet{vraga_assessing_2021} & 2021 & 916 & & $\bullet$ & & & &  & & $\bullet$ & & & $\bullet$ & & & & & & & $\bullet$ & & & & & & & & $\bullet$ & & $\bullet$ & & & $\bullet$ & & & & & $\bullet$ & & & & & & & $\bullet$ \\ \hline
\citet{vraga_i_2018} & 2018 & 1,384 & & $\bullet$ & & & & & & $\bullet$ & & & $\bullet$ & & & & & & & $\bullet$ & & & & & & & & & & $\bullet$ & & & $\bullet$ & & & & & $\bullet$ & & & & & & & $\bullet$ \\ \hline
% Authors 101-108
\citet{vraga_using_2017} & 2017 & 271 & & $\bullet$ & & & &  & $\bullet$ & $\bullet$ & & & $\bullet$ & & & & & & & $\bullet$ & & & & & & & & & & $\bullet$ & & & $\bullet$ & & & & & $\bullet$ & & & & & & & $\bullet$ \\ \hline
\citet{vraga_creating_2022} & 2022 & \parbox{0.5cm}{\vspace{.2\baselineskip}N1=1,207\\N2=603\vspace{.2\baselineskip}} & & $\bullet$ & & & & & & $\bullet$ & & & $\bullet$ & & & & & & & $\bullet$ & & & & & & & & $\bullet$ & & $\bullet$ & & & $\bullet$ & & & & & $\bullet$ & & & & & & & $\bullet$ \\ \hline
\citet{vraga_effects_2021} & 2021 & 1348 & & $\bullet$ & & & & & $\bullet$ & & & & & & & & $\bullet$ & & & $\bullet$ & & & & & & & & $\bullet$ & & $\bullet$ & & $\bullet$ & $\bullet$ & & & & & $\bullet$ & & & & & & & $\bullet$ \\ \hline
\citet{vraga_testing_2019} & 2019 & 406 & & $\bullet$ & & & & & $\bullet$ & & & & $\bullet$ & & & & & & & $\bullet$ & & & & & & & & & & $\bullet$ & & & & & & $\bullet$ & & $\bullet$ & & & & & & & $\bullet$ \\ \hline
\citet{vraga_testing_2020} & 2020 & 1,005 & & $\bullet$ & & & & &  & & & $\bullet$ & $\bullet$ & & & & & & & $\bullet$ & & & & & & & & $\bullet$ & $\bullet$ & $\bullet$ & & $\bullet$ & & & & $\bullet$ & & $\bullet$ & & & & & & & $\bullet$ \\ \hline
\citet{wahlheim_reminders_2020} & 2020 & 96 & $\bullet$ &  &  &  &  &  &  &  & $\bullet$ &  &  &  &  &  &  & $\bullet$ &  & $\bullet$ &  &  &  &  &  &  &  & $\bullet$ &  & $\bullet$ &  &  &  &  &  & $\bullet$ &  & $\bullet$ &  &  &  &  &  &  & $\bullet$ \\ \hline
\citet{waltenberger_reddit_2023} & 2023 & 9 &  &  & $\bullet$ &  & $\bullet$ &  &    &  &  & $\bullet$ & $\bullet$ &  &  &  &  &  &  &  & $\bullet$ & $\bullet$ &  &  &  &  &  &  &  & $\bullet$ & &  & $\bullet$ &  &  &  &  & $\bullet$ &  &  &  & $\bullet$ &  &  & \\ \hline
\citet{wang_countering_2021} & 2021 & 271 & & $\bullet$ & & & & & &  & & $\bullet$ & $\bullet$ & & & & & & & $\bullet$ & & & & & & & & & & $\bullet$ & & & & & & $\bullet$ & & $\bullet$ & & & & & & & $\bullet$ \\ \hline
\citet{wang_pm_2022} & 2022 & \parbox{0.5cm}{\vspace{.2\baselineskip}N1=601\\N2=1,060\vspace{.2\baselineskip}} &  & $\bullet$ &  & $\bullet$ &  &  & $\bullet$ &  &  & $\bullet$ & $\bullet$ & $\bullet$ &  &  &  &  & $\bullet$ & $\bullet$ &  &  &  &  &  &  &  &  &  & $\bullet$ &  &  & $\bullet$ &  &  &  &  & $\bullet$ &  &  &  &  &  &  & $\bullet$ \\ \hline
\citet{wang_rumorlens_2022} & 2022 & 1 & $\bullet$ &  &  &  &  &  &   &  & $\bullet$ &  & $\bullet$ &  &  &  &  &  &  &  &  &  &  &  &  &  & $\bullet$ &  &  &  & $\bullet$ &  &  &  & $\bullet$ &  &  &  & $\bullet$ &  &  &  & $\bullet$ &  &  \\ \hline
\citet{westbrook_mechanisms_2023} & 2023 & \parbox{0.5cm}{\vspace{.2\baselineskip}N1=125\\N2=138\\N3=251\vspace{.2\baselineskip}} &  & $\bullet$ &  &    &  &  &  &  & $\bullet$ &  &  & $\bullet$ &  &  &  &  &  & $\bullet$ &  &  &  &  &  &  &  &  &  & $\bullet$ &  &  &  &  &  & $\bullet$ &  & $\bullet$ &  &  &  &  &  &  & $\bullet$ \\ \hline
\citet{wijnker_debunking_2022} & 2022 & 441 &  & $\bullet$ &  &  &  &   &  &  & $\bullet$ &  &  &  &  &  &  & $\bullet$ & $\bullet$ & $\bullet$ & $\bullet$ &  & $\bullet$ &  &  &  &  &  &  & $\bullet$ &  &  & $\bullet$ &  &  &  &  & $\bullet$ &  &  &  &  &  &  & $\bullet$ \\ \hline
\citet{wood_promoting_2023} & 2023 & 2,257 &  & $\bullet$ &  &  &  &   &  &  & $\bullet$ &  &  & $\bullet$ &  &  &  &  &  & $\bullet$ &  &  &  &  &  &  &  &  &  & $\bullet$ &  & $\bullet$ &  &  &  & $\bullet$ &  & $\bullet$ &  &  &  &  &  &  & $\bullet$ \\ \hline
\citet{yong_explainability_2023} & 2023 & - &  &  &  &  &  & $\bullet$ &  &  & $\bullet$ &  &  &  & $\bullet$ &  &  &  &  &  & $\bullet$ &  & $\bullet$ &  &  &  &  &  &  & $\bullet$ &  &  & $\bullet$ &  &  &  &  & $\bullet$ &  &  &  &  &  &  & $\bullet$ \\ \hline
\citet{zade_tweet_2023} & 2023 & 21 &  &  &  &  & $\bullet$ &  &  & $\bullet$ &  &  & $\bullet$ &  &  &  &  &  &  &  & $\bullet$ &  &  &  &  &  & $\bullet$ &  &  & $\bullet$ &  &  & $\bullet$ &  &  &  &  & $\bullet$ &  &  &  &  &  &  & $\bullet$ \\ \hline
\citet{zhang_investigation_2022} & 2022 & - &  &  &  &  &  & $\bullet$ &  & &  & $\bullet$ & $\bullet$ &  &  &  &  &  & $\bullet$ & $\bullet$ &  & $\bullet$ &  &  &  &  & $\bullet$ &  &  &  & $\bullet$ &  & $\bullet$ &  &  &  &  & $\bullet$ &  &  &  &  &  &  & $\bullet$ \\ \hline
\citet{zhao_misinformation_2019} & 2019 & 252 & & $\bullet$ & & & &  & $\bullet$ & $\bullet$ & & $\bullet$ & $\bullet$ & & & & & & & $\bullet$ & & & & & & & & & & & $\bullet$ & & & & & & $\bullet$ & $\bullet$ & & & & & & & $\bullet$ \\ \hline
\citet{zheng_evaluating_2022} & 2022 & 222 &  & $\bullet$ &  &  &  &  &  &  & $\bullet$ &  &  &  &  &  &  & $\bullet$ &  &  &  &  & $\bullet$ &  &  &  & $\bullet$ &  &  &  & $\bullet$ &  & $\bullet$ &  &  &  &  & $\bullet$ &  &  &  &  &  &  & $\bullet$ \\ \hline
%\caption{Categorization of {\color{updated}163} publications regarding user intervention characteristics and methodological aspects.}
%\label{tab:sokfinaltable}\\
%}
\end{longtable}
}

\end{landscape}

%%%%%%%%%%%%
% Table effects
\newpage
\begin{landscape}
\fontsize{6.95}{1.81}\selectfont

\begin{longtable}{p{1.6cm} p{7cm} p{4cm} p{5cm}}
\caption{{\color{updated}Overview of effects and perceptions of reviewed misinformation interventions (conceptual studies without evaluation were excluded).}}
\label{tab:effectperceptions}\\
    \textbf{Source} & \textbf{Beneficial Effects} & \textbf{Beneficial Perceptions} & \textbf{Not effective / counterproductive}\\
\midrule
\endfirsthead
\multicolumn{4}{c}%
{Continued from previous page} \\
\toprule
    \textbf{Source} & \textbf{Beneficial Effects} & \textbf{Beneficial Perceptions} & \textbf{Not effective / counterproductive}\\
\midrule
\endhead
%\midrule
%\multicolumn{4}{r}{Continued on next page} \\
\endfoot
%\bottomrule
\endlastfoot
    \citet{agley_intervening_2021} & Exposure to infographics with scientific information slightly increases trust in science compared to exposure to control infographic. &  & Exposure to infographics with scientific information does not have direct or indirect effects on COVID-19 preventive behaviors.\\ \cline{1-4}
    \citet{aird_does_2018} & Exposure to fact-checks corrects beliefs and affects voters’ support when corrections outnumber affirmations compared to other correction ratios and for both sides of the political spectrum ($\eta^2$=0.13 (fact checks); $\eta^2$=0.01 (myth:fact ratio). &  &  \\\cline{1-4}
    \citet{almaliki_misinformation-aware_2019} &  & Users perceive interventions with gamification elements useful but preferences for elements vary. &  \\\cline{1-4}
    \citet{amin_visual_2021} & Interventions with Visual Selective Attention System can increase attentive behavior of COVID-19 misinformation sharing compared to pre-intervention (D-Scores similar to Cohen's d: Highest number of participants in category `Neutral/ No Preference' (D-score=-0.15 to D-score=0.15) &  &  \\\cline{1-4}
    \citet{andi_nudging_2020} & Social norm-based nudge decreases misinformation sharing behavior compared to non-application. &  &  \\\cline{1-4}
    \citet{ardevol-abreu_intentional_2020} &  &  & Warning labels to assess credibility are not regarded as central assessment measures by users. \\\cline{1-4}
    \citet{aslett_news_2022} &  &  & Providing dynamic, in-feed source reliability labels do not significantly improve news diet quality or reduce misperceptions ($<$0.08 change in SD of the pre-treatment measure). \\\cline{1-4}
    \citet{autry_correcting_2021} &  &  & Negated corrections and replacements lead to increased belief in misinformation for cases with no previous exposure to the target concept, relative to cases with exposure and cases with no treatment ($\eta^2$=0.22 (main effect of exposure); $\eta^2$=0.23(main effect of correction); $\eta^2$=0.18 (interaction between exposure and correction)). \\\cline{1-4}
    \citet{axelsson_learning_2021} & Observational learning and feedback as intervention tools increase user performance of credibility assessment compared to the non-treatment control group ($\eta^2$=0.043) &  &  \\\cline{1-4}
    \citet{ayoub_combat_2021} & Additional employment of SHapley Additive exPlanations (SHAP) in NLP misinformation detection model and SHAP combined with source and evidence information increases user trust in misinformation detection compared to presenting output text only. &  &  \\\cline{1-4}
    \citet{bachmann_studying_2023} & Fact-checks are similarly effective at reducing people's misperceptions across message formats (transparency elements, arousing visuals) (d=0.51 (Study 1) and d=0.38 (Study 2)) & & Compared to control groups without intervention, users exposed to political fact-checks trust news less and perceive the media as more biased, especially after reading corrections debunking pro-attitudinal misinformation.\\\cline{1-4}
    \citet{barman_does_2023} & Warning flags with and without explanation text from fact-checking websites reduce perceived accuracy of misinformation and intent to share. Explanatory texts could enhance the trustworthiness of the intervention. & &\\\cline{1-4}
    \citet{bhuiyan_nudgecred_2021} & Credibility nudges as browser extension improve user's skills to distinguish news tweets’ credibility compared to control group (d=0.296) &  &  \\\cline{1-4}
    \citet{bhuiyan_designing_2021} &  & Transparency cues (source and message credibility) on news websites increase consumer trust. &  \\\cline{1-4}
    \citet{bhuiyan_feedreflect_2018} & Attention and reflection nudges enhance users' credibility assessment compared to control group & Attention and reflection nudges enhance users' credibility assessment (reread and rethink news; use external sources; actively participate in assessment) compared to control group. &  \\\cline{1-4}
    \citet{bode_related_2015} & Exposure to corrective information decreases user misperceptions compared to pre-treatment and to the control group ($\eta^2$=0.052) &  &  \\\cline{1-4}
    \citet{bode_see_2018} & Interventions with algorithmic or social corrections are equally effective in health misinformation corrections compared to control conditions without intervention for high and low conspiracy belief individuals ($\eta^2$=0.046 (interventions overall); $\eta^2$=0.016 (comparison between algorithmic and social correction)) &  &  \\\cline{1-4}
    \citet{bozarth_wisdom_2023} & &Almost half of participants (moderators on Reddit) preferred cues over labels from expert fact-checkers as they can help discern user intent. A quarter distrusts professional fact-checkers.&\\\cline{1-4}
    \citet{brashier_timing_2021} & Debunking measures have a stronger long-term impact on users' fact-checking memory than prebunking, labeling, or no measures. &  &  \\\cline{1-4}
    \citet{buczel_how_2024} & Warning before misinformation reduces reliance on it in short-term in comparison to no warning. Warning after misinformation had no effect ($\eta^2$=0.05 (forwarning vs. retraction only)) & &Reliance on misinformation increased for over 7 days although the memory of retraction continued.\\\cline{1-4}
    \citet{capraro_i_2022} & Accuracy endorsement prompt nudge reduces fake news sharing but also increases sharing of real news compared to simple fake alert and no-nudge (f=0.129 (two nudges); f=0.125 (two nudges, different UI); f=0.129 (comparison between endorsing accuracy condition and accuracy salience condition)) &  &  \\\cline{1-4}
    \citet{caramancion_quantification_2022} & & &Preventive infographics have trivial to no effect on social media users\\\cline{1-4}
    \citet{chiang_impact_2022} & AI news source credibility system positively affects users’ information assessment and attitude towards media literacy learning. &  &  \\\cline{1-4}
    \citet{challenger_covid-19_2022} & Myth-busting formats, question-answer formats and fact-myth formats are more effective interventions than fact-only formats and control baseline in reducing COVID-19 misinformation agreement ratings. &  &  \\\cline{1-4}
    \citet{chen_examining_2022} & Intervention with narrative fear appeal messages are effective in promoting health experts to correct online health misinformation for the public. &  &  \\\cline{1-4}
    \citet{chen_visualbubble_2022} & Correct assessment of misinformation overall improved by VisualBubble. Participants became more willing to make assessments and more critical (effect sizes: Topic Filter: large (d=0.98 and 0.98); Opinion Filter: negligible (d=0.00) and medium (d=0.79); Source Filter: large (d=1.11 and 1.01)) & &Showed tendency to become over-skeptical\\\cline{1-4}
    \citet{clayton_real_2020} & Intervention with a general warning about misleading articles reduce the perceived accuracy of false headlines relative to a no-warning condition and `rated false' tag is more effective than `disputed' tag.(d=0.08 (general warning before seeing headlines); d=0.26 (`disputed' tag); d=0.38 (`rated false' tag)) &  &  \\\cline{1-4}
    \citet{craig_one_2023} & Corrective infographic improved rating of misinformation as untruthful and reduced reported willingness to share it. Debunking may be short-lived if followed by misinformation. Effect can be maintained in presence of further corrective information (e.g., $\eta^2$=0.150, 0.109 and 0.079) & &\\\cline{1-4}
    \citet{dai_effects_2021} & Timing of misinformation correction interventions (pre/post exposure) and addition of coherence message (debiasing/no addition) impacts effectiveness ($\eta^2$=0.087 (post exposure); $\eta^2$=0.047 (debiasing message); $\eta^2$=0.163 (time lapse)) &  &  \\\cline{1-4}
    \citet{dai_ask_2022} & & Most participants indicate that counterfactual explanations can accurately explain why a piece of news is fake and results suggest that the approach generates the most helpful explanations compared to state-of-the-art methods (human evaluation based on survey with young, well-educated participants).&\\\cline{1-4}
    \citet{danry_wearable_2020} &  & Wearable AI system with explainable feedback enhances rationality in evaluating information in comparison to non-explainable AI and control group &  \\\cline{1-4}
    \citet{denner_effects_2023} & A single correction and repeated corrections significantly increased organizational trust compared with no correction & &Small negative effect of perceived persuasive intent on organizational trust after repeated corrections.\\\cline{1-4}
    \citet{desai_does_2023} & & &No evidence that corrections explaining the reason the misinformation was presented were more effective than a correction not accompanied by explanation\\\cline{1-4}
    \citet{dobber_effect_2023} & Red and orange traffic light labels placed concurrently with in contrast to prior to the start of a political advertisement significantly affect credibility perception. Direct-to-consumer labels can be effective but it depends on timing and position. & &\\\cline{1-4}
    \citet{domgaard_combating_2021} & Interventions with info graphs increase user ability to identify vaccine-related misinformation compared to text-only intervention and no intervention. &  &  \\\cline{1-4}
    \citet{drolsbach_diffusion_2023} & Community fact-checked misinformation is less viral and receives fewer retweets than non-misleading posts. & &\\\cline{1-4}
    \citet{duncan_whats_2020} & Credibility labels are effective on news validation when ideological perspective of the user match the ideology of the news brand but also in cases where they do not match. & & \\\cline{1-4}
    \citet{ecker_you_2020} & Corrections are generally effective at influencing inferential reasoning but narrative corrections are not more effective than non-narrative & & \\\cline{1-4}
    \citet{ecker_reminders_2017} & Corrections are more effective when they explicitly repeat the myth compared to corrections that do not repeat the misinformation ($\eta^2$=0.04 (memory); $\eta^2$=0.27 (inferential reasoning)) & & \\\cline{1-4}
    \citet{ecker_correcting_2011} & Strong corrections and cognitive load interventions, measured in different degrees of interventions or misinformation strength, can reduce (but never fully) the continued influence effect of strong misinformation, but even strong interventions are less effective on weak misinformation. ($\eta^2$=0.05 (strength of misinformation); $\eta^2$=0.41 (strength of correction); $\eta^2$=0.04 (strength of cognitive load on misinformation); $\eta^2$=0.07 (strength of cognitive load on correction);) & & \\\cline{1-4}
    \citet{ecker_can_2020} & Misinformation corrections do not lead to familiarity backfire effects but instead lead to corrective effect in both, audiences unfamiliar to a misinformation and audiences familiar to the topic (i.a., $\eta^2$=0.024 (false claim inference across all conditions: no-exposure/fact-check with and without cognitive load); $\eta^2$=0.004 (fact check condition without cognitive load)) & & \\\cline{1-4}
    \citet{feng_examining_2023} & Provenance has effect on credibility perception. Helped correct truth judgments towards deceptive media (qualitatively measured) & &Over-corrected in some cases and shifted away from truth in some non-deceptive media\\\cline{1-4}
    \citet{figl_symbol_2023} & All evaluated flags lead to reduced perceived credibility. The semantic priming effect of different warning symbols (e.g., stop symbol associated with stopping behavior) makes a difference. Stronger warnings may be required on smartphones than on PCs.& &\\\cline{1-4}
    \citet{folkvord_effect_2022} & Interventions with source information positively affect the critical news evaluation compared to a control group with no intervention (e.g., $\eta^2$=0.05 for vaccination misinformation) & &Inclusion of a protective warning message does not significantly affect critical evaluation (e.g., $\eta^2$<0.001 for vaccination and health insurance misinformation). \\\cline{1-4}
    \citet{freeze_fake_2021} & General misinformation warnings which also contain invalid instances, in contrast to valid-only instances and control with no intervention, lead to a discarding of authentic information and to increased memory uncertainty.  & & \\\cline{1-4}
    \citet{gao_label_2018} & & & Stance labels on political ideologies intensify readers' selective exposure (tendency to look for agreeable opinions), and lower the perception of extremeness and criticality of misinformation. Credibility labels only have a limited effect on reducing selective exposure and misinformation identification. \\\cline{1-4}
    \citet{gesser-edelsburg_correcting_2018} & Corrections of misinformation from health organizations are more effective for pro-vaccination as well as for vaccination-hesitant individuals when communication addresses full, transparent information and emotional aspects compared to `common' one-dimensional, partial responses. & Additional qualitative analysis reinforces quantitative findings. & \\\cline{1-4}
    \citet{grady_nevertheless_2021} & Misinformation warnings for political news are effective in short-term to correct beliefs and eliminate partisan bias but in long-term corrected beliefs weaken and biases return. & & \\\cline{1-4}
    \citet{grandhi_by_2021} &  & Users perceive trustworthiness indicators as useful for reducing uncertainty and for providing guidance on content interaction. & \\\cline{1-4}
    \citet{guess_digital_2020} & Digital media literacy interventions increase user ability to discern between correct information and misinformation compared to control group without intervention (d=0.2 (US-based study); d=0.11 (India-based study)) & \\\cline{1-4}
    \citet{guo_seeing_2023} & Specific contextual warnings for video-sharing platforms can alert users to be vigilant and are influenced by explicitness and risk level. In terms of accuracy judgment the interstitial warning and specific contextual warning were both considered effective.& &\\\cline{1-4}
    \citet{hameleers_separating_2022} & A combination of media literacy- and fact-checking interventions are most effective in lowering perceived accuracy of political misinformation, compared to each intervention separately and control group without intervention. & & \\\cline{1-4}
    \citet{hameleers_striking_2023} & General rather than issue-specific warnings about misinformation are more effective for participants with higher level of trust in the media. & &The prebunking exposure to different warning interventions did not influence the truth rating of factually accurate information or misinformation. Observed negative spillover effects of prebunking warnings on truth rating of accurate information.\\\cline{1-4}
    \citet{hameleers_picture_2020} & Multimodality (text-plus-visual) impacts credibility of disinformation but also of fact-checking interventions compared to disinformation and intervention with text-only and compared to control without intervention. & & Source type (ordinary citizen, news agency) does not influence credibility level \\\cline{1-4}
    \citet{heuer_comparative_2022} & Checklist with source labels is significantly better in influencing participants' performance on correct article ratings for the better.&Checklist that provides source labels was considered most helpful. The interactive checklist is perceived as more helpful than the written checklist.&\\\cline{1-4}
    \citet{horne_trustworthy_2019} & Soft information nudging/trust nudging has potential benefit of moving even extreme or conspiracy news consumers towards higher quality information (based on simulations) & & \\\cline{1-4}
    \citet{huang_when_2022} & Misinformation belief is impacted by the message format (narrative/non-narrative) and correction mechanism (social/algorithmic correction) ($\eta^2$=0.03 and 0.04 (message format);) & & \\\cline{1-4}
    \citet{irving_correcting_2022} & Correction reduces number of references to misinformation (medium-to-large effect size) and was remembered and recalled ($\delta$=0.64, 95\% BCI [0.28, 0.99] (medium-to-large)) & &\\\cline{1-4}
    \citet{jahanbakhsh_leveraging_2022} & Lightweight nudging interventions (checkboxes, checklists, free-text rationales) which provide accuracy assessment and rationale reduce misinformation sharing (but also sharing overall). & & \\\cline{1-4}
    \citet{jahanbakhsh_exploring_2021} & Users perceive incorporation of three new user affordances into social media as useful tools to independent, user-friendly misinformation combat. & Qualitative examples reinforce quantitative findings. & \\\cline{1-4}
    \citet{jahanbakhsh_exploring_2023} & Personalized AI impacts users' judgment and grows larger over time, but is reduced when users provide reasoning for their assessment (e.g., exp($\beta$)=1.60 for condition whether AI's prediction had a statistically significant effect on user agreeing with AI)& &\\\cline{1-4}
    \citet{jahanbakhsh_our_2022} & Users perceived value in browser extension that allows to change headlines and used it to make various changes. In follow-up study: substantial number of alternative headlines were preferred especially if bias was removed or deceptions were corrected. & &\\\cline{1-4}
    \citet{jahng_they_2021} & Discounting cues (`fake news' labels) in online comments negatively impact users' ability of veracity evaluation and increase need to authenticate information compared to control group without exposure to discounting cues. (i.a., $\eta^2$=0.041 (evaluation ability); $\eta^2$=0.057 (need to authenticate))) & & \\\cline{1-4}
    \citet{jennings_asymmetric_2021} & Partisan affiliations impact likeliness to belief in misinformation, particularly about opposing parties (i.a., $\eta^2$=0.13 (user partisanship (P) and party-affiliation of misinformation target(M)); $\eta^2$=0.01 (P, M and fact-check condition (F)) & & Overall, independent from partisan affiliation, fact check interventions do not improve information evaluation compared to cases without intervention. \\\cline{1-4}
    \citet{jia_understanding_2022} & Interventions with misinformation labels (algorithm, community, third-party fact-checker, and no label) reduce credibility of misinformation for liberal users independent of post-ideology while only algorithm labels are effective in reducing ideology-consistent misinformation for conservative users (and all label types for opposing-ideology misinformation). & & \\\cline{1-4}
    \citet{karduni_vulnerable_2019} & & Visual analytic systems are a helpful tool to support the investigation of misinformation on social media and to enhance traditional (media literacy education) strategies. & \\\cline{1-4}
    \citet{karduni_vulnerable_2019} & Corrections of health-related misinformation with additional use of images is more effective in correcting myth belief than without images ($\eta^2$=0.117) & & Image type (machine-technical image, expert image, diagram) does not influence persuasive effect.\\\cline{1-4}
    \citet{kim_combating_2019} & Source rating mechanisms are effective interventions to correct users beliefs, whereby expert rating and user article rating are more effective than user source rating. Low ratings and no-ratings have a disproportional stronger effect on user skepticism than high ratings on user trust. & & \\\cline{1-4}
    \citet{kim_controversy_2019} & Controversy score that provides additional information of opinions on topics and encourages further exploration can be a more effective tool to combat myth belief than approaches that seek to correct or standardize news opinions. & & \\\cline{1-4}
    \citet{kim_exoskeleton_2023} & & &No single strategy ((1) hiding content, allowing for explanations, and option to toggle view, (2) including an engagement option with the correction that allows for indicator details, (3) Placing agent next to share button that asks for accuracy and reasoning and presents statistics) was superior over the control. Study highlights necessity of transparency and clarity about intervention's logic and concerns about repeated exposure to misinformation and lack of user engagement.\\\cline{1-4}
    \citet{kim_eye_2021} & Humorous interventions increase user attention to relevant corrections of misinformation, but non-humorous interventions outperform humorous interventions via higher credibility ratings. ($\eta^2$=0.19 humor) & & \\\cline{1-4}
    \citet{kirchner_countering_2020} & Warning-based interventions significantly effect perceived news accuracy but explanation-based approaches are most effective. & Warning-based interventions (with additional explanations) are more effective in correcting user beliefs than less transparent methods such as reduced post size and fact-checks in related articles. & \\\cline{1-4}
    \citet{komendantova_value-driven_2021} & & Stakeholders (journalists/fact-checkers, policymakers, citizens) require design tools for mitigating misinformation and prioritise information regarding actors behind misinformation and tracing the life cycle of misinforming posts. The most valued features across groups relate to timing and flow of misinformation. & \\\cline{1-4}
    \citet{koch_effects_2023} & Warning labels reduced perceived credibility and lowered self-reported likelihood to amplify fake news (rather small effect). & &Removing social endorsement cues (e.g., engagement counts) did not have an effect. Did not find a positive effect of warning labels on users' likelihood to elaborate on the fake news post.\\\cline{1-4}
    \citet{kreps_covid-19_2022} & Compared to no intervention, `false' tags only have a small effect on users' accuracy assessments while journalistic fact-checks are more effective in reducing misperceptions as well as sharing (independent of partisanship). & & \\\cline{1-4}
    \citet{lee_correcting_2024} & & &Repeated exposure of myths within corrective information increased perceived familiarity about misinformation and increased misinformation credibility (partial $\eta^2$=.02 (effect of correction types on misinformation familiarity))\\\cline{1-4}
    \citet{lee_effect_2022} & Web add-on corrections generally decrease the belief in misinformation compared to no correction. For those who are motivated to use social media for specifically for social interaction, narrative corrections are most effective, compared to web add-on's and no corrections ($\eta^2$=0.025 (narrative correction for social interaction-motivated users)) & & Amongst users in general, narrative corrections are not more effective than web add-on corrections or no corrections.\\\cline{1-4}
    \citet{lee_user_2023} & Both commenting and AI fact-checking labels were effective at promoting positive attitudes toward vaccination compared to no intervention. Commenting intervention emerged as promising for suburban participants and the AI intervention was pronounced for urban populations ($\eta^2$=.03 (for difference in attitudes between three experimental groups)) & &Neither of the interventions showed salient effects with the rural population.\\\cline{1-4}
    \citet{liaw_maater_2013} & & The proposed system utilizes crowd-sourced corrections, such as in-line commentary and corrections which are ranked by the user to enhance comprehension of news.  &\\\cline{1-4}
    \citet{lim_effects_2023} & The intent to comment and share was significantly lower for posts with a generic warning label than unlabeled posts. The knowledge, source, and propagation labels encouraged sharing instead. Partisanship effects were observed across the labels (partial $\eta^2$=0.016 for effect of warning labels on sharing intentions and 0.0077 on commenting intention) & &\\\cline{1-4}
    \citet{liu_checking_2023} & No differences in effectiveness across fact-checking sources (professional fact-checkers, mainstream news outlets, social media platforms, AI, crowd-sourcing; $\eta^2$=0.01) but sources perceived as more credible are more effective & &\\\cline{1-4}
    \citet{lo_all_2021} & & Indicates effectiveness of an fake news intervention module that co-works with a news recommendation system and guides users towards verified news. \\\cline{1-4}
    \citet{lu_effects_2022} & & & AI label nudges people into aligning their veracity belief in the news with the AI model’s prediction regardless of its correctness compared to a control group (Control vs. AI-before: d= 0.17; Control vs. AI-after: d=0.15) \\\cline{1-4}
    \citet{martel_youre_2021} & & & Hedging corrections or providing increased explanatory depth in corrections of misinformation had no impact on engagement with corrective messages on social media.\\\cline{1-4}
    \citet{martino_prta_2020} & &The Prta system raises awareness about the use of propaganda techniques in the news, promoting media literacy and critical thinking.& \\\cline{1-4}
    \citet{mena_cleaning_2020} & A warning label was effective in reducing the intention of a user to share misinformation on Facebook compared to a user who did not see the warning.  (d=0.36) & & \\\cline{1-4}
    \citet{moon_how_2023-1} & AI and user consensus (vs. human experts) source labels reduced partisan-based motivated reasoning in assessing fact-checking message credibility ($\eta^2$=0.0018 for pattern of motivated reasoning varied by fact-checking sources) & &\\\cline{1-4}
    \citet{moravec_appealing_2020} & System 1 (automatic cognition) and System 2 (deliberate cognition) interventions both were effective and intervention combining both was twice as effective. & & \\\cline{1-4}
    \citet{nekmat_nudge_2020} & A fact-check alert was effective in reducing likelihood of sharing misinformation compared to non-exposure.  & & \\\cline{1-4}
    \citet{ozturk_combating_2015} & A textual counter presented to a rumor was effective in decreasing the likelihood of sharing a tweet compared to the rumor only and rumor with information condition. & & \\\cline{1-4}
    \citet{papakyriakopoulos_impact_2022} & Textual overlap in labels reduces user interactions and stronger rebuttals reduced toxicity in comments. & &Label placement did not change propensity of users to share and engage with labeled content but falsity of content did.\\\cline{1-4}
    \citet{park_experimental_2021} & When opposite fact-checking labels are shown, users who initially disapprove of a claim are less likely to change their views than those who initially approve of the same claim. & User interviews revealed that users are more likely to share claims with a Divided Evidence label than those with a Lack of Evidence label. & \\\cline{1-4}
    \citet{pasquetto_social_2022} & Audio files on WhatsApp were found to be more effective than text or video-based sources in correcting beliefs about misinformation and they were shared more frequently when communicated by someone close to the user.  & & \\\cline{1-4}
    \citet{pennycook_implied_2020} & Warnings were effective in a modest reduction in perceived accuracy of false headlines, particularly for politically concordant headlines, relative to a control condition. & & The presence of warnings caused untagged headlines to be seen as more accurate than in the control, even if they were false. \\\cline{1-4}
    \citet{pennycook_fighting_2020} & Simple accuracy reminders before sharing information on social media are effective in increasing truth discernment in participants' sharing intentions compared to a control group (d=0.142) & & \\\cline{1-4}
    \citet{pennycook_shifting_2021} & Shifting the attention of the users on the accuracy of information can encourage them to share higher quality news (e.g., Pearson's r=0.71/0.67/0.61) & & \\\cline{1-4}
    \citet{pillai_explaining_2023} & Participants were less likely to share false headlines in the explain prompt condition compared to control group (exceeded the necessary number of participants according to a priori power analysis; $\eta^2$=.03) & &\\\cline{1-4}
    \citet{pluviano_misinformation_2017} & & & Displaying a myth about vaccines causing autism alongside a factual correction resulted in an increase in belief in the myth over a 7 day time period (partial $\eta^2$=0.175)\\\cline{1-4}
    \citet{porter_factual_2022} & Corrections eliminate effects of misinformation on beliefs about vaccine. Effect is robust to formatting changes in the presentation of corrections. Corrections without any formatting modifications are effective at reducing false beliefs with formatting variations playing a very minor role (fact-checks increase accuracy by 0.41 scale points on a four-point scale regardless of formatting; modifications to formatting increase accuracy only by 0.03 points.) & &\\\cline{1-4}
    \citet{porter_political_2022} & Fact-checks are effective in increasing factual accuracy on realistic simulations of social media platforms (Study 1 Correction Effect d=0.55; Study 2 d=0.79) & & \\\cline{1-4}
    \citet{pretus_imisleadingi_2024} & Adding a misleading count next to the like count reduced participants' reported likelihood to share inaccurate information by 25\% compared to control condition. It was five times more effective as an accuracy nudge (misleading count compared to no intervention: d=0.20; misleading count compared to accuracy nudge: d=0.13). & &\\\cline{1-4}
    \citet{qian_fighting_2023} & Active interventions significantly increased intention of using reverse image search tools compared to passive interventions and a control group. & & Neither passive nor active interventions had an effect on credibility judgment or misinformation discernment.\\\cline{1-4}
    \citet{rich_correcting_2020} & & & When correcting misinformation, there was no evidence that the time of correction mattered for the efficacy of the correction and the participants corrected beliefs were not durable (durability of corrected belief $\eta^2$=0.43; time of correction $\eta^2$=0.02)\\\cline{1-4}
    \citet{ruffin_explaining_2022} & Simply highlighting and explaining manipulation in photos was not always effective but when it was, it did help make users less agreeing with intended messages (e.g., $\beta$=-0.58 of linear regression model for explaining the manipulation versus seeing the original image). & &Intervention was not always effective. Explanation had negative effect on feeling/sentiment toward the subject/image\\\cline{1-4}
    %\citet{safieddine_corporate_2016} & & Collaboration among organizations guided by standards is necessary to effectively use available tools and methods for text and image authentication.  &\\\cline{1-4}
    \citet{sakhnini_review_2022} & & Fact-checking apps should be sensitive to age-related, personal, and political biases & \\\cline{1-4}
    \citet{saltz_encounters_2021} & & Findings suggest strong emotional reactions to misinformation labels in general, which are perceived as overly paternalistic, biased, and punitive. & \\\cline{1-4}
    \citet{sangalang_potential_2019} & Narrative correctives (with or without emotional ending) can effectively reduce misinformation beliefs, while emotional corrective endings are better at correcting attitudes. & & \\\cline{1-4}
    \citet{schaewitz_combating_2020} & Detailed corrections presented alongside disinformation are more effective in better remembering facts compared to simple corrections ($\eta^2$=0.02)  & & The influence of detailed corrections on personal beliefs regarding the topic of the disinformation is counterproductive as more details in the correction seem to raise readers’ concerns when corrections are presented together with the disinformation.\\\cline{1-4}
    \citet{scharrer_watch_2022-1} & Warnings on top of a scientific message made laypeople hesitant about uncritically and confidently accepting the message as true. Participants agreed less with the claims and deemed the text to be less credible than without the warning ($\eta^2$=0.48) & &Warnings cannot reduce or prevent boost in persuasiveness of easily understandable misinformation.\\\cline{1-4}
    \citet{schmid_digital_2022} & & A web app based on Social Network Analysis could effectively provide an overview of potentially misleading vs. non-misleading content on Twitter, which can be explored by users and enable foundational learning. & \\\cline{1-4}
    \citet{schmid_benefits_2022} & Text-based refutations effectively reduced belief in misinformation and immunized participants against impact in short-time (final power of 94.5\% was reached to detect a small effect size. Credibility judgment after 2 months was slightly lower (d=0.04)) & Unintended effects: lacking effect on intentions, backfire-effects among religious groups, biased judgments when omitting information about vaccine side effects\\\cline{1-4}
    \citet{seo_trust_2019} & Machine-Learning-Graph warning, indicating Source Reliability, Content Truthfulness and Picture/Video Truthfulness, was effective increased participants’ sensitivity in differentiating fake from real news. ($\eta^2$=0.018) & & \\\cline{1-4}
    \citet{sharevski_i_2023} & & &Warning labels as visual frictions are not accessible for low vision or blind users.\\\cline{1-4}
    \citet{sharevski_meaningful_2022} & & SPAM warning tags are promising and increase trust in soft moderation. Text-only variant tells participants more of what is going on and a text-and-flag variant gives more specifics and is tougher to refute as a large visual cue. Warning tag with improbable interpretation of facts (FFS) gave convincing options for users to pick why the context is fitting to the misinformation tweet. Left- and right-leaning participants positively rated the intervention.&\\\cline{1-4}
    \citet{sharevski_two_2021} & & & The utterance of a warning cover before a Tweet containing valid information about COVID-19 vaccines by Alexa will not reduce the perceived accuracy of the spoken back Tweet’s content relative to a no warning cover condition (d=0.018) \\\cline{1-4}
    \citet{sherman_designing_2021} & A combination of expert and user insights is effective in defining interpretable warnings and design guidelines for communicating the provenance of video content to end-users. & & results raise concerns around the potential for users to overgeneralize misinformation warnings regarding video or text information \\\cline{1-4}
    \citet{smith_correcting_2019} & Corrective mock Facebook news feeds were effective in reducing belief in neuroscience myths when shown immediately after the misinformation for those who held incorrect beliefs at pretest. & & If participants held correct beliefs at pretest, a single exposure to misinformation (even when immediately corrected) was enough to have a negative impact on their beliefs.\\\cline{1-4}
    \citet{song_fighting_2022} & Image-only modality triggered significantly lower levels of message elaboration and heightened message credibility and increased engagement intentions (effect of evidence type on self-reported message elaboration: $\eta^2$=0.01. Effect of presentation mode on message elaboration: $\eta^2$=0.02) & Presence of statistical evidence in assertions reduced message elaboration and effects of message in correcting misperceptions, decreased perceived message credibility and lowered intentions to further engange with and disseminate the corrective message.\\\cline{1-4}
    %\citet{sotirakou_toward_2022} & & Gamification within a crowd-sourcing platform might motivate digital readers to correct disinformation, reducing the time and cost for verifying false content and minimizing its societal impact. & \\\cline{1-4}
    \citet{sullivan_leveraging_2019} & & & Libraries were not effective in correcting misconceptions about the flu vaccine through comments on social media.\\\cline{1-4}
    \citet{tanaka_exposure_2019} & Objective countermessages reduced belief in rumors and subjective countermessages strengthened false beliefs (e.g., $\eta^2$=0.02. Post-hoc power analysis revealed adequate G*Power >0.80 at medium to large effect size levels). & &Subjective countermessages even strengthened false beliefs\\\cline{1-4}
    \citet{tanaka_toward_2013} & Displaying criticism of false information prior to rumors during a disaster response is effective in increasing proportion of responses aimed at stopping the spread of rumors compared to displaying the criticism after the rumor. & & \\\cline{1-4}
    \citet{tao_hope_2023} & All three types of corrections improved belief accuracy. Corrections incorporating hope appeals showed enhanced effectiveness when threat information was present in comparison to absent hope appeals (Power analysis reveals study can detect small effect sizes (f=0.11) with power of 80\%. Hope appeal when threat was present versus absent: $\eta^2$=0.01) & &\\\cline{1-4}
    \citet{thornhill_digital_2019} & BalancedView, a proof-of-concept that shows news stories relevant to a tweet suggests that nudging users by providing context information may change the behavior of them towards that of informed news readers. & & \\\cline{1-4}
    \citet{tseng_investigating_2022} & Corrective information in the form of text, images, or videos is effective in reducing participants' perceived credibility and potential action for misinformation, with videos being particularly effective in correcting text-based misinformation. & & \\\cline{1-4}
    \citet{tsipursky_fighting_2018} & The Pro-Truth Pledge (PTP) has been shown to effectively reduce the sharing of misinformation and encourage truthful behavior on social media (d=-1.93). & & \\\cline{1-4}
    \citet{tulin_beyond_2024} & Truth sandwich in fact-check had indirect benefits such as more positively perceived intentions of fact-checkers and less reactance to reading subsequent fact-checks compared to classic fact check that repeats false claim (small effect sizes, e.g., for classic fact-check: $\beta$=-.13 & &Truth sandwich was not effective in correcting false beliefs but had indirect benefits.\\\cline{1-4}
    \citet{tully_mobilizing_2020} & Users tend to provide accurate information in corrections, particularly after viewing other corrections. However, users are generally unlikely to respond to tweets containing misinformation (exposure to corrections: $\eta^2$=0.001; tone of corrections: $\eta^2$=0.010). & & neither exposure to corrections nor tone of corrections increases the self-reported likelihood of responding to the misinformation tweet as compared to the misinformation-only condition\\\cline{1-4}
    \citet{tully_designing_2020} & News literacy messages alter misinformation perceptions, however not with a single message (e.g., partial $\eta^2$=.0.006 for NL tweet leading participants to rate stories overall as less credible than texting tweet) & & \\\cline{1-4}
    \citet{van_der_meer_can_2023} & & &Warnings can prime general distrust in authentic news.\\\cline{1-4}
    \citet{van_der_meer_seeking_2020} & Corrective information is effective in debunking misinformation, and factual elaboration compared to simple rebuttal stimulates intentions to take protective action, with government agencies and news media being more effective in improving belief accuracy compared to social peers. & & \\\cline{1-4}
    \citet{velasco_factit_2023} & &The browser extension that allows to insert text and creates a (binary) feedback based on logistic regression was rated highly acceptable in terms of functionality, reliability, usability, efficiency, maintainability, and portability.&\\\cline{1-4}
    \citet{velez_latino-targeted_2023} & Fact-checks undo effects of misinformation on beliefs (large and significant effect of over .26 scale points change). No Backfire effect was observed. & &\\\cline{1-4}
    \citet{von_der_weth_nudging_2020} & Nudging users toward more conscious posting and sharing behavior by using linguistic analysis to infer the factuality of content and the credibility of sources is effective in reducing the reach and speed of spread of misinformation. & &\\\cline{1-4}
    \citet{vraga_assessing_2021} & User corrections of a meme containing misinformation are effective in reducing the credibility assessment of the misinformation post ($\eta^2$=0.077) and misperceptions ($\eta^2$=0.088) & &Exposure to news literacy messages did not enhance the effectiveness of corrective responses or boost NL attitudes and may have generated cynicism.\\\cline{1-4}
    \citet{vraga_i_2018} & Social corrections providing a source are effective compared to not giving a source (partial $\eta^2$=0.035) & & \\\cline{1-4}
    \citet{vraga_using_2017} & Misinformation correction by expert group is effective without loosing the groups credibility and trustworthiness in the context of a health topic (misinformation correction: partial $\eta^2$=0.009; trustworthiness: partial $\eta^2$=0.001; credibility: partial $\eta^2$=0.004) & &misinformation corrections of a single user is not effective\\\cline{1-4}
    \citet{vraga_creating_2022} & Expert organizations can be effective in successfully correcting misinformation on social media on two controversial health topics & &\\\cline{1-4}
    \citet{vraga_effects_2021} & User corrections in real-time partially reduce the effect of misinformation videos on beliefs (partial $\eta^2$=0.03 compared to no intervention) but not on intentions. & &\\\cline{1-4}
    \citet{vraga_testing_2019} & Logic-based and humor-based rhetorical corrections reduce misperceptions only for some topics (partial $\eta^2$=0.013). & &\\\cline{1-4}
    \citet{vraga_testing_2020} & Logic-focused (before and after misinformation) and fact-focused (after the misinformation) corrections reduce misperceptions, with logic-focused corrections appearing to reduce the credibility of misinformation and fact-focused corrections being more credible. & &\\\cline{1-4}
    \citet{wahlheim_reminders_2020} & Reminders of misinformation are effective to diminish the negative effects of fake-news exposure short-term (d=0.29) & &\\\cline{1-4}
    \citet{waltenberger_reddit_2023} & Contextualizing user profiles with data from previous contributions helped users contextualize posts, identify political tendencies, distinguish humor from problematic mindsets (qualitatively measured) & &\\\cline{1-4}
    \citet{wang_countering_2021} & One sided narrative messages are more effective then two-sided ones for correcting misinformation on e-cigarettes & &Effect disappeared if participants had smoked e-cigarettes before\\\cline{1-4}
    \citet{wang_pm_2022} & Participants accept unwelcome fact-checks on Facebook but welcome fact-checks on Line (private messaging app). Fact-checks help increase media literacy in open platforms and hamper media literacy in private messaging apps. & &\\\cline{1-4}
    \citet{westbrook_mechanisms_2023} & External correction (news source labeling misinformation as false) influences perceptions of misinformation source. Perceptions of the misinformation source can cause changes in belief in misinformation. (a priori power analysis allowed for desired power of 0.8) & &\\\cline{1-4}
    \citet{wijnker_debunking_2022} &All investigated correction methods for misleading graphs were effective for debunking misinformation directly after correction and reduces over time. Showing an accurate alternative graph was more effective than visual cues or text-based warning cues to activate graph literacy or warning messages for possible deceit.& &\\\cline{1-4}
    \citet{wood_promoting_2023} & Debunking messages of healthcare professional lead to increase in beliefs about risks of vaccines in the UK but not the US. Messages from political authorities and discrediting messages had no effect. There is a joint importance of message source and messaging strategy regarding effectiveness of debunking (e.g., debunking by health experts reduced belief that vaccines cause severe side effects by 0.19 points on Likert scale) & &\\\cline{1-4}
    \citet{zade_tweet_2023} &  &Tweet trajectory (e.g., unfamiliar activity invokes skepticism in following network) and contextual cues (e.g., profile description helps infer purpose of account) helped support users in assessing credibility (qualitatively evaluated). &\\\cline{1-4}
    \citet{zhao_misinformation_2019} & Participants exhibit a more positive attitude towards corrective messages and have higher vaccination certainty when such messages are present across multiple social media platforms, as opposed to only one platform. & & \\\cline{1-4}
    \citet{zhang_investigation_2022} & Concise corrections are more effective than exhaustive ones. Graphical explanation has small positive effect (e.g., Spearman's $\rho$=0.126). & &Textual explanations for why misinformation is wrong do not significantly affect effectiveness. Warnings in a tough tone make corrections worse. Textual and graphic warnings have negative associations with correction effectiveness.\\\cline{1-4}
    \citet{zheng_evaluating_2022} &Explanatory annotations and interactive linking in misinformation combining text and visualizations can significantly lower perceived credibility (e.g., d=-0.367). The effect to raise awareness is limited/marginal while linking was more effective than annotation (e.g., d=-0.367) & &\\
    
% Add more rows as needed
\end{longtable}

\end{landscape}

%%
%% End of file `sample-manuscript.tex'.

\end{document}